\documentclass[manuscript]{aastex62}
\usepackage[encapsulated]{CJK}
\usepackage{subfigure}
\usepackage{mathrsfs}
\usepackage{amsmath}
\usepackage{epstopdf}
\usepackage{longtable}
\usepackage{booktabs}
\usepackage{tablefootnote}
\usepackage{threeparttable}
\usepackage{multirow}
\usepackage{hyperref} %\usepackage{threeparttable}
\def\etal {et al.~}

\newbox\grsign \setbox\grsign=\hbox{$>$} \newdimen\grdimen \grdimen=\ht\grsign
\newbox\laxbox \newbox\gaxbox
\setbox\gaxbox=\hbox{\raise.5ex\hbox{$>$}\llap
	{\lower.5ex\hbox{$\sim$}}}\ht1=\grdimen\dp1=0pt
\setbox\laxbox=\hbox{\raise.5ex\hbox{$<$}\llap
	{\lower.5ex\hbox{$\sim$}}}\ht2=\grdimen\dp2=0pt
\shorttitle{Light Deflection on Astrometry}
\shortauthors{Li \etal}

\begin{document}
\begin{CJK*}{UTF8}{gbsn}
\title{The Effect of Light Deflection by Solar System Objects on High-Precision SKA Astrometry}
		
\correspondingauthor{Yingjie Li}
\email{liyj@pmo.ac.cn, xuye@pmo.ac.cn}
		
\author{Yingjie Li}\affiliation{Purple Mountain Observatory, Chinese Academy of Sciences, Nanjing 210023, China}
\author{Ye Xu}\affiliation{Purple Mountain Observatory, Chinese Academy of Sciences, Nanjing 210023, China}
\affiliation{University of Science and Technology of China, Hefei, Anhui 230026, China}
	
\author{Shaibo Bian}
\affiliation{Purple Mountain Observatory, Chinese Academy of Sciences, Nanjing 210023, China}
\affiliation{University of Science and Technology of China, Hefei, Anhui 230026, China}
		
\author{ZeHao Lin}
\affiliation{Purple Mountain Observatory, Chinese Academy of Sciences, Nanjing 210023, China}
\affiliation{University of Science and Technology of China, Hefei, Anhui 230026, China}		
		
\author{JingJing Li}
\affiliation{Purple Mountain Observatory, Chinese Academy of Sciences, Nanjing 210023, China}
		
\author{DeJian Liu}
\affiliation{Purple Mountain Observatory, Chinese Academy of Sciences, Nanjing 210023, China} 
\affiliation{University of Science and Technology of China, Hefei, Anhui 230026, China}  
		
\author{Chaojie Hao}
\affiliation{Purple Mountain Observatory, Chinese Academy of Sciences, Nanjing 210023, China}	
\affiliation{University of Science and Technology of China, Hefei, Anhui 230026, China}    	
		
\begin{abstract}
We have computed the deflection angles caused by 195 objects in the solar system, including 177 satellites and eight asteroids. Twenty-one satellites and six asteroids can bend light from distant compact extragalactic sources by more than 0.1 $\mu$as, and fourteen satellites and the asteroid Ceres can deflect light by more than 1.0 $\mu$as. We calculated the zones and durations of perturbations posed by the gravitational fields of five planets (excluding Earth, Jupiter, and Saturn), Pluto, and Ceres, where the perturbations would affect astrometry measured with the Squared Kilometre Array (SKA). Perturbed zones with deflection angles larger than 0.1 and 1.0 $\mu$as appear as ribbons. Their widths range from dozens of degrees for Uranus, Neptune, and Venus to several degrees or less for other objects at 0.1 $\mu$as, and from $\sim$ 16$^{\circ}$ for Venus to several degrees or less for other objects at 1.0 $\mu$as. 
From the calculated perturbation durations, the influence of the gravitational fields of selected objects can be divided into four levels: hardly affect SKA astrometry (I), may have little effect (II), may have a great effect (III) on single-epoch astrometry, and may greatly affect both single- and multi-epoch astrometry (IV). The objects corresponding to these levels are Ceres (I), Pluto (II), Mercury and Mars (III), and other objects (IV).
\end{abstract}
\keywords{gravitation — quasars — relativity —
			techniques: interferometric}

\section{Introduction} 

Testing the theory of general relativity (GR) by observing the light deflection by the Sun  \citep{Dyson+1920} made GR famous and initiated the ``modern era'' of astronomy \citep{Johnson1983}. 
GR has been tested over 100 years, especially in terms of gravitational light deflection, relativistic perihelion advance, and gravitational redshift. Complementary to light deflection is the Shapiro time delay \citep[see][]{Shapiro1964}. Celestial bodies in the solar system (which define Barycentric Celestial Reference System, BCRS) act as gravitational sources and can play an important role in GR tests \citep{Ivanitskaia+1986, Soffel1989, Damour+1991, Damour+1992, Damour+1993, Damour+1994, Hees+2014, Will2014, Crosta+2017, Ni2017, Bernus+2019, Crosta2019}. One important parameter to be tested is the parameterized post-Newtonian (PPN) parameter, $\gamma$, which is measuring the amount of the curvature produced by a unit mass-energy and is equal to unity in GR \citep{Damour1989, Poisson-Will2014, Will2014}. The accuracy of $\gamma$ obtained by measurements has rapidly improved in recent years \citep[for more details see the review by][]{Will2015}. The highest accuracy of $\gamma$ is $\sim 2\times10^{-5}$, determined via data obtained with the Cassini spacecraft \citep{Shapiro1964,Bertotti+2003}. The latest accuracy of $\gamma$ obtained using the very long baseline interferometry (VLBI) technique has reached $\sim 9\times10^{-5}$ \citep{Titov+2018}. Planets, such as Jupiter, have also been used to test GR \citep[e.g.,][]{Fomalont-Kopeikin2003, Fomalont-Kopeikin2008, Abbas+2022, Li+2022}. Light deflection and the underlying theory (e.g., relativistic gravity, gravitational lensing theory, etc.) have become important tools for both astronomy and cosmology \citep{Will2015}.

Measurements of $\gamma$ with high accuracy are intertwined with high-precision astrometry. Obtaining a high accuracy of $\gamma$ benefits from the rapid improvement of the relevant technologies and instruments. Ultra-precise ($\sim$ 1 $\mu$as or higher) and ultra-sensitive \citep[e.g., several hundred nJy/beam @ 9.2 GHz for an 8 hr exposure with the Square Kilometre Array, SKA,][]{Bonaldi+2021} astrometry have been the immediate goals for SKA \citep{Braun+2015}, the next-generation Very Large Array \citep[ngVLA,][]{Murphy+2018}, and their pathfinders \citep{Rioja-Dodson2020}, and also Gaia Mission from space \citep{Vecchiato+2003, Butkevich+2022}. Such improvements together with other laboratories and space experiments (e.g., Solar System Odyssey, \citealt{Christophe+2009}; the series of Astrodynamical Space Test of Relativity using Optical Devices, ASTROD, \citealt{Ni1998, Ni2009}; Astrometric Science and Technology Roadmap for Astrophysics, ASTRA, \citealt{Gai+2020}; Transiting Exoplanet Survey Satellite mission, TESS, \citealt{Gai+2022}, etc.; see more experiments in \citealt{Ni2017}) have the potential to yield another 3--4 orders of magnitude of precision in testing GR \citep{Ni2017}. %Behind the glitter, the light deflection caused by the objects in the solar system is a huge challenge in astrometry.
Development of theoretical study of gravity is in turn vital to future ultra-precise and ultra-sensitive astrometry \citep{Sekido-Fukushima2006, Pertit-Luzum2010, Li+2022}.

High-precision astrometry has played a leading role in revealing the spiral structure of the Milky Way at both radio \citep[e.g.,][]{Xu+2006, Reid+2009arm, Xu+2016, Reid+2019} and optical \citep[e.g.,][]{Xu+2018raa, Xu+2018, Xu+2021, Poggio+2021} wavelengths, and in producing the 3rd realization of the International Celestial Reference Frame \citep[ICRF3,][]{Charlot2020} and the Gaia DR2/(E)DR3 Celestial Reference Frame \citep[Gaia-CRF2/CRF3,][]{Gaia-Collaboration+2018,Gaia-Collaboration+2022b}. As the astrometric precision is stepping into 10 $\mu$as or better (e.g., the recorded parallax precision has reached $\pm$ 3 $\mu$as by \citealt{Zhang+2013} using the Very Long Baseline Array and the predicted parallax precision is 10 $\mu$as for GAIA data release 4, \citealt{Gaia-Collaboration+2022}), the revealed structures of the Milky Way would be extended to distant regions beyond $\sim$ 10 kpc. High-precision astrometry is also helpful for studying notable VLBI-Gaia positional offsets \citep[i.e., the offsets between measured optical and radio positions;][]{Mignard+2016, Kovalev+2017, Petrov+2017, Charlot2020}.
High-precision astrometry can also shed light on other broad sources and problems in astrophysics, e.g., providing % resolved images of 
electromagnetic counterparts of gravitational wave sources, monitoring the orbit of binaries and analyzing their physical parameters, investigating the evolution of galaxies, etc. \citep[see][]{Reid+2019baas}.

High-precision astrometry suffers from the effects posed by the gravitational fields of objects in the solar system. For instance, the predicted deflection angle (according to the predictions of GR) caused by the planets, Pluto, and even large satellites exceeds several $\mu$as and can even reach a dozen of mas \citep[see][]{Fienga+2011, Hees+2014a, Bertone+2014, Crosta+2015, Crosta+2017}. The critical angular distances between compact extragalactic sources (CESs, hereafter) and solar system objects (e.g., planets and the Moon), at which the deflection angles are still $\geq$ 1 $\mu$as, are larger than several arcminutes or even tens of degrees or more \citep[e.g.,][]{Crosta-Mignard2006, Li+2022}. 

As a pilot work, we have evaluated the effect of light deflection caused by objects in the solar system on high-precision SKA astrometry. This evaluation may be helpful in saving computing resources for the vast data volumes expected to be collected by SKA \citep[e.g.,][]{Quinn+2015}. In addition, this work may contribute to the selection of interesting region and time to test GR or PPN parameters via light deflections or the Shapiro time delay, especially when there are unknown weak sources that can only be detected by the ultra-sensitive SKA \citep[e.g.,][]{Bonaldi+2021}. Those interesting region and time may also be helpful in broader researches \citep{Li+2022}, e.g., measuring additional Shapiro delay relative to post-Newtonian corrections \citep[see][]{Will2014}, testing and developing multi-plane lensing effects \citep{Subramanian-Chitre1984, Erdl-Schneider1993, Ramesh+2021}, higher-order (e.g., second-order) PPN formalisms \citep{Crosta-Mignard2006, Kopeikin-Makarov2007}, etc. 

The remainder of this paper is organized as follows. Section \ref{sec-delfection-all-solar-system} overviews the deflection angle caused by 195 objects in the solar system, using the sample given in Section \ref{sec-sample}. In Section \ref{sec:impact-regions-start}, we discuss the zones and durations of perturbations posed by the gravitational fields of objects that revolve around the Sun and evaluate the impact of those objects on SKA astrometry. Finally, we summarize our conclusions in
Section \ref{sec-summary}.

\section{Sample}\label{sec-sample}

The sample includes 195 objects in the solar system; i.e., the Sun, all eight planets, Pluto, 177 satellites (including the Moon), and eight asteroids with $GM > 0.1$ km$^3$ s$^{-2}$, where $G$ and $M$ are the gravitational constant and the mass of the solar system object, respectively. The parameters of the Sun were obtained from IAU 2015 Resolution B3 \citep[see][]{IAU2015}. \footnote{See \url{https://www.iau.org/administration/resolutions/general_assemblies/}}The physical and orbital parameters of the planets, Pluto, satellites, and asteroids were obtained from the Jet Propulsion Laboratory (JPL, hereafter). \footnote{Planets and Pluto: physical parameters see \url{https://ssd.jpl.nasa.gov/?planet_phys_par}; orbital parameters see \url{https://ssd.jpl.nasa.gov/planets/approx_pos.html}. Satellites: physical parameters see \url{https://ssd.jpl.nasa.gov/?sat_phys_par}; and orbital parameters see \url{https://ssd.jpl.nasa.gov/?sat_elem}. The parameters of the asteroids were obtained from \url{https://ssd.jpl.nasa.gov/tools/sbdb_query.html}. The references therein are not listed here because they are very numerous, and we kindly advise the reader to the consult the various references found on the aforementioned websites.}We replaced the orbital parameters of the satellites (other than the Moon) with that of their host planets or Pluto because the ratios of the semi-major axis of the orbital revolution for these satellites and that of the corresponding planets or Pluto, $R_{\mathrm{sp}}$, are small; i.e., all of them are $\lesssim 0.04$ (see Table \ref{tab:original-para}). The other parameters required in this work are also listed in Table \ref{tab:original-para}.

\section{Deflection Angle Caused by Solar System Objects}\label{sec-delfection-all-solar-system}

\subsection{Orbits of the Planets}\label{sec:orbit}

To obtain the positions of the planets, Pluto, and asteroids, it is convenient to use the heliocentric plane orbital coordinate system. Assuming that a planet moves along a fixed elliptical orbit, the position of the planet with eccentric anomaly, $E$, is
\begin{equation}\label{equ:orbit}
	x = a(\cos E - e); \;\; y = a \sqrt{1 - e^2} \sin E; \;\; z = 0,
\end{equation}
where $a$ is the semi-major axis, $e$ is the eccentricity (see Table \ref{tab:original-para}), and $z = 0$ represents that the planet is located in its orbital plane. The calculated coordinates can be transformed into the coordinates in the ecliptic coordinate system ($x_{\mathrm{ecl}}$, $y_{\mathrm{ecl}}$, $z_{\mathrm{ecl}}$) by
\begin{equation}\label{equ:ecl} 
	\left\{
	\begin{aligned}
		x_{\mathrm{ecl}} & = (\cos \omega \cos \Omega - \sin \omega \sin \Omega \cos I) \;\; x & + & \;\;(- \sin \omega \cos \Omega - \cos \omega \sin \Omega \cos I) \;\;y \\ 
		y_{\mathrm{ecl}} & = (\cos \omega \sin \Omega + \sin \omega \cos \Omega \cos I) \;\;x & + & \;\; (- \sin \omega \sin \Omega + \cos \omega \cos \Omega \cos I) \;\; y, \\
		z_{\mathrm{ecl}} & = (\sin \omega \sin I) \;\; x & + & \;\; (\cos \omega \sin I) \;\; y
	\end{aligned}
	\right.
\end{equation}
%\begin{equation}\label{equ:ecl} 
%	\left\{
%	\begin{aligned}
%		x_{\mathrm{ecl}} & =  (\cos \omega \cos \Omega - \sin \omega \sin \Omega \cos I) \;\; x  \\
%		 &  + (- \sin \omega \cos \Omega - \cos \omega \sin \Omega \cos I) \;\;y \\ 
%		y_{\mathrm{ecl}} & =  (\cos \omega \sin \Omega + \sin \omega \cos \Omega \cos I) \;\;x \\
%		&  + (- \sin \omega \sin \Omega + \cos \omega \cos \Omega \cos I) \;\; y, \\
%		z_{\mathrm{ecl}} & =  (\sin \omega \sin I) \;\; x \\
%		&  + (\cos \omega \sin I) \;\; y
%	\end{aligned}
%	\right.
%\end{equation}
where $I$, $\omega$, and $\Omega$ are the inclination, perihelion, and longitude of the ascending node (see Table \ref{tab:original-para}), respectively. These three parameters are valid for the time interval 1800 AD--2050 AD. The computations below are based on the rough orbital determinations derived using the above formulas.

\subsection{Light Deflection and Its Impact}\label{sec:deflection}

\subsubsection{Calculation of Light Deflection}

Light will bend when it passes a massive body. When the impact parameter, $b$, is far less than the distance from the body to both the Earth (observer) and target source, the deflection angle, $\alpha$, can be calculated as follows \citep{Hosokawa+1993}
\begin{equation}\label{equ:alpha1}
	\alpha = (1+\gamma)\frac{2GM}{c^2 b},
\end{equation}
where $\gamma$ is the PPN parameter, which is $= 1$ for GR. When $b$ does not satisfy the described condition above, the expression of $\alpha$ \citep{Cowling1984, Will1993, Ni2017} is
\begin{equation}\label{equ:alpha2}
	%\alpha=\frac{4GM}{c^2 b}
	\alpha  = (1+\gamma)\frac{GM}{c^2 b}(\cos \theta_1 - \cos \theta_0),
\end{equation}
where $\theta_0$ (and $\theta_1$) is the angle between the light propagation vector and the CES \citep[and the Earth; see schematic diagrams of light deflection under the gravitational field of a lens in Figure 3 of][]{Li+2022}. We denote Equation (\ref{equ:alpha1}) as ``Simplified'' and Equation (\ref{equ:alpha2}) as ``Normal'' because Equation (\ref{equ:alpha2}) shows a wider range of applications. $\theta_0 \approx 180^{\circ}$ if the observed target is a distant CES, $b = r \sin \beta$ \citep[see][]{Cowling1984} and $\theta_1 \approx \alpha + \beta$, where $r$ is the distance from the observer to the celestial body, and $\beta$ is the angle between the CES (in the absence of gravitational or aberrational bending) and the celestial body as seen by the observer. When $\theta_1$ approaches 0$^{\circ}$, Equation (\ref{equ:alpha2}) approaches Equation (\ref{equ:alpha1}). When the celestial body is a solar system object, it is valid to replace $\theta_1$ with $\beta$ under a precision of $0.1$ $\mu$as \citep[see][]{Li+2022}. 

The maximum and minimum distances (i.e., $r_\mathrm{max}$ and $r_\mathrm{min}$, respectively) from the Earth to the celestial body are calculated based on the orbit determination in Section \ref{sec:orbit}. The corresponding angular radii of a celestial body seen from the Earth are $\Theta_{\mathrm{min}}$ and $\Theta_{\mathrm{max}}$ (see Table \ref{tab:betas}), respectively, assuming that the physical size of the celestial body is its mean radius. For satellites other than the Moon, their values of $r_\mathrm{max}$, $r_\mathrm{min}$, $\Theta_{\mathrm{min}}$, and $\Theta_{\mathrm{max}}$ are replaced with the corresponding values of their primary component (including Pluto). 

In this work, every solar system object is regarded as an isolated individual lens. The reasons are as follows. Firstly, in the process of the data correlator, a necessary step that all radio interference data should be conducted, %time delay that includes general relativistic delay would be corrected. T
the total corrected general relativistic time delay is the linear superposition of the time delay of each individual lens in the solar system \citep[see][]{Pertit-Luzum2010}. Secondly, it is showed that the effect of uncertainty of $\beta$ (assumed to be 10 mas, which can be regarded as the combined deflection angle caused by other objects) would be rapidly decreased to less than 0.1 $\mu$as with the increase of $\beta$ for Saturn and Jupiter \citep[see figure 7 of][]{Li+2022}. These facts indicate that it is reasonable to regard every solar system object as an isolated individual lens. The combined deflection angle may not affect the results of individual, and could be a linear superposition of the deflection angle \citep[as a vector, e.g.,][]{Li+2022} of each individual lens in this work. Further study considering multi-plane lensing effects would be conducted to accurately determine the range of application of ``linear superposition of the deflection angle''.

\subsubsection{Maximum Deflection Angle}\label{sec-alpha-max}

The maximum values of $\alpha$ calculated by Equation (\ref{equ:alpha1}), where $b$ is the mean radius of the celestial body (see Table \ref{tab:original-para}), are consistent with those calculated by \citet{Crosta-Mignard2006}. All planets can deflect light by up to $\sim$ 82.0 $\mu$as or more. Pluto can also bend light by $\sim$ 7.0 $\mu$as. The value of $\alpha_{\max}$ for asteroid Ceres is 1.2 $\mu$as. This asteroid is the only one with $\alpha_{\max} > 1.0$ $\mu$as. The number of asteroids with $\alpha_{\max} > 0.1$ $\mu$as is six. 

Large satellites can also bend light by several or dozens of $\mu$as. The number of satellites with $\alpha_{\max} > 1.0$ $\mu$as is 14, and they are the Moon, Ganymede, Callisto, Io, Europa (four satellites of Jupiter), Titan, Rhea, Iapetus, Dione (four satellites of Saturn), Titania, Oberon, Ariel, Umbriel (four satellites of Uranus), Triton (a satellite of Neptune), and Charon (a satellite of Pluto). The number of satellites with $\alpha_{\max} > 0.1$ $\mu$as is 21, i.e., the Moon, four, seven, five, three, and one satellite(s) of Jupiter, Saturn, Uranus, Neptune, and Pluto, respectively. $R_{\mathrm{sp}} < 3\%$ and $< 0.2\%$ for satellites other than the Moon with $\alpha_{\max}$ $\geq$ 0.1 and $\geq$ 1.0 $\mu$as, respectively. Therefore, replacing $r_{\mathrm{max}}$ and $r_{\mathrm{min}}$ of these satellites with their primaries does not affect the conclusions of this work. 

\subsubsection{Impact Ranges}\label{sec-alpha-beta-impact}

\begin{figure*}[!htb]
	\centering
	%	\subfigbottomskip=-0.3cm
	%	\subfigcapskip=-0.3cm
	\subfigure[Jupiter with Maximum Distance from Earth]{\includegraphics[width=0.49\textwidth]{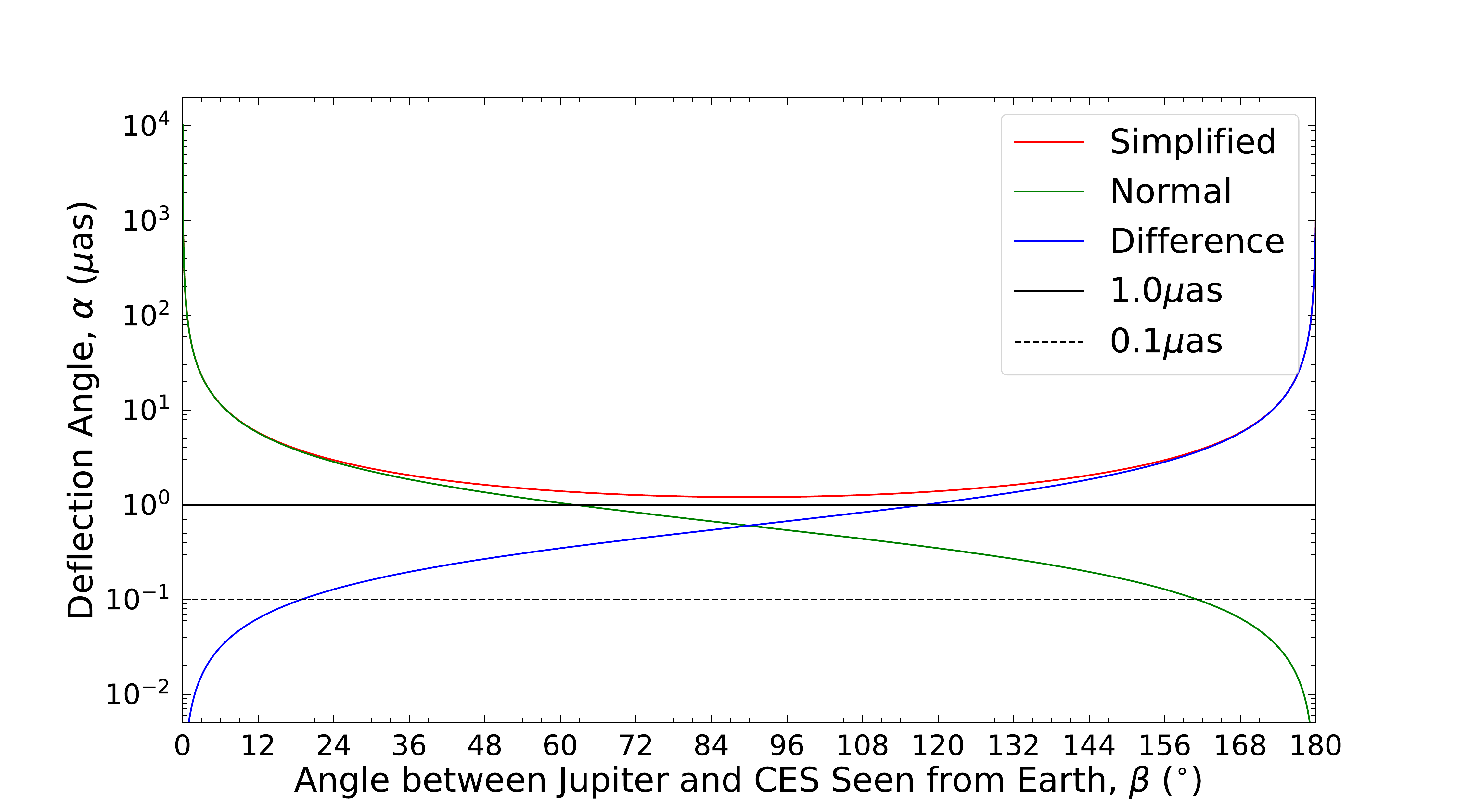}}
	\subfigure[Jupiter with Minimum Distance from Earth]{\includegraphics[width=0.49\textwidth]{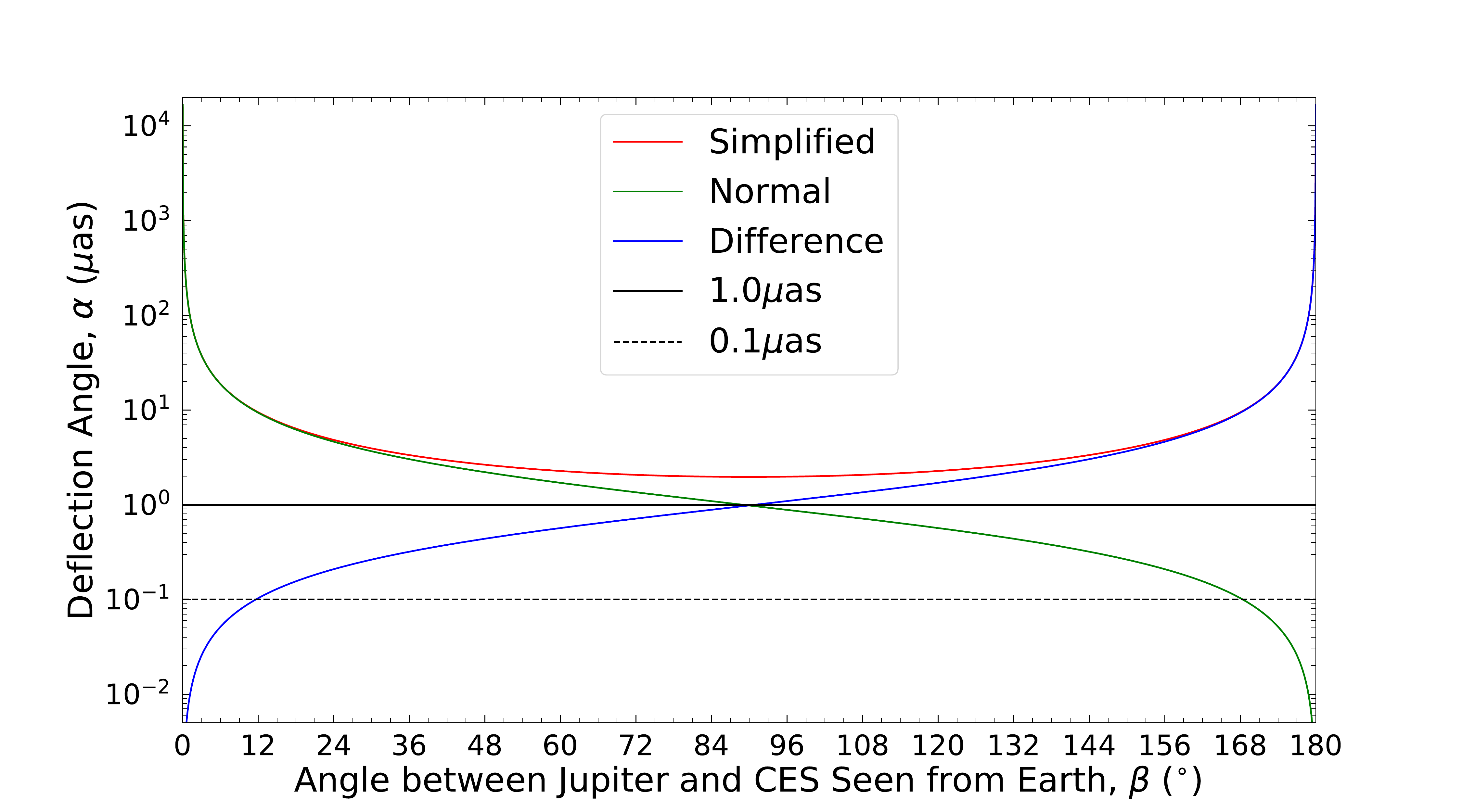}}	
	\subfigure[Mars with Maximum Distance from Earth]{\includegraphics[width=0.49\textwidth]{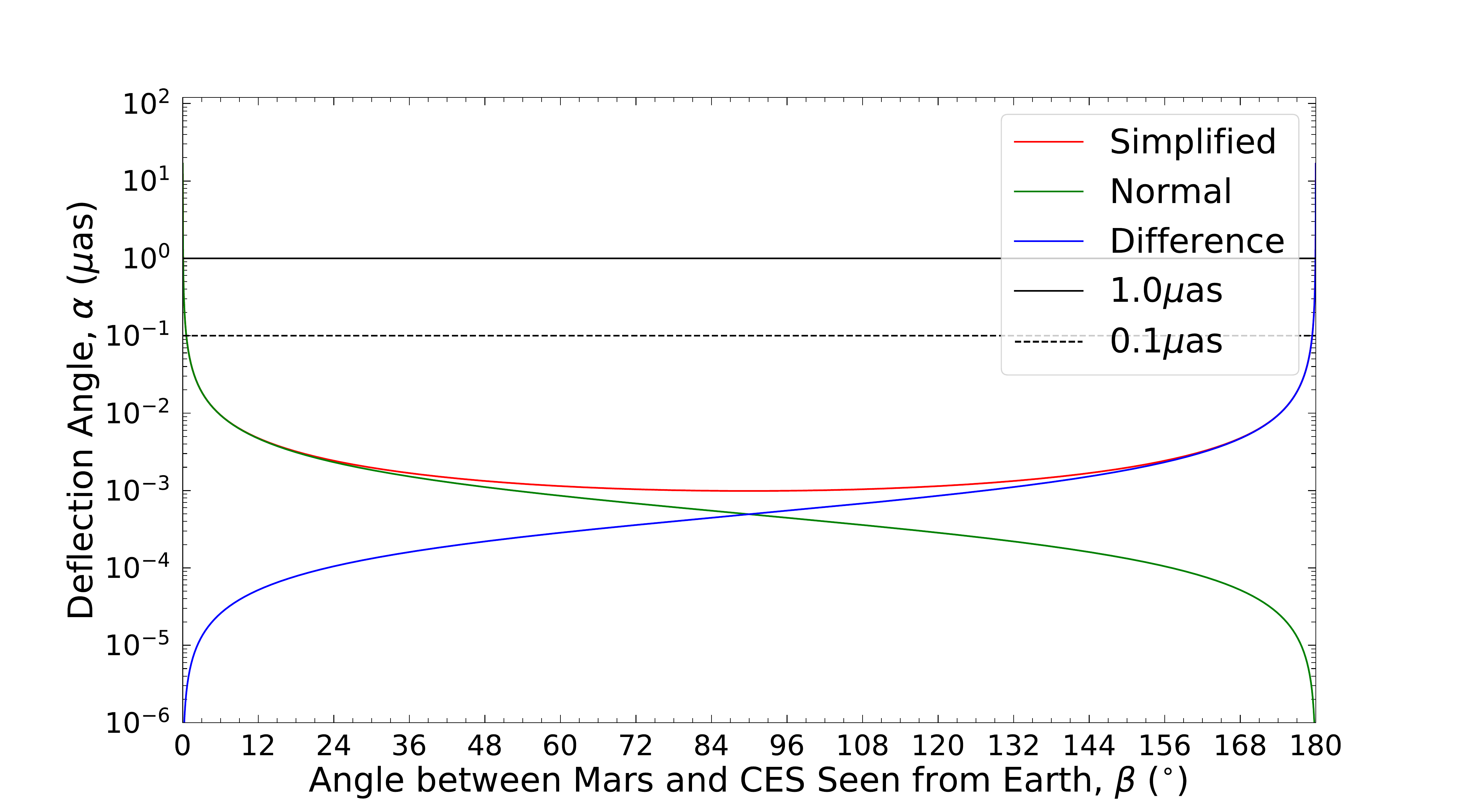}}
	\subfigure[Mars with Minimum Distance from Earth]{\includegraphics[width=0.49\textwidth]{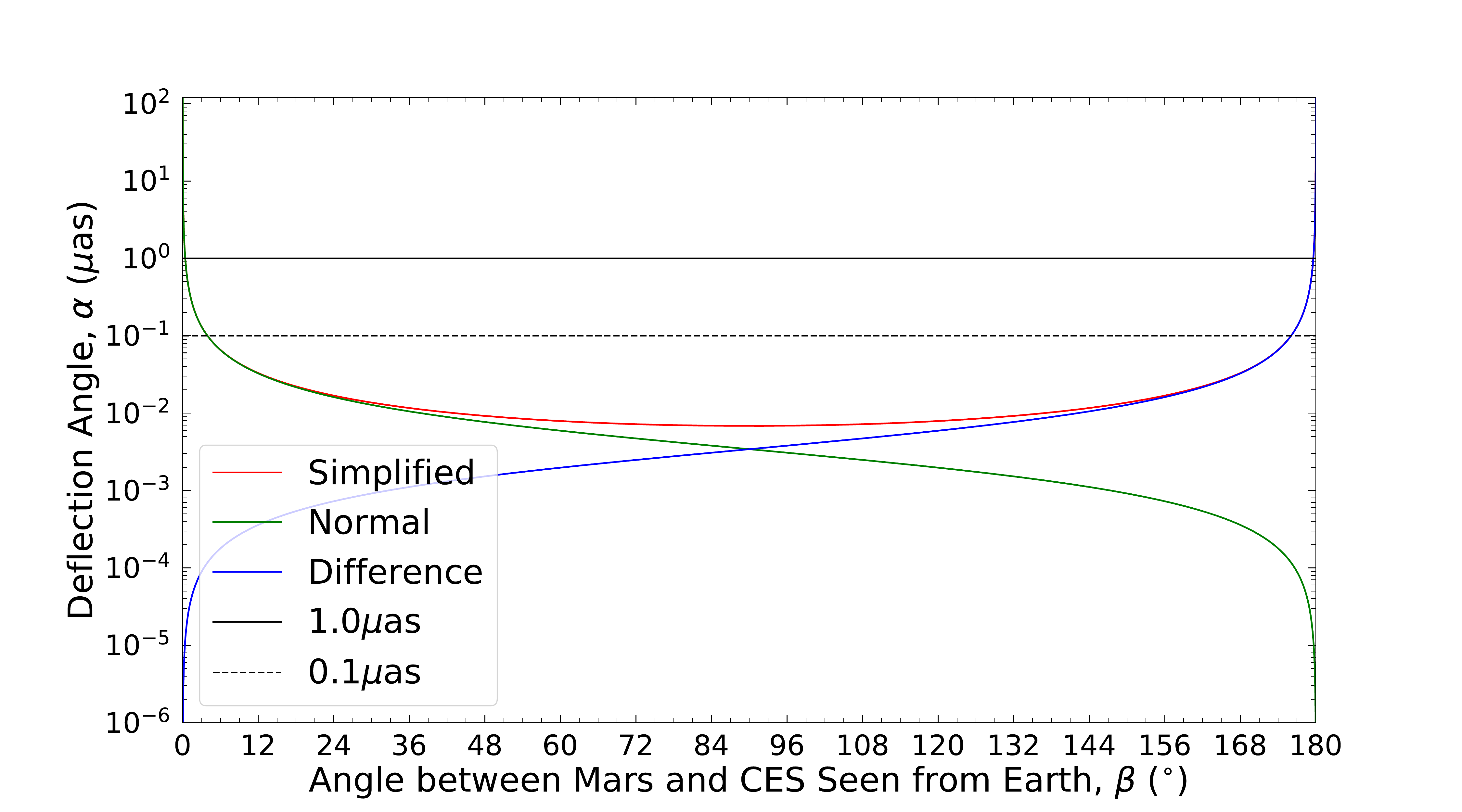}}
	\subfigure[Sun with Minimum Distance from Earth]{\includegraphics[width=0.49\textwidth]{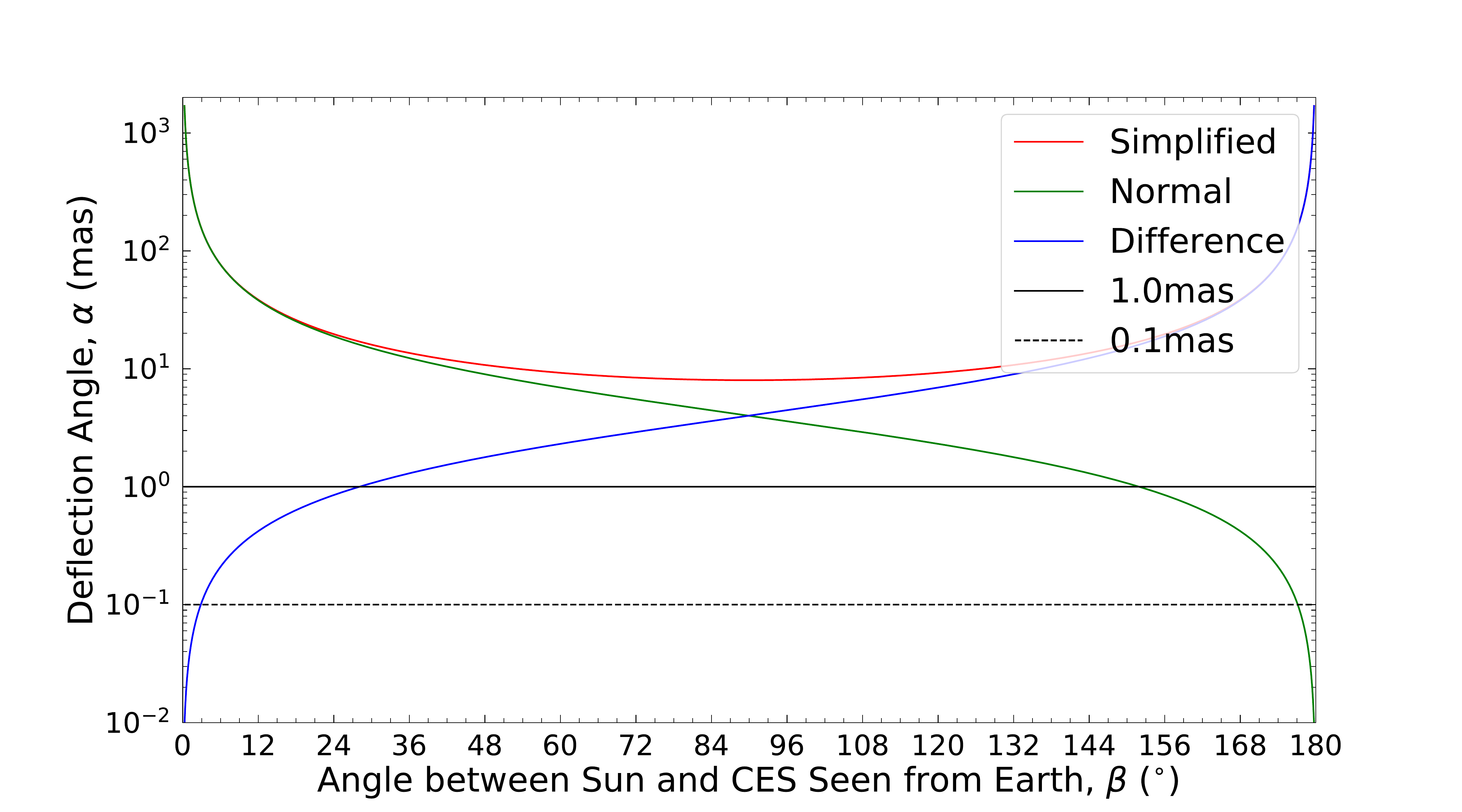}}	
	\subfigure[Sun with Maximum Distance from Earth]{\includegraphics[width=0.49\textwidth]{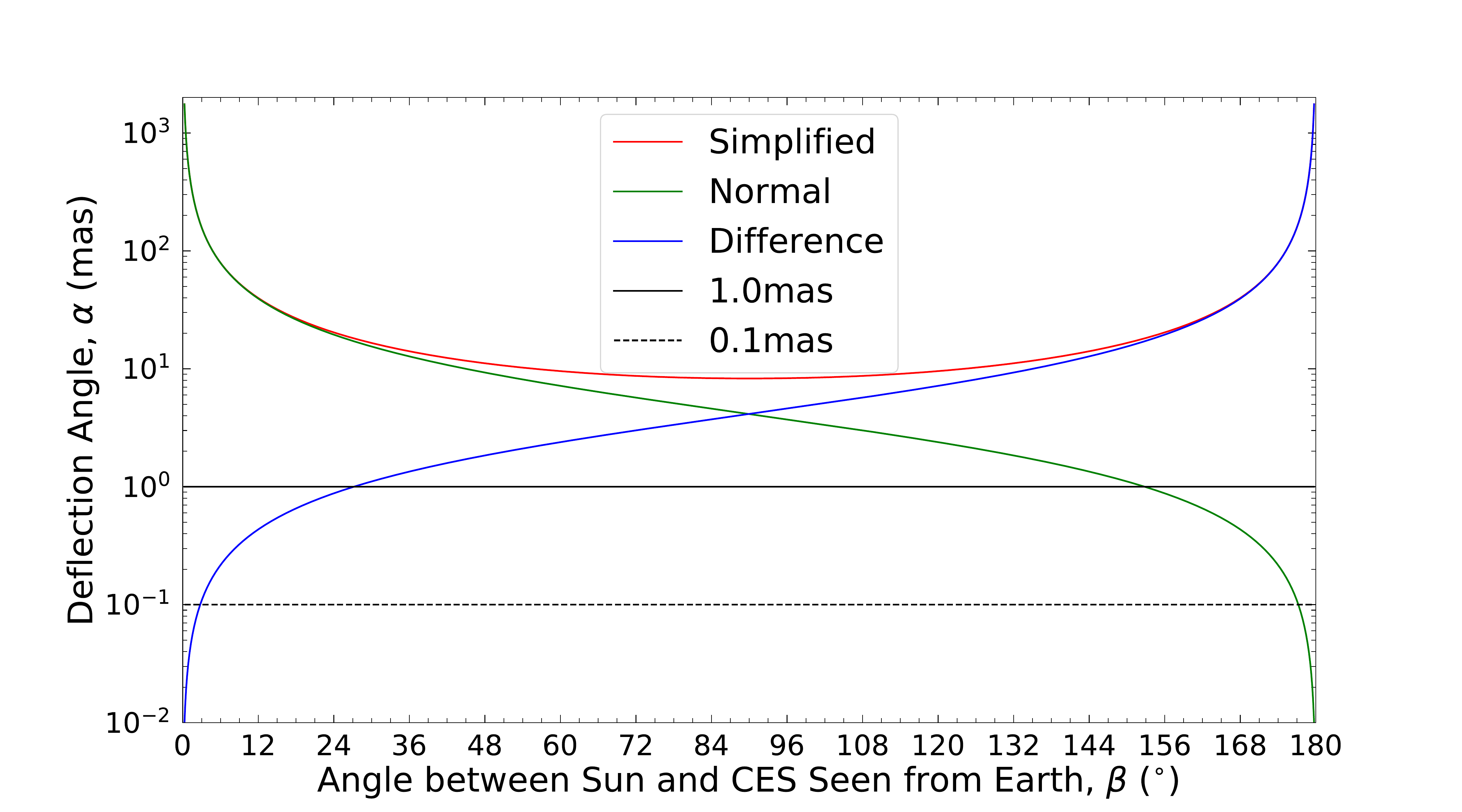}}	
	\caption{$\alpha$ as a function of $\beta$ under $r_\mathrm{max}$ (left) and $r_\mathrm{min}$ (right) for Jupiter (top), Mars (middle), and the Sun (bottom), respectively. The red and green lines present the results computed from Equations (\ref{equ:alpha1}) and (\ref{equ:alpha2}), respectively, and the blue line shows the difference between the two results (see Section \ref{sec-alpha-beta-validity}). The solid and dashed black lines denote $\alpha$ values of 1.0 and 0.1 $\mu$as, respectively, for Jupiter and Mars, and 1.0 and 0.1 mas for the Sun, respectively.}
	\label{fig-alpha-position-rely}
\end{figure*}

Figure \ref{fig-alpha-position-rely} presents $\alpha$ as a function of $\beta$ for three objects, i.e., Jupiter, Mars, and the Sun under both $r_\mathrm{max}$ (left) and $r_\mathrm{min}$ (right). The critical value of $\beta$ where $\alpha$ is equal to a given value (e.g., 1.0 $\mu$as) under a given condition (e.g., $r_\mathrm{min}$), can be regarded as the impact range of an event of light deflection, and thus is denoted as $\beta$-impact for short hereafter. For instance, $\beta$-impact for $\alpha$ = 1.0 $\mu$as under $r_\mathrm{min}$, denoted as $\beta_{1.0, \mathrm{min}}$, indicates that a celestial body can deflect light by amount of 1.0 $\mu$as up to $\leq \beta_{1.0, \mathrm{min}}$. This naming convention also applies to other given values of $\alpha$ or conditions; e.g., $\beta_{1.0, \mathrm{max}}$, $\beta_{0.1, \mathrm{min}}$, and $\beta_{0.1, \mathrm{max}}$ represent the critical values of $\beta$ when the deflection angle equals 1.0 $\mu$as under $r_{\mathrm{max}}$, 0.1 $\mu$as under $r_{\mathrm{max}}$, and $r_{\mathrm{min}}$, respectively; and $\beta_{1.0}$ and $\beta_{0.1}$ denote the critical values of $\beta$ when the deflection angle equals 1.0 and 0.1 $\mu$as, respectively.

Table \ref{tab:betas} lists $\beta$-impact for more planets, satellites, and asteroids, where only objects with $\alpha_\mathrm{max} > 0.005$ $\mu$as and $\beta_{0.1, \mathrm{max}} > \Theta_\mathrm{max}$ have been presented. The results of $\beta$-impact for the planets, Pluto, and the Moon are consistent with those reported by \citet{Crosta-Mignard2006}. 

\begin{deluxetable*}{lccccccccccc}
	\tablecolumns{12}
	%	\tablenum{1}
	\tabletypesize{\footnotesize}
	\setlength\tabcolsep{3pt}
	%	\tablewidth{0pt}
	\renewcommand{\arraystretch}{1.2}
	\tablecaption{$\beta$-Impact and $\beta$-Validity \label{tab:betas}}
	%\begin{tabular}
	\tablehead{
		\colhead{Objects} & \colhead{Index} & \colhead{$\Theta_{\mathrm{min}}$} & \colhead{$\Theta_{\mathrm{max}}$}　& \colhead{$\beta_\mathrm{1.0, min}$} & \colhead{$\beta_\mathrm{1.0, max}$} & \colhead{$\beta_\mathrm{0.1, min}$} &  \colhead{$\beta_\mathrm{0.1, max}$} & \colhead{$\beta_\mathrm{d, 1.0, min}$}  & \colhead{$\beta_\mathrm{d, 1.0, max}$} & \colhead{$\beta_\mathrm{d, 0.1, min}$} & \colhead{$\beta_\mathrm{d, 0.1, max}$} \\
		\colhead{} & \colhead{} & \colhead{($''$)} & \colhead{($''$)} & \colhead{($''$)} & \colhead{($''$)} & \colhead{($''$)} & \colhead{($''$)} & \colhead{($^{\circ}$)} & \colhead{($^{\circ}$)} & \colhead{($^{\circ}$)} & \colhead{($^{\circ}$)}
	}
	\startdata
	Sun	&	1	&	943.45 	&	975.52 	&	647896 	&	647900 	&	647989 	&	647990 	&	0.26 	&	0.27 	&	0.26 	&	0.27 	\\
	Mercury	&	2	&	2.32 	&	6.13 	&	191 	&	507 	&	1920 	&	5078 	&	179.95 	&	179.86 	&	179.47 	&	178.59 	\\
	Venus	&	3	&	4.81 	&	31.58 	&	2368 	&	15555 	&	23661 	&	148812 	&	179.34 	&	175.68 	&	173.43 	&	138.66 	\\
	Moon	&	5	&	883.38 	&	987.00 	&	22859 	&	25534 	&	208923 	&	228881 	&	173.65 	&	172.91 	&	121.97 	&	116.42 	\\
	Mars	&	6	&	1.75 	&	12.16 	&	203 	&	1410 	&	2035 	&	14102 	&	179.94 	&	179.61 	&	179.43 	&	176.08 	\\
	Jupiter	&	9	&	14.93 	&	24.39 	&	223618 	&	320605 	&	580120 	&	606206 	&	117.88 	&	90.94 	&	18.86 	&	11.61 	\\
	Ganymede	&	10	&	0.56 	&	0.92 	&	19 	&	31 	&	194 	&	316 	&	179.99 	&	179.99 	&	179.95 	&	179.91 	\\
	Callisto	&	11	&	0.51 	&	0.84 	&	14 	&	23 	&	141 	&	230 	&	180.00 	&	179.99 	&	179.96 	&	179.94 	\\
	Io	&	12	&	0.39 	&	0.64 	&	11 	&	19 	&	116 	&	191 	&	180.00 	&	179.99 	&	179.97 	&	179.95 	\\
	Europa	&	13	&	0.33 	&	0.54 	&	5 	&	10 	&	62 	&	103 	&	180.00 	&	180.00 	&	179.98 	&	179.97 	\\
	Saturn	&	77	&	7.28 	&	9.99 	&	43350 	&	59304 	&	334988 	&	398610 	&	167.96 	&	163.53 	&	86.95 	&	69.27 	\\
	Titan	&	78	&	0.32 	&	0.44 	&	9 	&	13 	&	102 	&	140 	&	180.00 	&	180.00 	&	179.97 	&	179.96 	\\
	Rhea	&	79	&	0.10 	&	0.13 	&	NaN	&	NaN	&	1 	&	2 	&	180.00 	&	180.00 	&	180.00 	&	180.00 	\\
	Iapetus	&	80	&	0.09 	&	0.13 	&	NaN	&	NaN	&	1 	&	1 	&	180.00 	&	180.00 	&	180.00 	&	180.00 	\\
	Dione	&	81	&	0.07 	&	0.10 	&	NaN	&	NaN	&	NaN	&	1 	&	180.00 	&	180.00 	&	180.00 	&	180.00 	\\
	Uranus	&	139	&	1.66 	&	2.02 	&	3477 	&	4241 	&	34691 	&	42269 	&	179.03 	&	178.82 	&	170.36 	&	168.26 	\\
	Titania	&	140	&	0.05 	&	0.06 	&	NaN	&	NaN	&	1 	&	1 	&	180.00 	&	180.00 	&	180.00 	&	180.00 	\\
	Oberon	&	141	&	0.05 	&	0.06 	&	NaN	&	NaN	&	1 	&	1 	&	180.00 	&	180.00 	&	180.00 	&	180.00 	\\
	Neptune	&	167	&	1.08 	&	1.18 	&	2762 	&	3001 	&	27582 	&	29965 	&	179.23 	&	179.17 	&	172.34 	&	171.68 	\\
	Triton	&	168	&	0.06 	&	0.06 	&	NaN	&	NaN	&	5 	&	6 	&	180.00 	&	180.00 	&	180.00 	&	180.00 	\\
	Pluto	&	182	&	0.03 	&	0.06 	&	NaN	&	NaN	&	2 	&	3 	&	180.00 	&	180.00 	&	180.00 	&	180.00 	\\
	Ceres	&	188	&	0.16 	&	0.41 	&	NaN	&	NaN	&	1 	&	4 	&	180.00 	&	180.00 	&	180.00 	&	180.00 	\\
	Pallas	&	189	&	0.09 	&	0.31 	&	NaN	&	NaN	&	NaN	&	1 	&	180.00 	&	180.00 	&	180.00 	&	180.00 	\\	
	\enddata
	\tablecomments{
		$\beta$-Impact is the impact range of an event of light deflection (see Section \ref{sec-alpha-beta-impact}); $\beta$-Validity is the range of application of Equation (\ref{equ:alpha1}) (see Section \ref{sec-alpha-beta-validity}); 
		NaN represents that $\beta_\mathrm{1.0, min}$ or $\beta_\mathrm{0.1, min}$ being less than $\Theta_{\mathrm{min}}$, or $\beta_\mathrm{1.0, max}$ or $\beta_\mathrm{0.1, max}$ being less than $\Theta_{\mathrm{max}}$.
	}
\end{deluxetable*}

The Sun's gravitational field could bend light in the direction of $\beta \in [\Theta, 180 - \Theta]$ (where $\Theta$ is the apparent radius of the Sun) with a deflection angle $>$ 10 $\mu$as. Above all, Jupiter can deflect light by an amount of 0.1 $\mu$as up to $\leq 168.4^{\circ}$ (i.e., $\beta_{0.1,\mathrm{max}} \approx 168.4^{\circ}$), and by an amount of 1.0 $\mu$as up to $\leq 89.1^{\circ}$ (i.e., $\beta_{1.0,\mathrm{max}} \approx 89.1^{\circ}$; see Figure \ref{fig-alpha-position-rely}). Such large values of $\beta_{0.1,\mathrm{max}}$ and $\beta_{1.0,\mathrm{max}}$ make Jupiter an unavoidable issue in astrometry, like the Sun. For Saturn, the values of $\beta_{0.1,\mathrm{max}}$ and $\beta_{1.0,\mathrm{max}}$ are $\approx 110.7^{\circ}$ and $\approx 16.5^{\circ}$, respectively. For Mercury, Venus, Mars, Uranus, and Neptune, the values of ($\beta_{0.1,\mathrm{max}}$, $\beta_{1.0,\mathrm{max}}$) are $\approx$ ($1.4^{\circ}$, $0.1^{\circ}$), ($41.3^{\circ}$, $4.3^{\circ}$), ($3.9^{\circ}$, $0.4^{\circ}$), ($11.7^{\circ}$, $1.2^{\circ}$), and ($8.3^{\circ}$, $0.8^{\circ}$), respectively. 

Moon's effect is comparable to that of the planets, i.e., ($\beta_{0.1,\mathrm{max}}$, $\beta_{1.0,\mathrm{max}}$) $\approx$ ($63.6^{\circ}$, $7.1^{\circ}$) (see Table \ref{tab:betas}). For Ganymede, ($\beta_{0.1,\mathrm{max}}$, $\beta_{1.0,\mathrm{max}}$) $\approx$ (316$''$, 31$''$). Eleven satellites (including the Moon) can bend light by 0.1 $\mu$as, and six can bend light by 1.0 $\mu$as, as long as the corresponding $\beta$ is $\geq 1''$. The $\beta$-impacts of Pluto and several asteroids (i.e., Ceres and Pallas) are comparable to those of large satellites.

The aforementioned values of ($\beta_{0.1,\mathrm{max}}$, $\beta_{1.0,\mathrm{max}}$) highlight the challenge of conducting high-precise astrometry, especially by using next-generation observatories. 
% The fast moving of the Moon relative to the Earth is a more intractable challenges for high-precise astrometry \citep[e.g.,][]{Kopeikin-Gwinn2000, Kopeikin+2011, Kopeikin-Makarov2021}. 

\subsubsection{Validity of the Simplified Equation}\label{sec-alpha-beta-validity}

Equation (\ref{equ:alpha1}) is much simpler relative to Equation (\ref{equ:alpha2}), therefore it may be helpful in saving computing resources by using Equation (\ref{equ:alpha1}). The difference between Equations (\ref{equ:alpha2}) and (\ref{equ:alpha1}) is presented to judge the range of application of Equation (\ref{equ:alpha1}) (see Figure \ref{fig-alpha-position-rely}). The critical value of $\beta$ where the difference of $\alpha$ calculated from the two equations is equal to a given value (e.g., 1.0 $\mu$as) under a given condition (e.g., $r_\mathrm{min}$) can be regarded as the validity of Equation (\ref{equ:alpha1}) (which is much simpler relative to Equation (\ref{equ:alpha2})), and is denoted as $\beta$-validity for short hereafter. The naming convention is similar to that for $\beta$-impact, and denoted as, e.g., $\beta_{\mathrm{d, 1.0, min}}$, $\beta_{\mathrm{d, 1.0, max}}$, $\beta_{\mathrm{d, 0.1, min}}$, and $\beta_{\mathrm{d, 0.1, max}}$.

The Sun and Mars can be seen as two extremes (see Figure \ref{fig-alpha-position-rely}). For the Sun, Equation (\ref{equ:alpha1}) is invalid even when light grazes its surface under a precision of 1.0 $\mu$as. For Jupiter and Saturn (see Table \ref{tab:betas}), $\beta_{\mathrm{d, 0.1, max}}$ is less than 90$^{\circ}$, but $\beta_{\mathrm{d, 1.0, max}}$ exceeds 90$^{\circ}$. For other objects, including the other five planets, Pluto, and all satellites and asteroids, $\beta_{\mathrm{d, 0.1, max}}$, $\beta_{\mathrm{d, 0.1, min}}$, $\beta_{\mathrm{d, 1.0, max}}$ and $\beta_{\mathrm{d, 1.0, min}}$ all exceed 90$^{\circ}$, indicating that Equation (\ref{equ:alpha1}) is valid even under a precision of 0.1 $\mu$as when $\beta \leq 90^{\circ}$. If $\beta_{\mathrm{d, 0.1, max}}$ or $\beta_{\mathrm{d, 1.0, max}}$ exceeds $90^{\circ}$ when $\beta > 90^{\circ}$, no correction for a positional offset caused by the gravitational fields of the corresponding objects is required even under a precision level of 0.1 $\mu$as. 

\section{Discussion}\label{sec:impact-regions-start}

\subsection{Maximum Deflection Angle and Earth-Lens Distance on Impact Range of Light Deflection}\label{sec:imapact-regins-beta}

\begin{figure*}[!htb]
	\centering
%	\subfigbottomskip=-0.3cm
%	\subfigcapskip=-0.3cm
	\subfigure[$\beta_{1.0,\mathrm{min}}$ with Maximum Distance]{\includegraphics[width=0.49\textwidth]{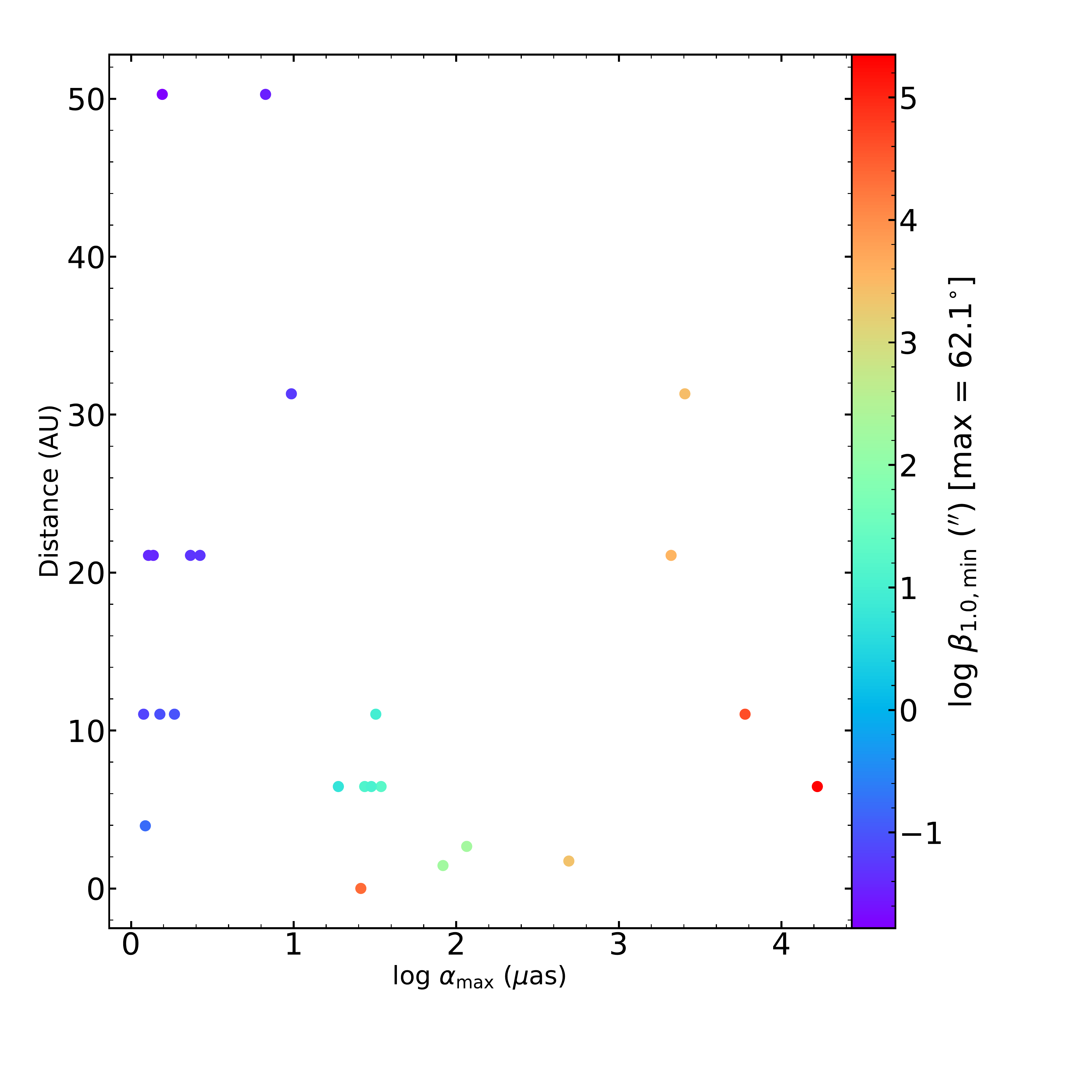}}
	\subfigure[$\beta_{1.0,\mathrm{max}}$ with Minimum Distance]{\includegraphics[width=0.49\textwidth]{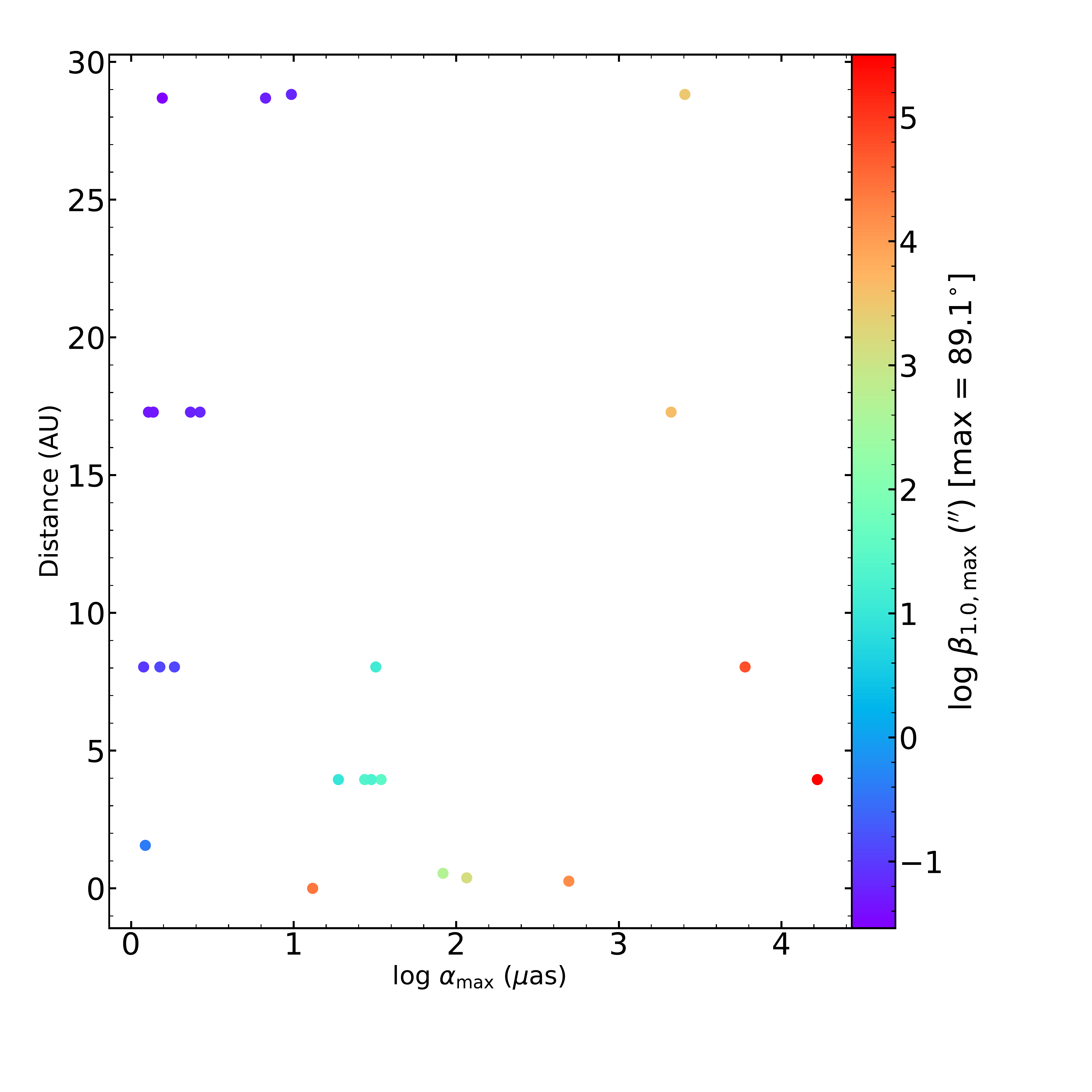}}	
	\subfigure[$\beta_{0.1,\mathrm{min}}$ with Maximum Distance]{\includegraphics[width=0.49\textwidth]{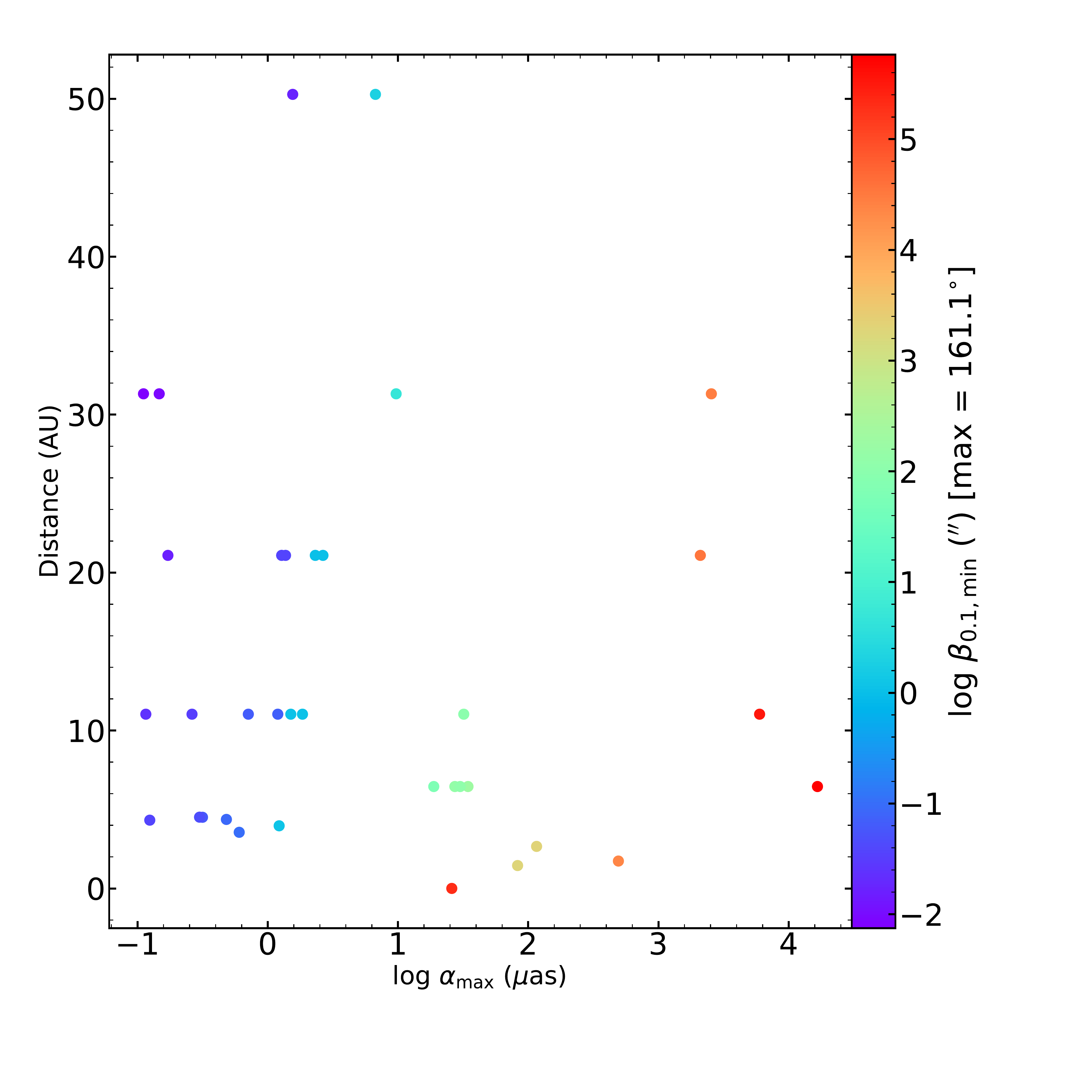}}
	\subfigure[$\beta_{0.1,\mathrm{max}}$ with Maximum Distance]{\includegraphics[width=0.49\textwidth]{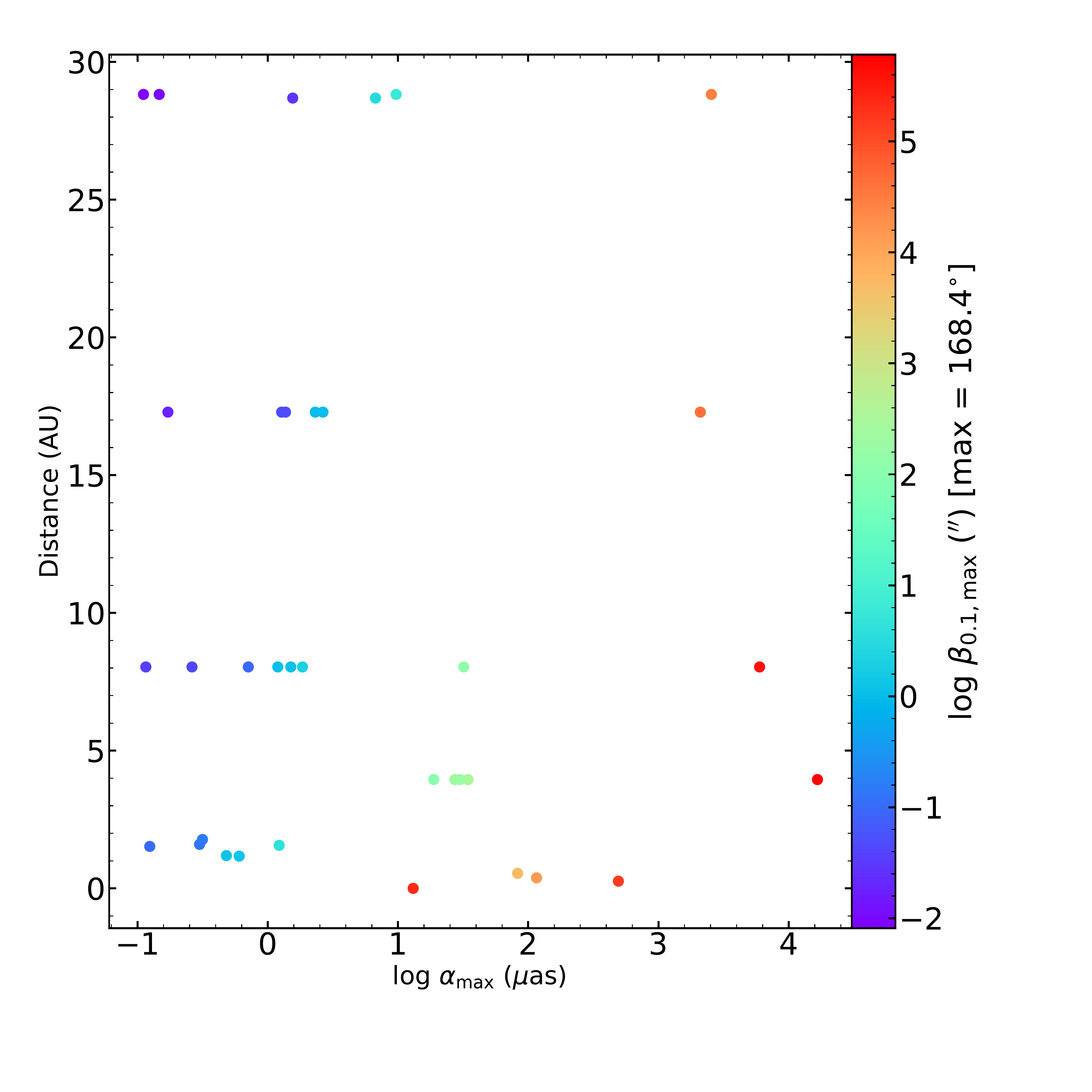}}	
	\caption{The value of $\beta$-impact (i.e., $\beta_{1.0,\mathrm{min}}$, $\beta_{1.0,\mathrm{max}}$, $\beta_{0.1,\mathrm{min}}$, and $\beta_{0.1,\mathrm{max}}$) as a function of $\alpha_{\mathrm{max}}$ and the distances from the celestial bodies to the Earth. Only sources with $0.1 \leq \alpha_{\mathrm{max}} < 17\,000.0$ $\mu$as and the corresponding values of $\beta$-impact $\geq 1''$ are included.}
	\label{fig-beta-impact-relation}
\end{figure*}

The value of $\beta$-impact is related to $\alpha_{\mathrm{max}}$ and the distance from the celestial body to the Earth (see Section \ref{sec:deflection}). Figure \ref{fig-beta-impact-relation} presents the relationship among $\alpha_{\mathrm{max}}$, distance, and the values of $\beta$-impact (i.e., $\beta_{1.0,\mathrm{min}}$, $\beta_{1.0,\mathrm{max}}$, $\beta_{0.1,\mathrm{min}}$, and $\beta_{0.1,\mathrm{max}}$). 
The results indicate that the values of $\beta$-impact increase with an increase of $\alpha_{\mathrm{max}}$ and a decrease of distance.

\subsection{Zones of Perturbation}\label{sec:imapact-regions-coverage}

This section roughly estimates the specific zones in which light from CESs would be deflected by a given angle (i.e., 1.0 and 0.1 $\mu$as) by Mercury, Venus, Mars, Uranus, Neptune, Pluto, and Ceres. The $\beta_\mathrm{0.1, max}$ values of these sources are all $\geq 1''$. Jupiter and Saturn are not included because their values of $\beta_\mathrm{0.1, min}$ are larger than $90^{\circ}$. So heavy is their impact on light from CESs, like the Sun, that they should be taken into consideration in most cases. For satellites, two orbital centers (both the Sun and primary components including Pluto) should be considered. In addition, the perturbed zones for satellites other than the Moon are buried within those for their primary components. For the Moon, $\beta_\mathrm{0.1, min}$ is about $58^{\circ}$. The impact of the Moon is comparable to that of Saturn. Therefore, the perturbed zones for the many satellites are not included in this work. We use the position of the Earth-Moon system as the position of the Earth, and assume that the planets, Pluto, and asteroids move along a fixed elliptical orbit around the Sun.

We assume that the CES, the Earth and the celestial body (e.g., planets, Pluto or asteroids) all exist along a single line. The position of a CES can be calculated by assuming two arbitrary values of $E$ for the Earth and the celestial body, respectively; see the definition of $E$ in Equation (\ref{equ:orbit}). The distance from the Earth to a CES is fixed to be $10^{14}$ times larger than the distance from the Earth to the celestial body. This corresponds to a distance from the Earth to a fiducial CES of $\sim$ 500 Mpc if the distance from the Earth to the celestial body is $\sim$ 1 au. If the distance from the Earth to the CES fluctuates between 1 Mpc and 500 Mpc, the corresponding fluctuation of the computed position of the CES is about dozens of $\mu$as. This computed position is used to calculate the deflection angle caused by the gravitational field of solar system objects. The positional uncertainty of CESs of dozens of $\mu$as does not affect the computed deflection angle under precision of 0.1 $\mu$as. Therefore, this fluctuation does not affect the conclusions of this work. 

The coverage area of the perturbed zone corresponding to $\beta_{1.0}$ (i.e., the zone where light from a CES would be deflected by a celestial body by 1.0 $\mu$as) is a circle, at the given positions of the Earth and the celestial body; i.e., two given values of $E$. The center of this circle is the computed CES position projected on the sky as seen from Earth (coinciding with the projected point on the sky of the trajectory of the corresponding celestial body), and the radius is $\beta_{1.0}$. Similarly, the perturbed zone for $\beta_{0.1}$ is also a circle with the same center, but its radius is $\beta_{0.1}$.

The ecliptic coordinate system was used to compute the positions of the Earth, celestial bodies, and CESs. These positions can be transformed into the J2000 equatorial coordinate system as follows:
\begin{equation}\label{equ:ecl} 
	\left\{
	\begin{aligned}
		x_\mathrm{eq} & = x_\mathrm{ecl} \\ 
		y_\mathrm{eq} & = \cos \varepsilon\; y_\mathrm{ecl} - \sin \varepsilon\; z_\mathrm{ecl}, \\
		z_\mathrm{eq} & = \sin \varepsilon\; y_\mathrm{ecl} + \cos \varepsilon\; z_\mathrm{ecl}
	\end{aligned}
	\right.
\end{equation}
where $\varepsilon = 23.43928^{\circ}$ is the obliquity of the ecliptic. 

\begin{figure*}[!htb]
	\centering
%	\subfigbottomskip=-0.5cm
	\subfigcapskip=-0.3cm
	\subfigure[Mercury]{\includegraphics[width=0.38\textwidth]{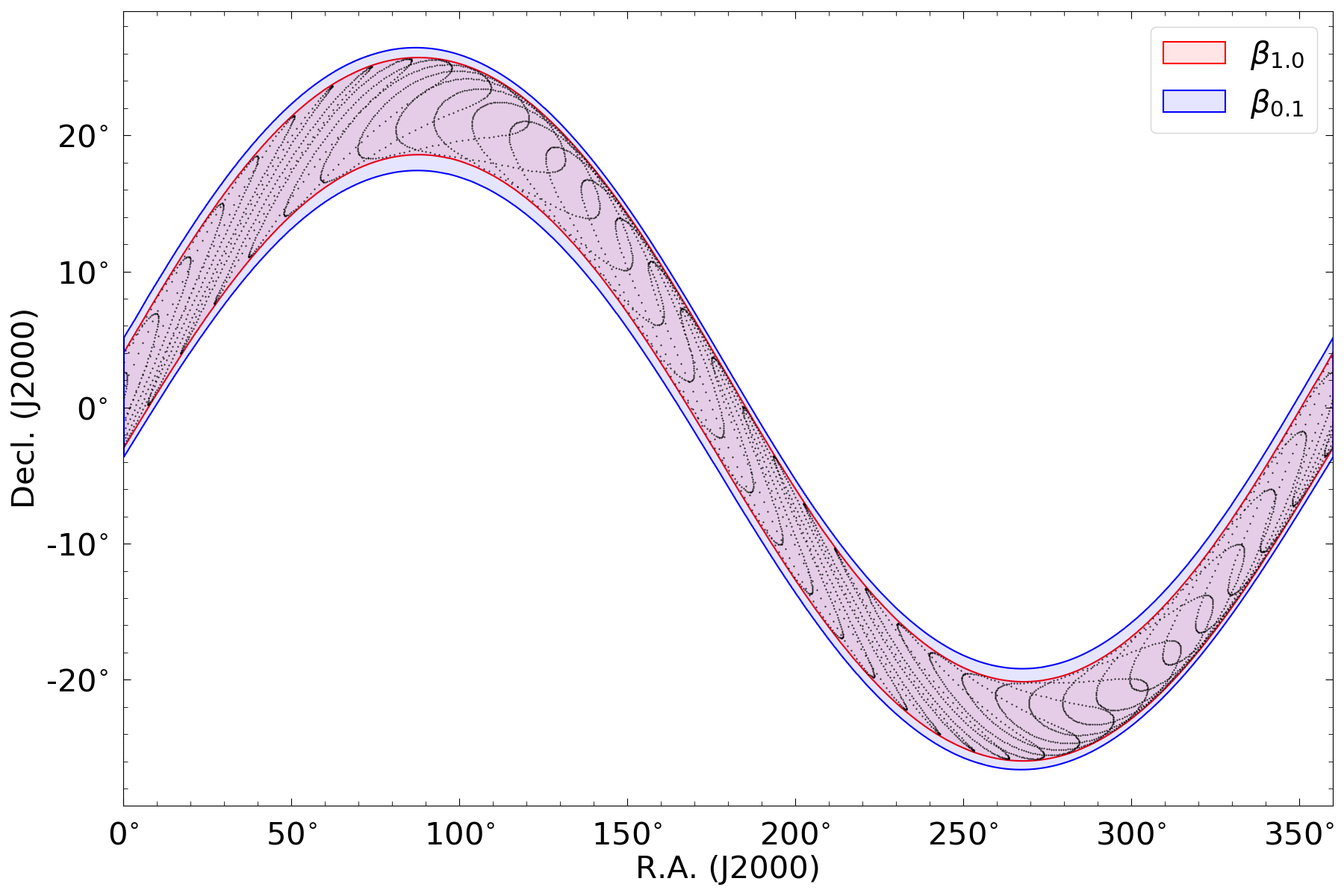}}
    \subfigure[Venus]{\includegraphics[width=0.38\textwidth]{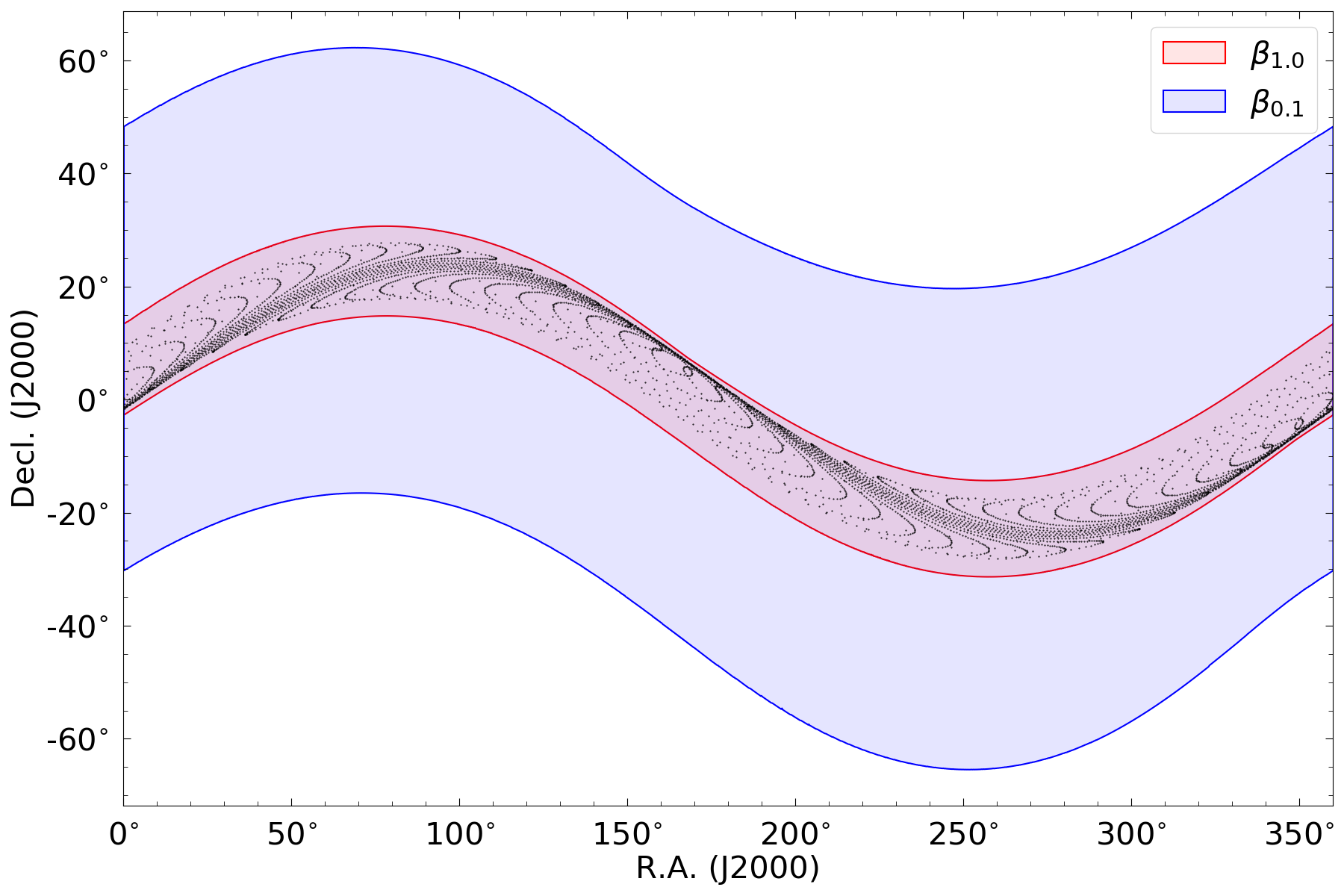}}
    \subfigure[Mars]{\includegraphics[width=0.38\textwidth]{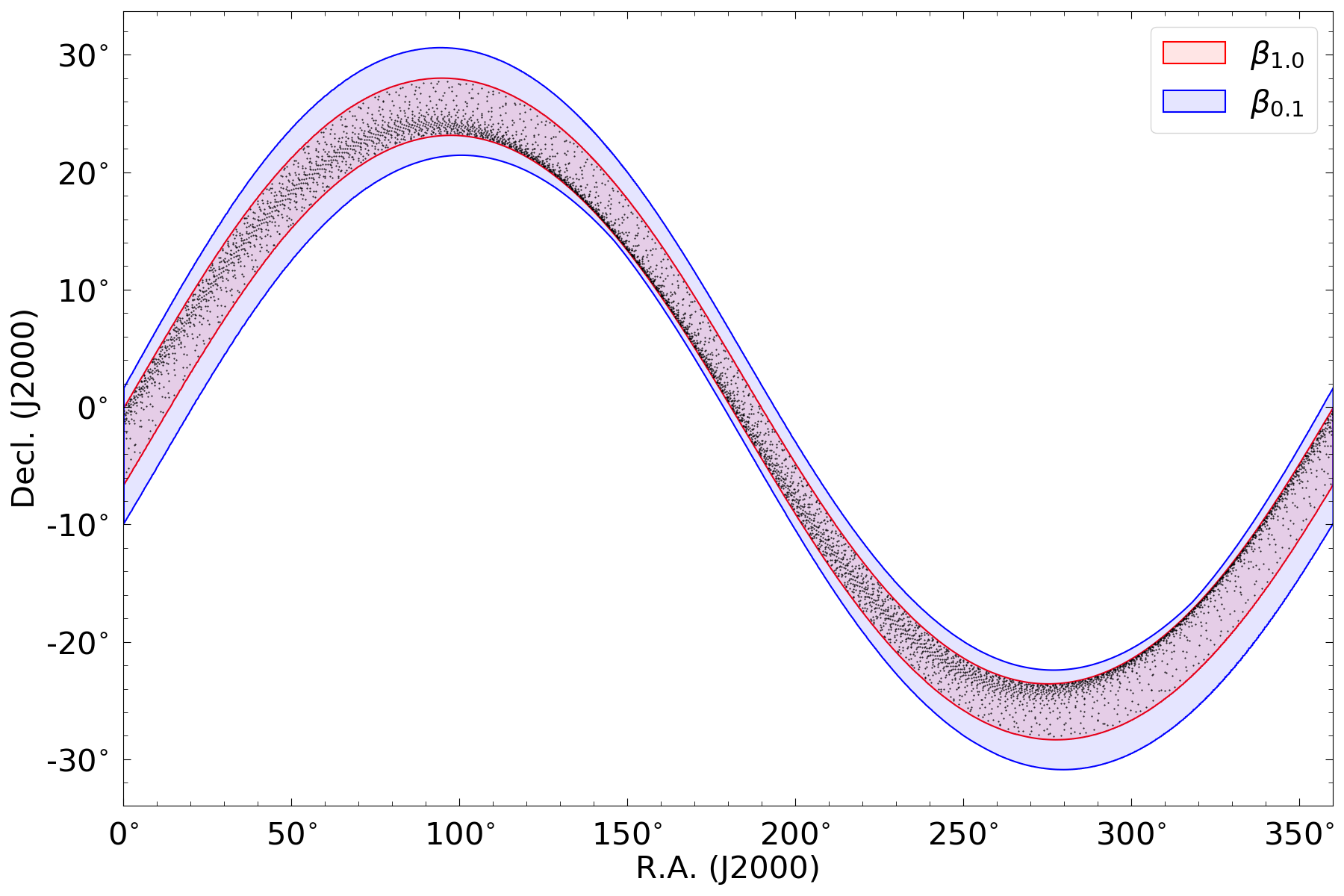}}
    \subfigure[Uranus]{\includegraphics[width=0.38\textwidth]{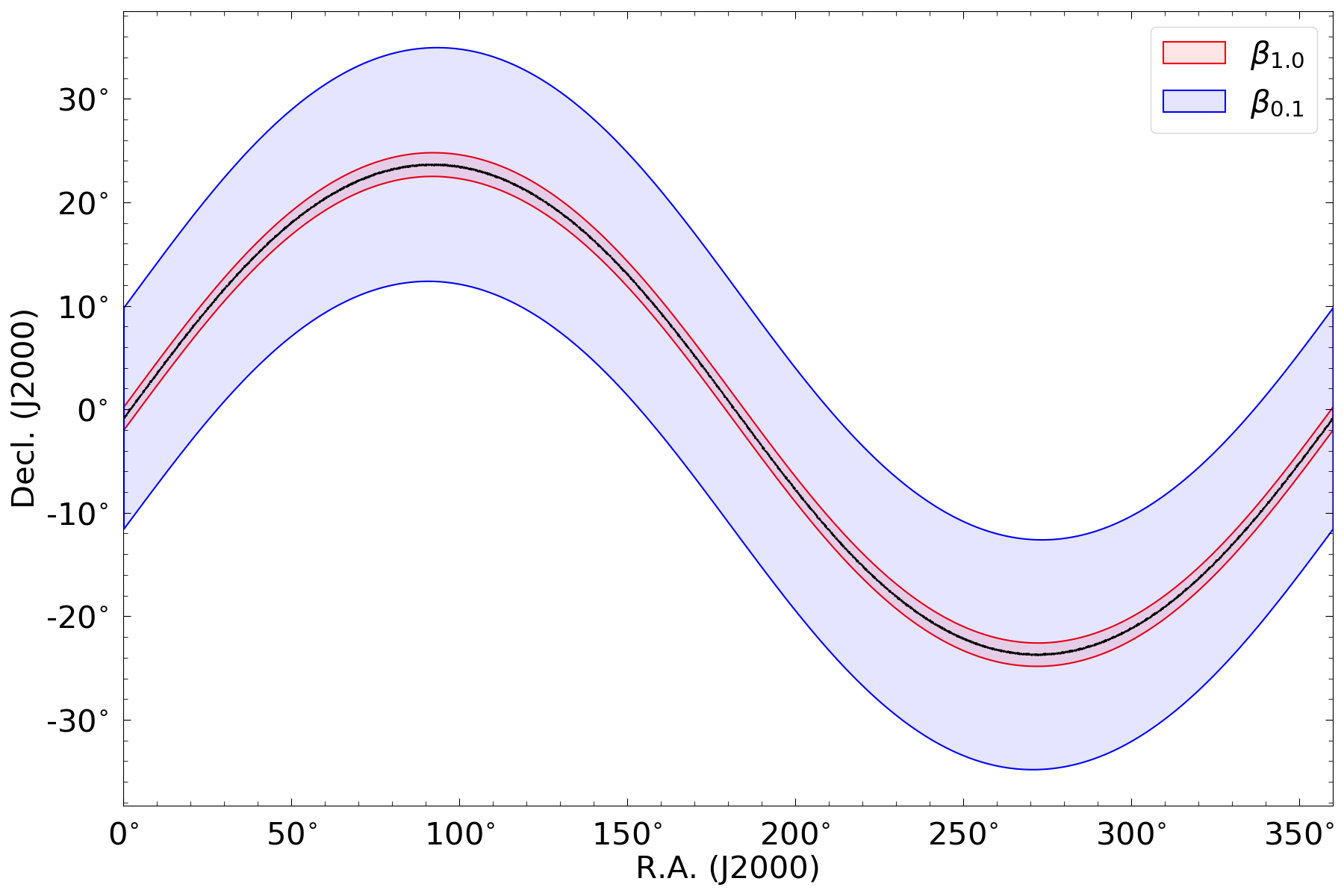}}
    \subfigure[Neptune]{\includegraphics[width=0.38\textwidth]{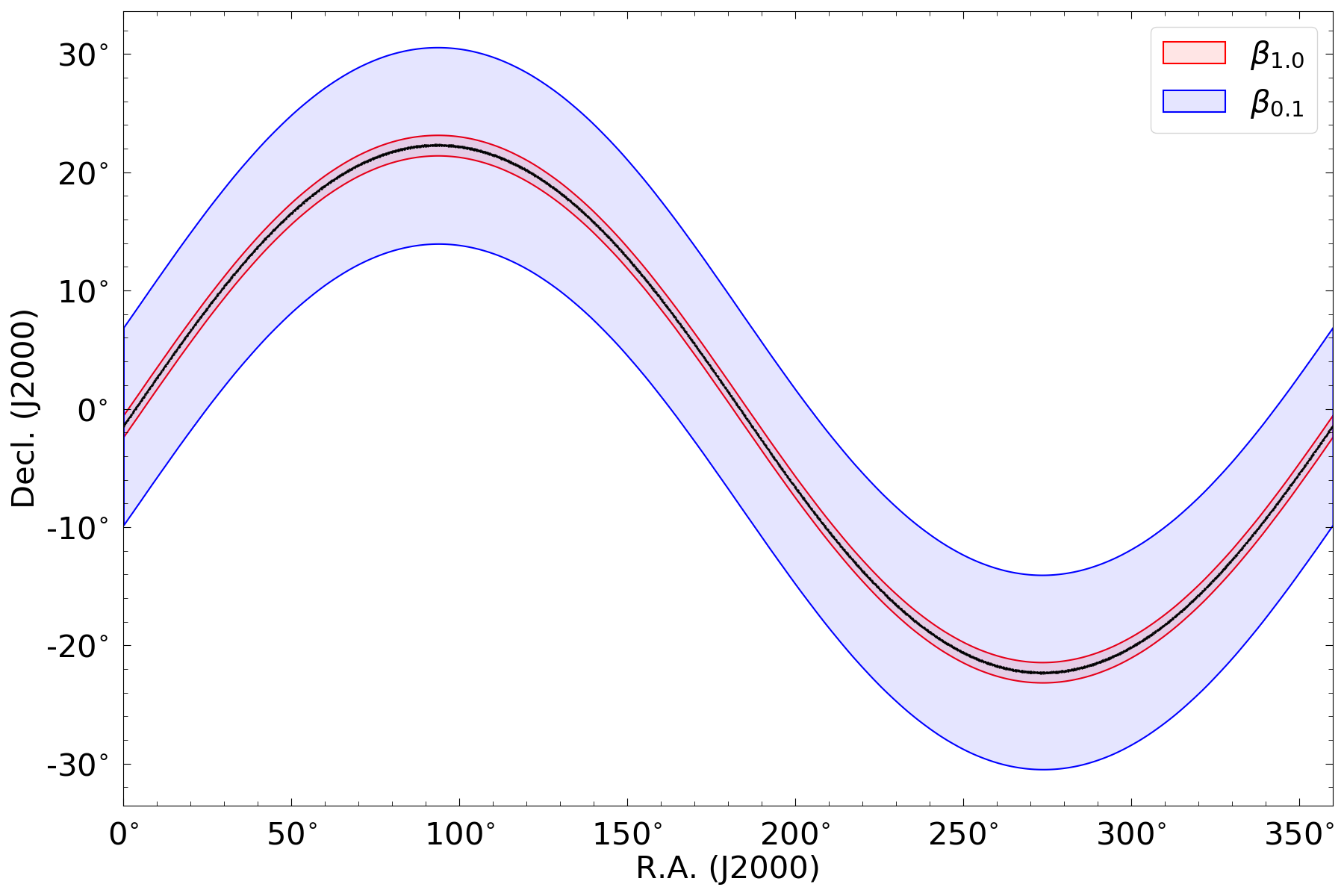}}
    \subfigure[Pluto]{\includegraphics[width=0.38\textwidth]{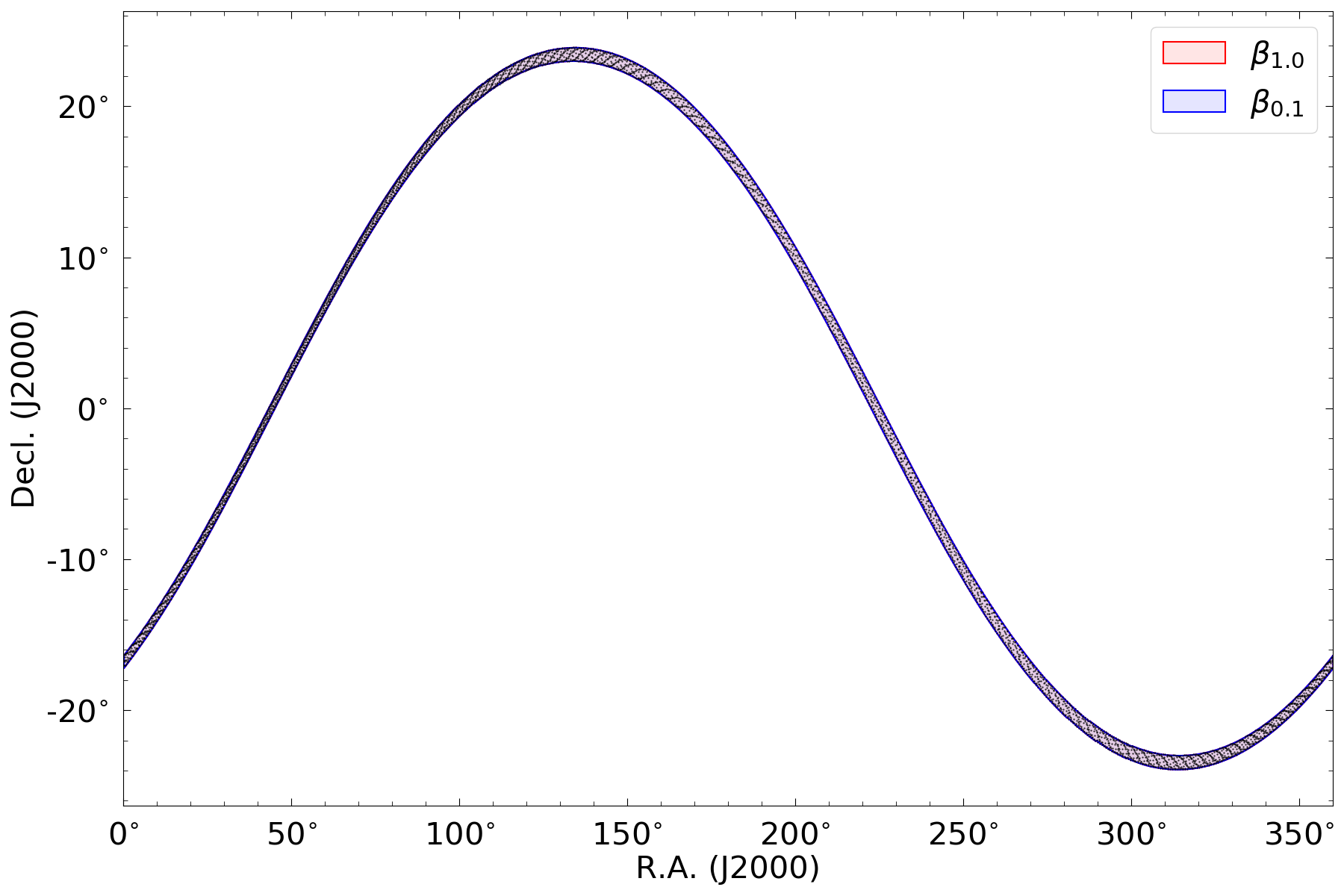}}
    \subfigure[Ceres]{\includegraphics[width=0.38\textwidth]{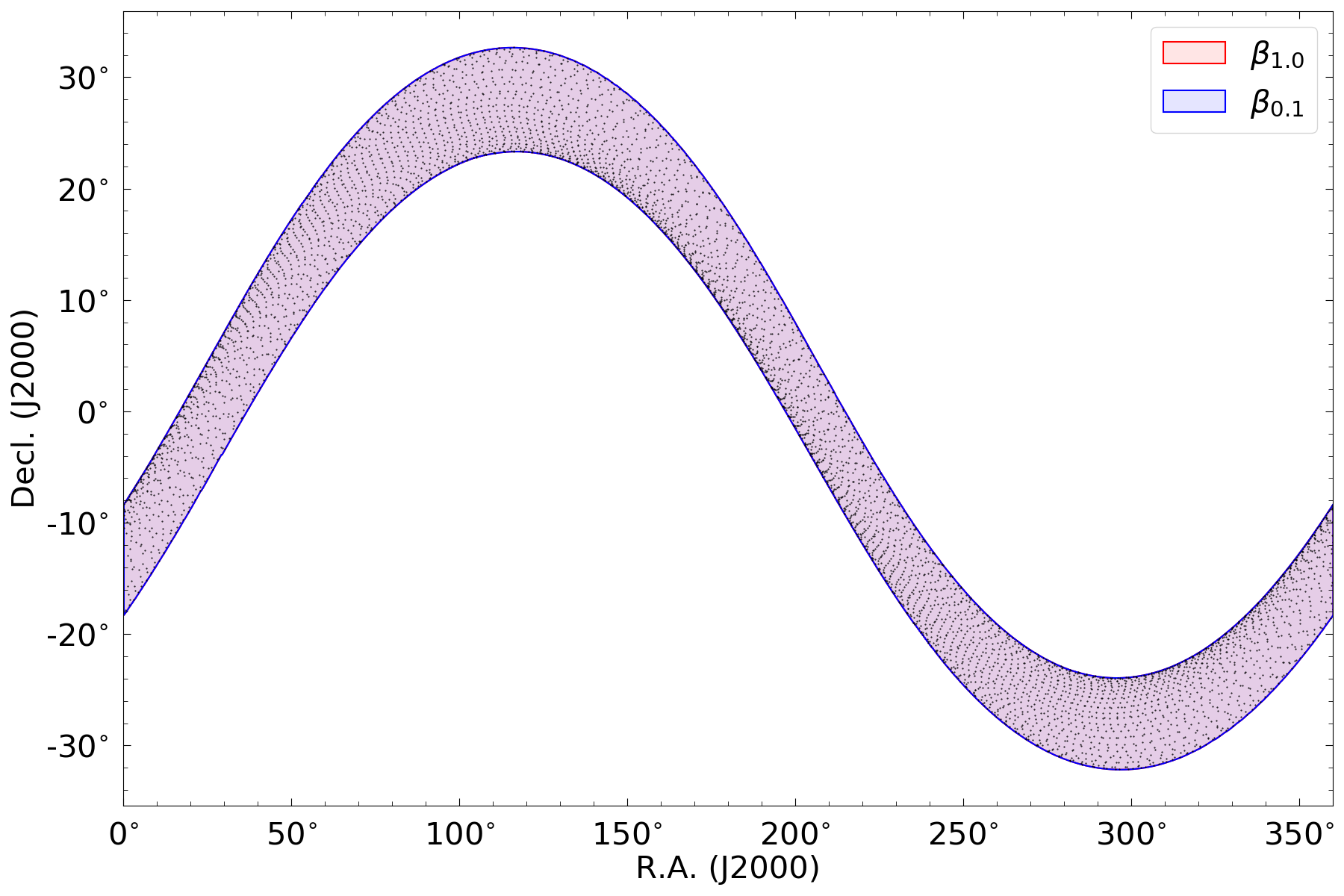}}	
    \caption{Whole perturbed zone for $\beta_{1.0}$ (red) and $\beta_{0.1}$ (blue). The corresponding celestial bodies are denoted in the subhead. The black points show the projected points on the sky of the trajectory of the corresponding celestial body.}
	\label{fig-coverage}
\end{figure*}

We have computed the whole zone of perturbation for both $\beta_{1.0}$ and $\beta_{0.1}$ (see Figure \ref{fig-coverage}). The perturbed zones for $\beta_{1.0}$ and $\beta_{0.1}$ are ribbons and their widths are denoted as $W_{1.0}$ and $W_{0.1}$, respectively. For Pluto, $W_{1.0}$ and $W_{0.1}$ are less than $\sim$ 1$^{\circ}$ (see Figure \ref{fig-coverage} and Table \ref{tab:coverage}). For Mercury, Mars, and Ceres, $W_{1.0}$ and $W_{0.1}$ are similar and their values are $\sim$ 4$^{\circ}$--9$^{\circ}$, $\sim$ 2$^{\circ}$--6$^{\circ}$, and $\sim$ 5$^{\circ}$--10$^{\circ}$, respectively. The values of $W_{1.0}$ for Uranus and Neptune are $\sim$ 3$^{\circ}$ or less, and close to that of Pluto. However, considering $\beta_{\mathrm{0.1}}$ of $\sim$ 10$^{\circ}$ for Uranus and Neptune, their values of $W_{0.1}$ range from $\sim$ 10$^{\circ}$ to $\sim$ 23$^{\circ}$, which are larger than those for Mercury, Mars, and Ceres. In particular, $W_{0.1}$ for Venus is close to $\sim$ 80$^{\circ}$ and $W_{1.0}$ also reaches $\sim$ 13$^{\circ}$--16$^{\circ}$. These results imply that the size of the perturbed zone is negatively correlated with the distance from the Earth to the celestial body and positively correlated with the value of $\beta_{\mathrm{0.1}}$ or $\beta_{\mathrm{1.0}}$. 

\begin{deluxetable*}{cccccccc}
	\tablecolumns{8}
	%	\tablenum{1}
%	\setlength\tabcolsep{10pt}
	%\tablewidth{20pt}
	%\renewcommand{\arraystretch}{1.2}
	\tablecaption{Widths of the Ribbon-like Perturbed Zones \label{tab:coverage}}
	%\begin{tabular}
	\tablehead{
		\colhead{Quantity} & \colhead{Mercury} & \colhead{Venus} & \colhead{Mars}　& \colhead{Uranus} & \colhead{Neptune} & \colhead{Pluto} & \colhead{Ceres}
	}
	\startdata
	$W_{1.0}$ & $\sim$ 4$^{\circ}$--7$^{\circ}$ & $\sim$ 13$^{\circ}$--16$^{\circ}$ & $\sim$ 2$^{\circ}$--5$^{\circ}$ & $\sim$ 2$^{\circ}$--3$^{\circ}$ & $\sim$ 1$^{\circ}$--2$^{\circ}$ & $\lesssim$ 1$^{\circ}$ & $\sim$ 5$^{\circ}$--10$^{\circ}$ \\
	$W_{0.1}$ & $\sim$ 5$^{\circ}$--9$^{\circ}$ & $\sim$ 66$^{\circ}$--80$^{\circ}$ & $\sim$ 4$^{\circ}$--6$^{\circ}$ & $\sim$ 14$^{\circ}$--23$^{\circ}$ & $\sim$ 10$^{\circ}$--17$^{\circ}$ & $\lesssim$ 1$^{\circ}$ & $\sim$ 5$^{\circ}$--10$^{\circ}$ \\
	\enddata
	\tablecomments{
		See the detailed distribution of whole perturbed zone in Figure \ref{fig-coverage}.
	}
\end{deluxetable*}

\subsection{Duration of Perturbation }\label{sec:imapact-regions-duration}

It is interesting to investigate the durations when light from CESs can be deflected by celestial bodies by 1.0 and 0.1 $\mu$as, referred to as $\tau_{1.0}$ and $\tau_{0.1}$, respectively. Similar to the perturbed zones, we only focused on calculating the values for Mercury, Venus, Mars, Uranus, Neptune, Pluto, and Ceres. These bodies have been classified into three categories: planets inside and outside the Earth's orbit and others (i.e., Pluto and Ceres).

\subsubsection{Trajectory of the Earth and Celestial Bodies}\label{sec:imapact-regions-duration:introduction}

It is necessary to determine the positions of the Earth and the celestial body in the solar system at any given time. For celestial bodies inside and outside the Earth's orbit, two different methods are used to obtain the positions of the Earth and the celestial body. For celestial bodies inside the Earth's orbit, we divide the orbits of a given celestial body into segments, and compute the movement time in each segment. The relative position of the Earth is determined by using this movement time. The reason for choosing the given celestial body as a reference body is that the angle variation of the Earth should not exceed 180$^{\circ}$ if that of the given gravitational body is $\leq 180^{\circ}$, because the angular speed of the Earth is slower. Therefore, in the progress of summing the movement time for all eligible segments (i.e., the segments where the deflection angle is larger than 1.0 or 0.1 $\mu$as), the movement time for any segment should not be double-counted when the angular change of the given celestial body is within $\pm$ 180$^{\circ}$. For celestial bodies outside the Earth's orbit, the selected reference body is the Earth for similar reasons.

Specifically, for the computation of planets inside and outside the Earth's orbit, $E$ is divided into 129\,601 parts for the reference bodies; i.e., the resolution of $E$ is about 10$''$. This resolution corresponds to a mean time resolution of $\sim$ 1.0 and $\sim$ 2.5 min for Mercury and Venus (inside the Earth's orbit), and $\sim$ 4.1 min for Mars, Uranus, and Neptune (outside the Earth's orbit), respectively. When computing the values for the category of others (i.e., Pluto and Ceres, also outside the Earth's orbit), $E$ is divided into 5\,184\,001 parts, i.e., the resolution of $E$ is about 0.25$''$. That is because $\beta_{\mathrm{0.1, min}}$ is a few arcseconds for Pluto and Ceres. The corresponding mean time resolution for Pluto and Ceres is about 6.0 s.

As a pilot work, we divided the orbit of a given celestial body into 1000 points (this is enough to show details; see Figures \ref{fig-times-Mercury}--\ref{fig-times-Ceres}) as the initial positions, $\theta_{\mathrm{CES}}$, i.e., $\theta_{\mathrm{CES}}$ = $\frac{360^{\circ}}{1000} \times 0, 1, 2, ..., 1000$. For each point, we calculate the value of $E$ for the Earth corresponding to the minimum distance from the Earth to the given gravitational body. We select one of these values of $E$, and set $E + 0^{\circ}, 10^{\circ},..., 180^{\circ}$ as the initial position of the Earth, $\phi_{\mathrm{CES}}$. The number of total initial combinations of a given celestial body and the Earth, $(\theta_{\mathrm{CES}}, \phi_{\mathrm{CES}})$, is 19\,000. The same method as presented in Section \ref{sec:imapact-regions-coverage} is used to determine the position of a CES under $(\theta_{\mathrm{CES}}, \phi_{\mathrm{CES}})$, where the position of a CES remains unchanged during the calculation for a given initial combination of $(\theta_{\mathrm{CES}}, \phi_{\mathrm{CES}})$. We then calculate the duration of perturbation where the angle between the CES and the given celestial body as seen from the Earth is less than $\beta_{1.0}$ and $\beta_{0.1}$, i.e., $\tau_{1.0}$ and $\tau_{0.1}$, respectively.

\subsubsection{Timescale Criteria Potentially Affecting SKA Astrometry}

Here, we assume that the flux density of a CES is 0.15 Jy/beam @ 9200 MHz \citep[the median flux density in the X band for a short baseline is $\sim$ 0.15 Jy from ICRF3; see][]{Charlot2020}, and 0.30 Jy/beam @ 1400 and 560 MHz (the median flux density in the S band is $\sim$ 0.23 Jy from ICRF3). The sensitivity of SKA is assumed to be 0.43 $\mu$Jy/beam @ 9200 MHz, 0.71 $\mu$Jy/beam @ 1400 MHz, and 2.88 $\mu$Jy/beam @ 560 MHz for an exposure time of 8 hr, and the full width at half-maximum size in the three wavebands is assumed to be 0.09$''$, 0.6$''$, and 1.5$''$, respectively \citep[see][]{Bonaldi+2021}. 

The integration time needed to achieve a positional accuracy of 1.0 $\mu$as should be $\sim$ 8.0 min @ 9200 MHz, $\sim$ 4.0 hr @ 1400 MHz, and $\sim$ 414.7 hr @ 560 MHz (this one may be meaningless), respectively. These three timescales can be regarded as the timescale criteria that the predicted event potentially affects SKA astrometry under a precision level of 0.1 $\mu$as.
%Therefore the timescale criteria that the predicted event potentially affects SKA astrometry under a precision level of 0.1 $\mu$as are set to $\sim$ 8.0 min @ 9200 MHz, $\sim$ 4.0 hr @ 1400 MHz, and $\sim$ 414.7 hr @ 560 MHz, respectively. 
These three criteria are denoted as $\tau_{c,0.1,9200}$, $\tau_{c,0.1,1400}$, and $\tau_{c,0.1,560}$ (see Table \ref{tab:timescale}), respectively. 
The three timescale criteria corresponding to a precision level of 1.0 $\mu$as are smaller than those of 0.1 $\mu$as by a factor of 100, and are denoted as $\tau_{c,1.0, 9200}$, $\tau_{c,1.0,1400}$, and $\tau_{c,1.0, 560}$, respectively. When the duration exceeds 48.0 hr (except for the case at 560 MHz), the events can affect SKA astrometry over multiple epochs. We denote this criterion as $\tau_{c,m}$.

\begin{deluxetable*}{ccccccc}
	\tablecolumns{7}
	%	\tablenum{1}
	\setlength\tabcolsep{10pt}
	%\tablewidth{20pt}
	%\renewcommand{\arraystretch}{1.2}
	\tablecaption{Timescale Criteria Potentially Affecting SKA Astrometry \label{tab:timescale}}
	%\begin{tabular}
	\tablehead{
		\colhead{$\tau_{c,0.1, 9200}$} & \colhead{$\tau_{c,0.1, 1400}$} & \colhead{$\tau_{c,0.1, 560}$\tablenotemark{$\dagger$}} & \colhead{$\tau_{c,1.0, 9200}$}　& \colhead{$\tau_{c,1.0, 1400}$} & \colhead{$\tau_{c,1.0, 560}$} & \colhead{$\tau_{c,m}$} \\
		\colhead{(min)} & \colhead{(hr)} & \colhead{(hr)} & \colhead{(sec)} & \colhead{(min)} & \colhead{(hr)} & \colhead{(hr)}
	}
	\startdata
	8.0 & 4.0 & 414.7 & 4.8 & 2.4 & 4.1 & 48.0 \\
	\enddata
	\tablenotetext{\dagger}{This value may be meaningless.}
\end{deluxetable*}

\subsubsection{Category of Planets inside the Earth's Orbit}\label{sec:imapact-regions-duration:inside}

\begin{figure*}[!htb]
	\centering
%	\subfigbottomskip=-0.3cm
	\subfigcapskip=-0.3cm
	\subfigure[$\tau_{1.0}$ as a function of $\theta_{\mathrm{CES}}$ and $\phi_{\mathrm{CES}}$]{\includegraphics[height=0.36\textwidth]{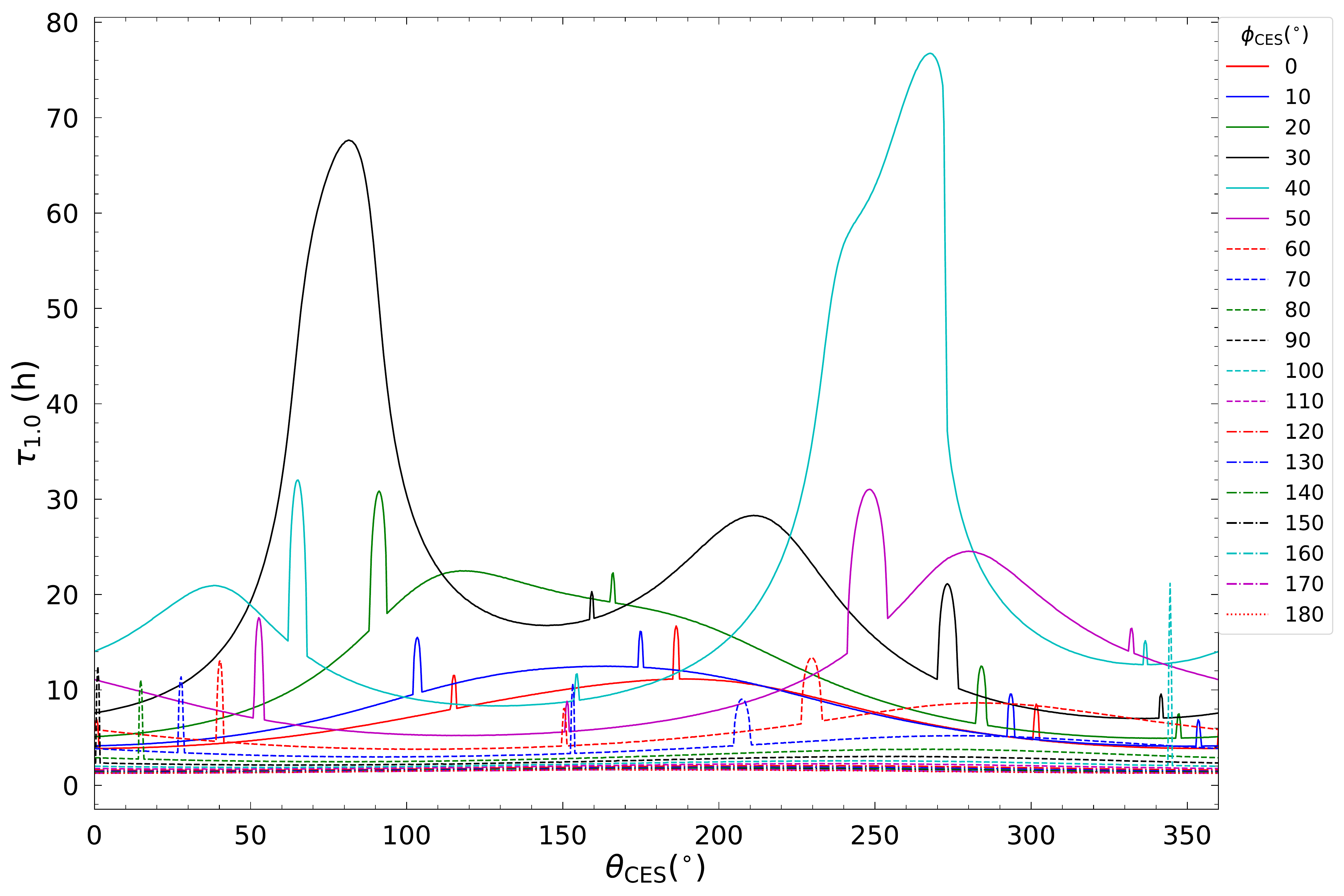}}
	\subfigure[Distribution of $\tau_{1.0}$]{\includegraphics[height=0.36\textwidth]{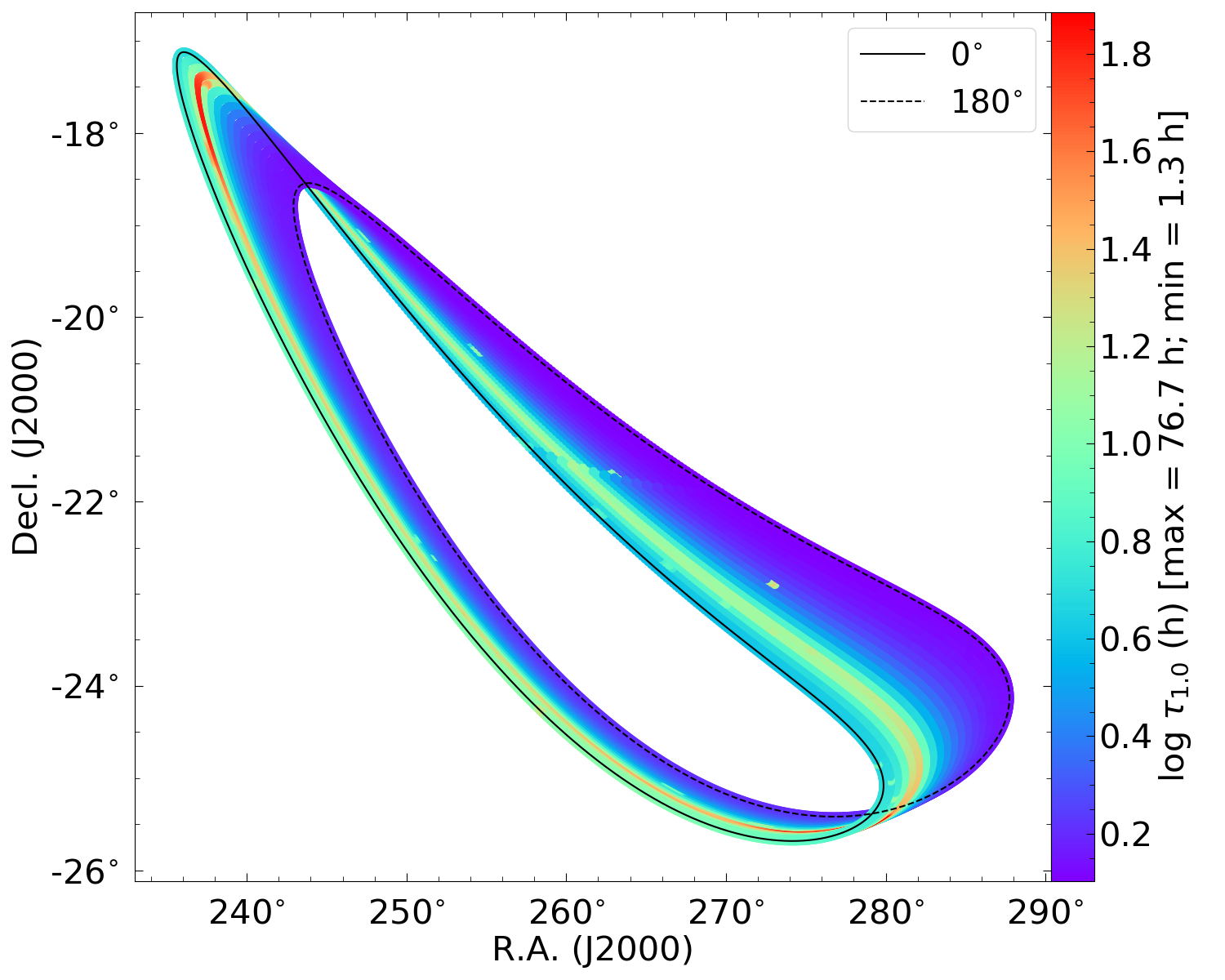}}	
	\subfigure[$\tau_{0.1}$ as a function of $\theta_{\mathrm{CES}}$ and $\phi_{\mathrm{CES}}$]{\includegraphics[height=0.36\textwidth]{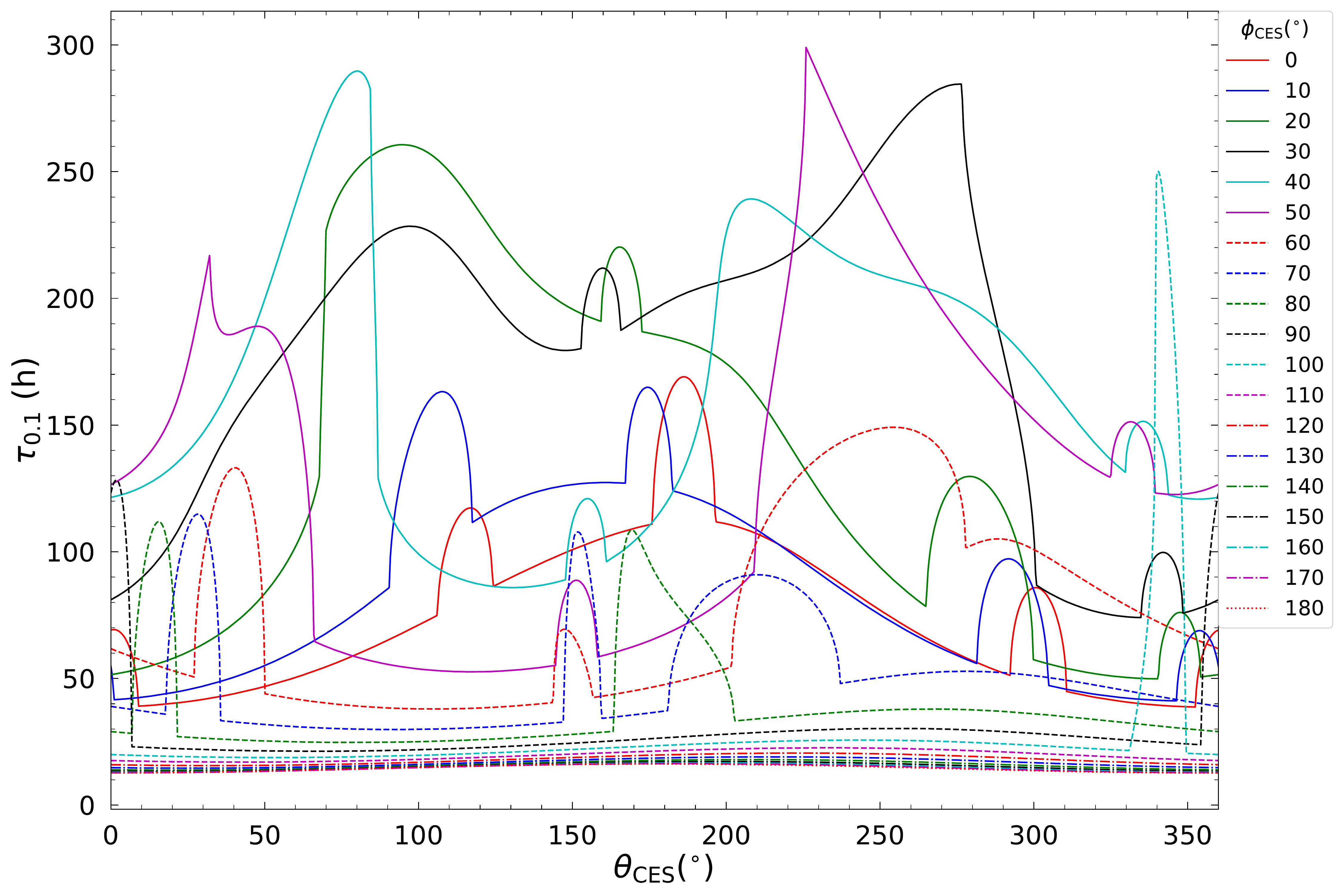}}
	\subfigure[Distribution of $\tau_{0.1}$]{\includegraphics[height=0.36\textwidth]{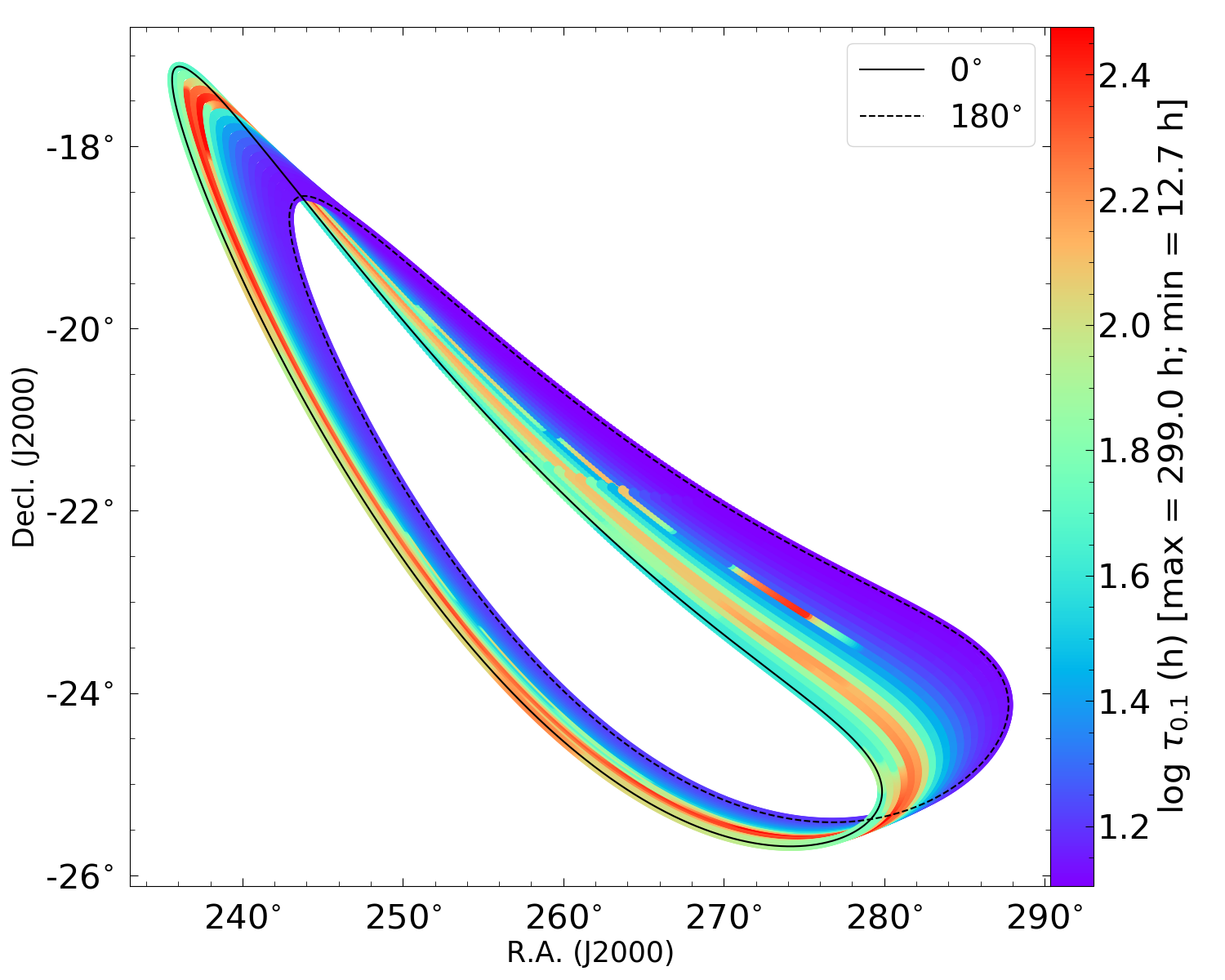}}
	\caption{$\tau_{1.0}$ and $\tau_{0.1}$ for Mercury. The value of $\phi_{\mathrm{CES}}$ is labeled in the legend; see the main text for $\theta_{\mathrm{CES}}$ and $\phi_{\mathrm{CES}}$.}
	\label{fig-times-Mercury}
\end{figure*}

\begin{figure*}[!htb]
	\centering
%	\subfigbottomskip=-0.3cm
	\subfigcapskip=-0.3cm
	\subfigure[$\tau_{1.0}$ as a function of $\theta_{\mathrm{CES}}$ and $\phi_{\mathrm{CES}}$]{\includegraphics[height=0.36\textwidth]{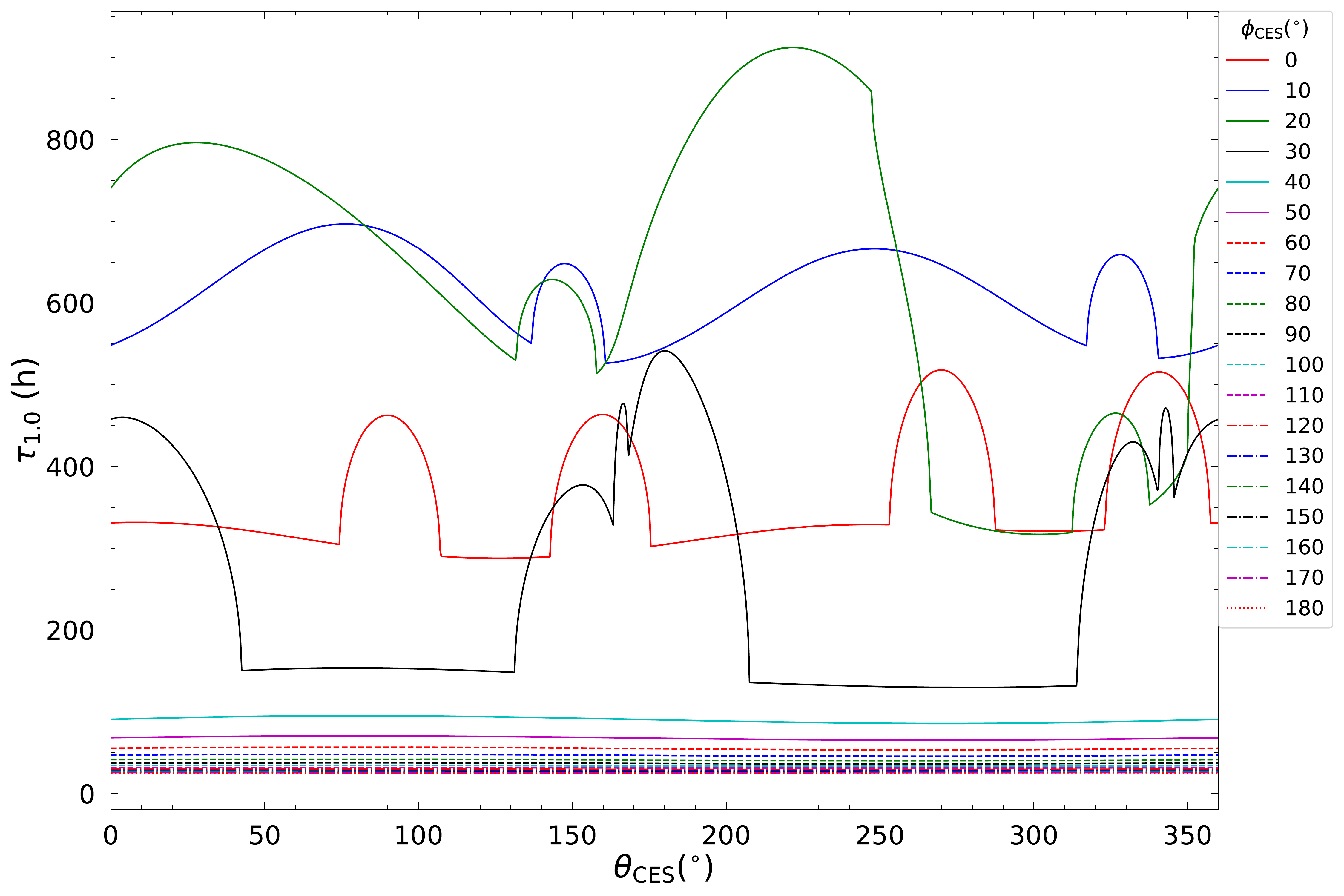}}
	\subfigure[Distribution of  $\tau_{1.0}$]{\includegraphics[height=0.363\textwidth]{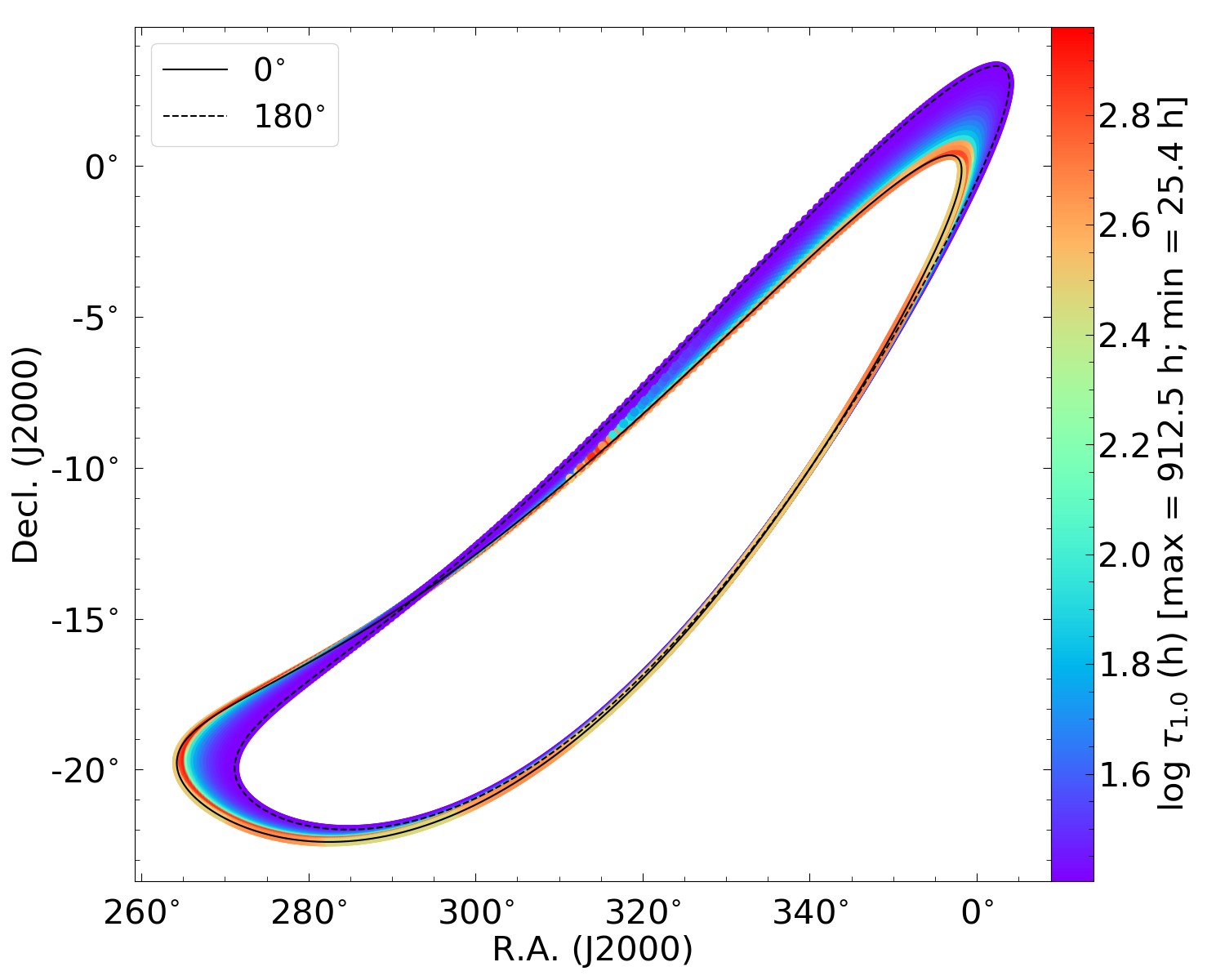}}	
	\subfigure[$\tau_{0.1}$ as a function of $\theta_{\mathrm{CES}}$ and $\phi_{\mathrm{CES}}$]{\includegraphics[height=0.36\textwidth]{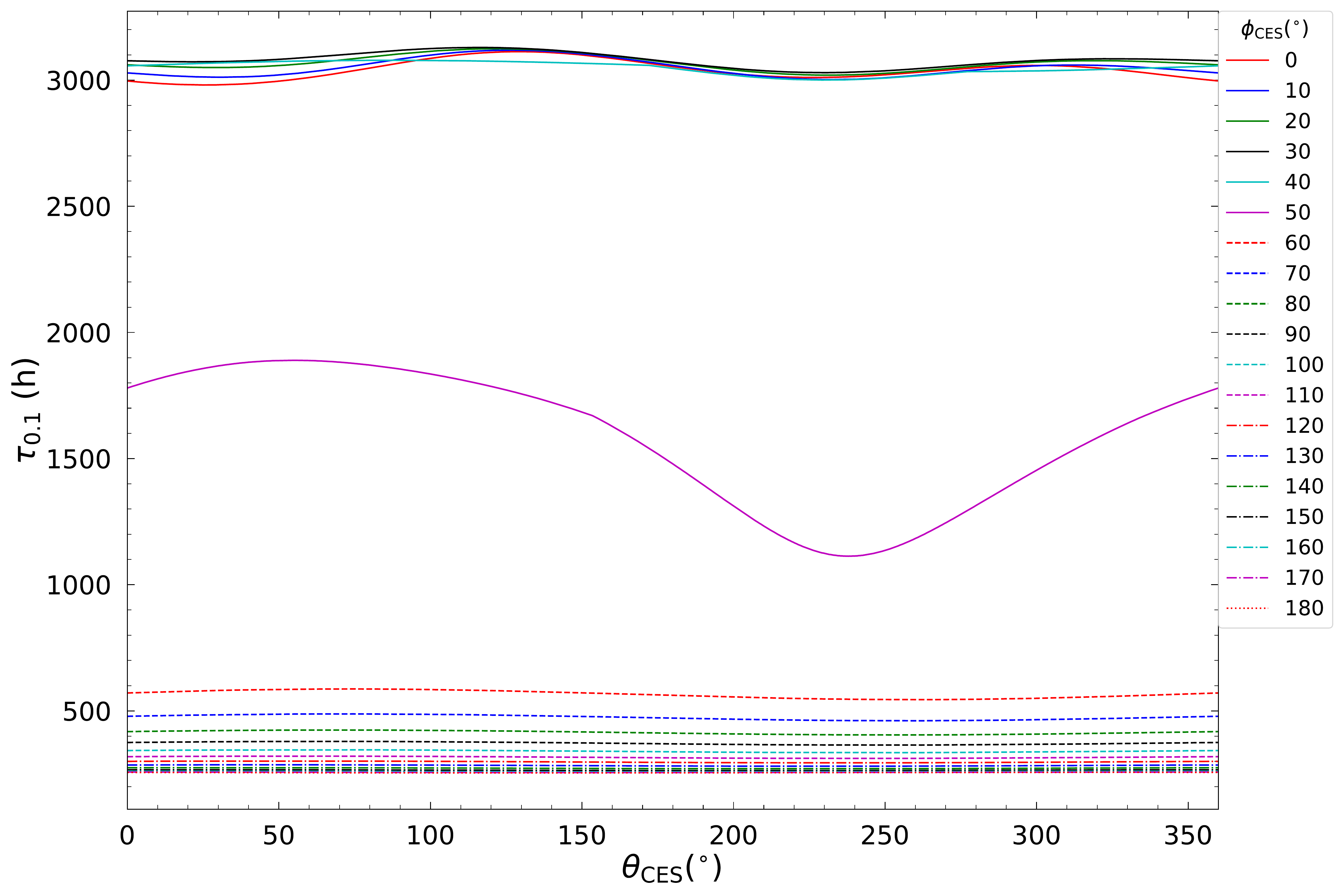}}
	\subfigure[Distribution of $\tau_{0.1}$]{\includegraphics[height=0.363\textwidth]{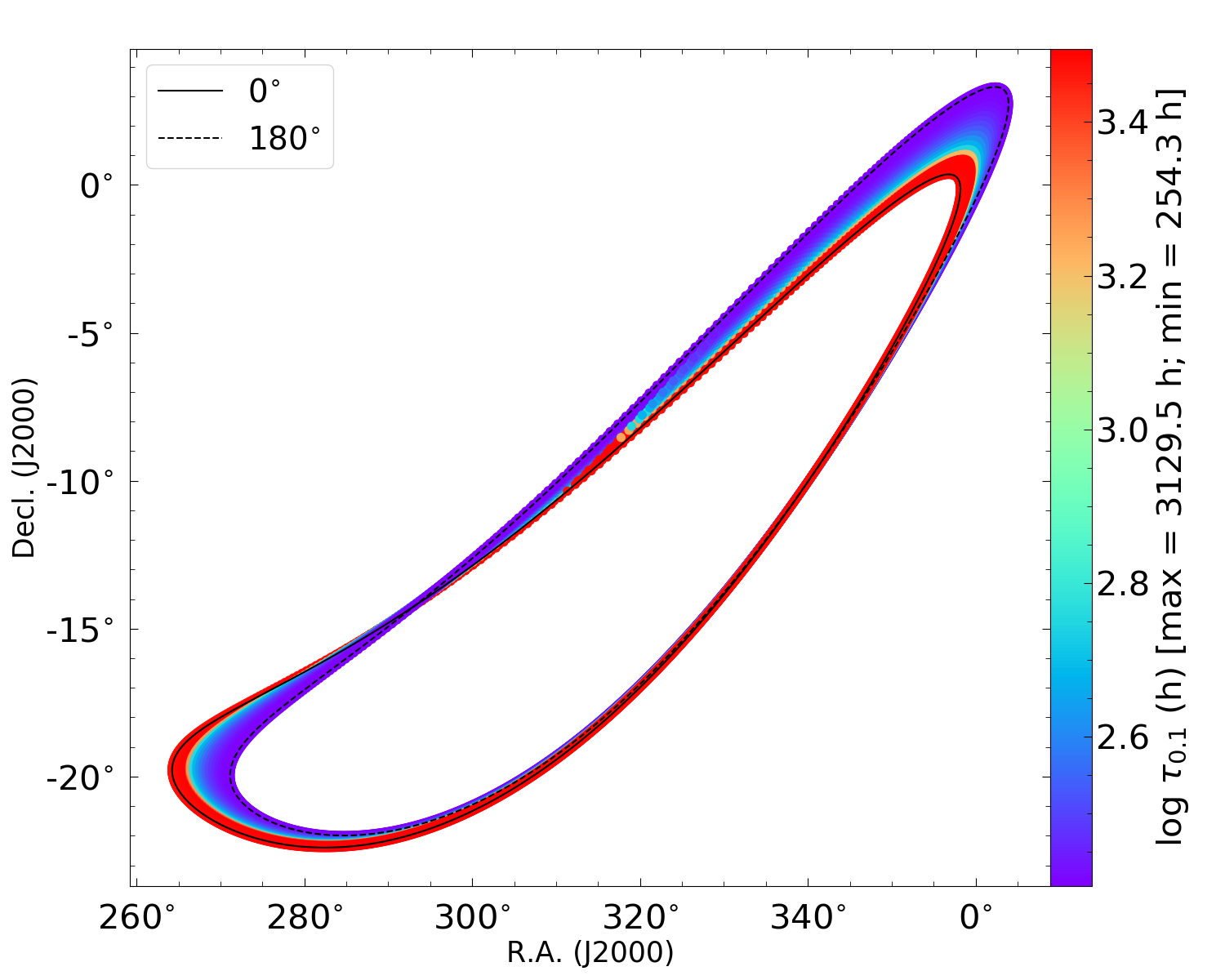}}
	\caption{$\tau_{1.0}$ and $\tau_{0.1}$ for Venus. The description of each map refers to Figure \ref{fig-times-Mercury}.}
	\label{fig-times-Venus}
\end{figure*}

This category includes Mercury and Venus. The computed positions in this work in this category manifest as a group of closed patterns (see Figures \ref{fig-times-Mercury} and \ref{fig-times-Venus}).

Figure \ref{fig-times-Mercury} shows the distributions of $\tau_{1.0}$ and $\tau_{0.1}$ (right panel) and they as functions of $(\theta_{\mathrm{CES}}, \phi_{\mathrm{CES}})$ (left panel) for Mercury. The minimum value of $\tau_{1.0}$ is 1.3 hr, which is larger than $\tau_{c,1.0,9200}$ and $\tau_{c,1.0,1400}$. This indicates that the gravitational field of Mercury affects SKA astrometry over single-epoch observations at 9200 and 1400 MHz under a precision level of 1.0 $\mu$as. This conclusion also applies to the case for a precision level of 0.1 $\mu$as.  $\tau_{1.0}$ is larger than 48.0 hr in a few cases, indicating that the gravitational field of Mercury hardly affects SKA astrometry over multi-epoch observations under a precision level of 1.0 $\mu$as. However, $\tau_{0.1}$ is larger than 48.0 hr in most cases, indicating that the gravitational field of Mercury is likely going to affect SKA astrometry over multi-epoch observations under a precision level of 0.1 $\mu$as. Table \ref{tab:duration-effect} summarizes our conclusions. It is worth noting that large values of $\tau_{1.0}$ and $\tau_{0.1}$ do not appear at $\phi_{\mathrm{CES}} = 0$ (i.e., the minimum distance from the Earth to Mercury), but at $\phi_{\mathrm{CES}}$ being about 20$^{\circ}$--40$^{\circ}$ and about 20$^{\circ}$--50$^{\circ}$, respectively. 

\begin{deluxetable*}{lccccccc}
	\tablecolumns{8}
	%	\tablenum{1}
%	\tabletypesize{\footnotesize}
%	\setlength\tabcolsep{3pt}
	%	\tablewidth{0pt}
%	\renewcommand{\arraystretch}{1.2}
	\tablecaption{Effect of Celestial Bodies on SKA Astrometry \label{tab:duration-effect}}
	%\begin{tabular}
	\tablehead{
		\colhead{Objects} & \colhead{Index} & \colhead{Precision Level} & \multicolumn{3}{c}{Single-Epoch} & & \colhead{Multi-epoch} \\
		\cline{4-6}
		\colhead{} & \colhead{} & \colhead{($\mu$as)} & \colhead{9200 MHz} & \colhead{1400 MHz} & \colhead{560 MHz} & & \colhead{} 
	}
	\startdata
	Mercury & 2 & 1.0 & Y & Y & N  & & N? \\
	        &   & 0.1 & Y & Y & N\tablenotemark{$\dagger$}  & & Y?  \\
	Venus   & 3 & 1.0 & Y & Y & Y  & & Y? \\
	        &   & 0.1 & Y & Y & Y?\tablenotemark{$\dagger$} & & Y  \\
	Mars    & 6 & 1.0 & Y & Y & Y? & & N? \\
	        &   & 0.1 & Y & Y & N?\tablenotemark{$\dagger$} & & Y? \\
	Uranus  &139& 1.0 & Y & Y & Y  & & Y  \\
	        &   & 0.1 & Y & Y & Y\tablenotemark{$\dagger$}  & & Y  \\
	Neptune &167& 1.0 & Y & Y & Y  & & Y  \\
            &   & 0.1 & Y & Y & Y\tablenotemark{$\dagger$}  & & Y  \\
    Pluto   &182& 1.0 & N & N & N  & & N  \\
            &   & 0.1 & Y & N?& N\tablenotemark{$\dagger$}  & & N? \\
    Ceres   &188& 1.0 & N & N & N  & & N  \\
            &   & 0.1 & N?& N?& N\tablenotemark{$\dagger$}  & & N  \\ 
	\enddata
	\tablenotetext{\dagger}{This result may be meaningless.}
	\tablecomments{Y = affects SKA astrometry, Y? = likely affects SKA astrometry, N? = hardly affects SKA astrometry, N = does not affect SKA astrometry.}
\end{deluxetable*}

From Figure \ref{fig-times-Venus} it can be seen that the effect of Venus is larger than that of Mercury. The minimum values of $\tau_{1.0}$ and $\tau_{0.1}$ are 25.4 and 254.3 hr, respectively. The maximum values of $\tau_{1.0}$ and $\tau_{0.1}$ are both larger than 912.5 hr (i.e., exceeding one month). Large values of $\tau_{1.0}$ and $\tau_{0.1}$ appear at $\phi_{\mathrm{CES}} \sim 0^{\circ}$--30$^{\circ}$ and $\sim 0^{\circ}$--40$^{\circ}$, respectively. Therefore, for Venus, its gravitational field can largely affect SKA astrometry over both single-epoch and multi-epoch observations (see the summary in Table \ref{tab:duration-effect}). 

\subsubsection{Category of Planets outside the Earth's Orbit}\label{sec:imapact-regions-duration:outside}

\begin{figure*}[!htb]
	\centering
%	\subfigbottomskip=-0.3cm
%	\subfigcapskip=-0.3cm
	\subfigure[$\tau_{1.0}$ as a function of $\theta_{\mathrm{CES}}$ and $\phi_{\mathrm{CES}}$]{\includegraphics[height=0.36\textwidth]{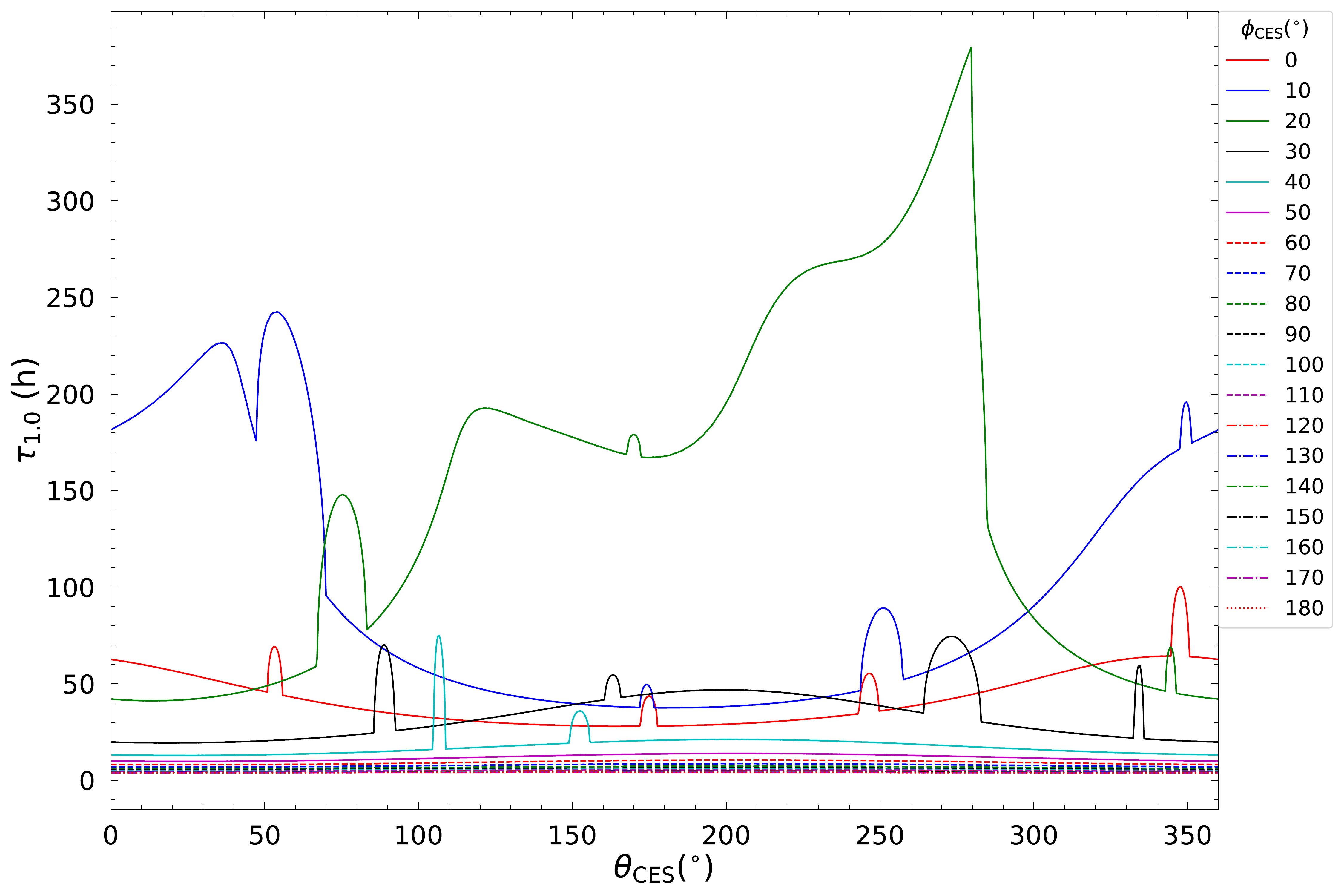}}
	\subfigure[Distribution of $\tau_{1.0}$]{\includegraphics[height=0.36\textwidth]{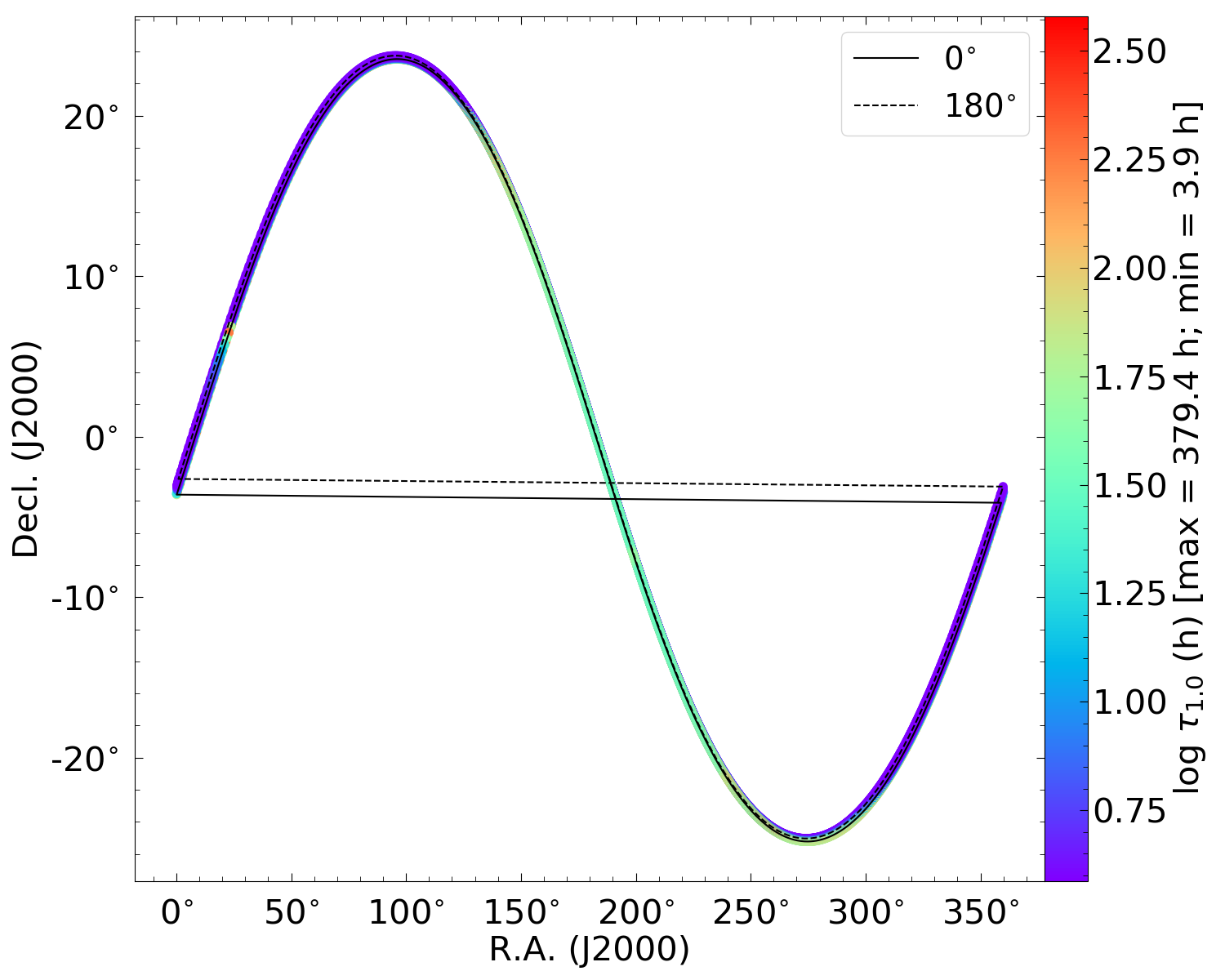}}	
	\subfigure[$\tau_{0.1}$ as a function of $\theta_{\mathrm{CES}}$ and $\phi_{\mathrm{CES}}$]{\includegraphics[height=0.354\textwidth]{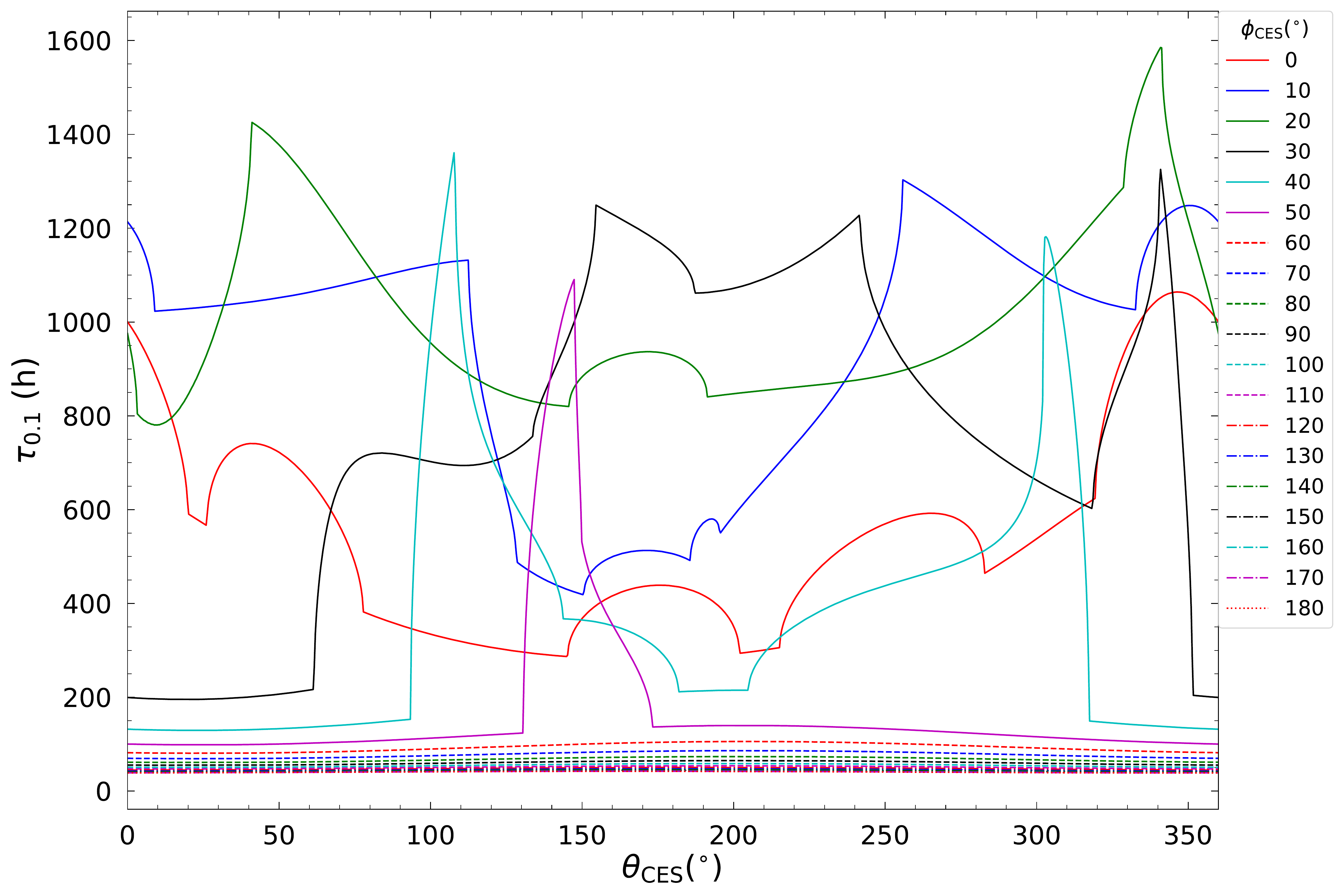}}
	\subfigure[Distribution of $\tau_{0.1}$]{\includegraphics[height=0.364\textwidth]{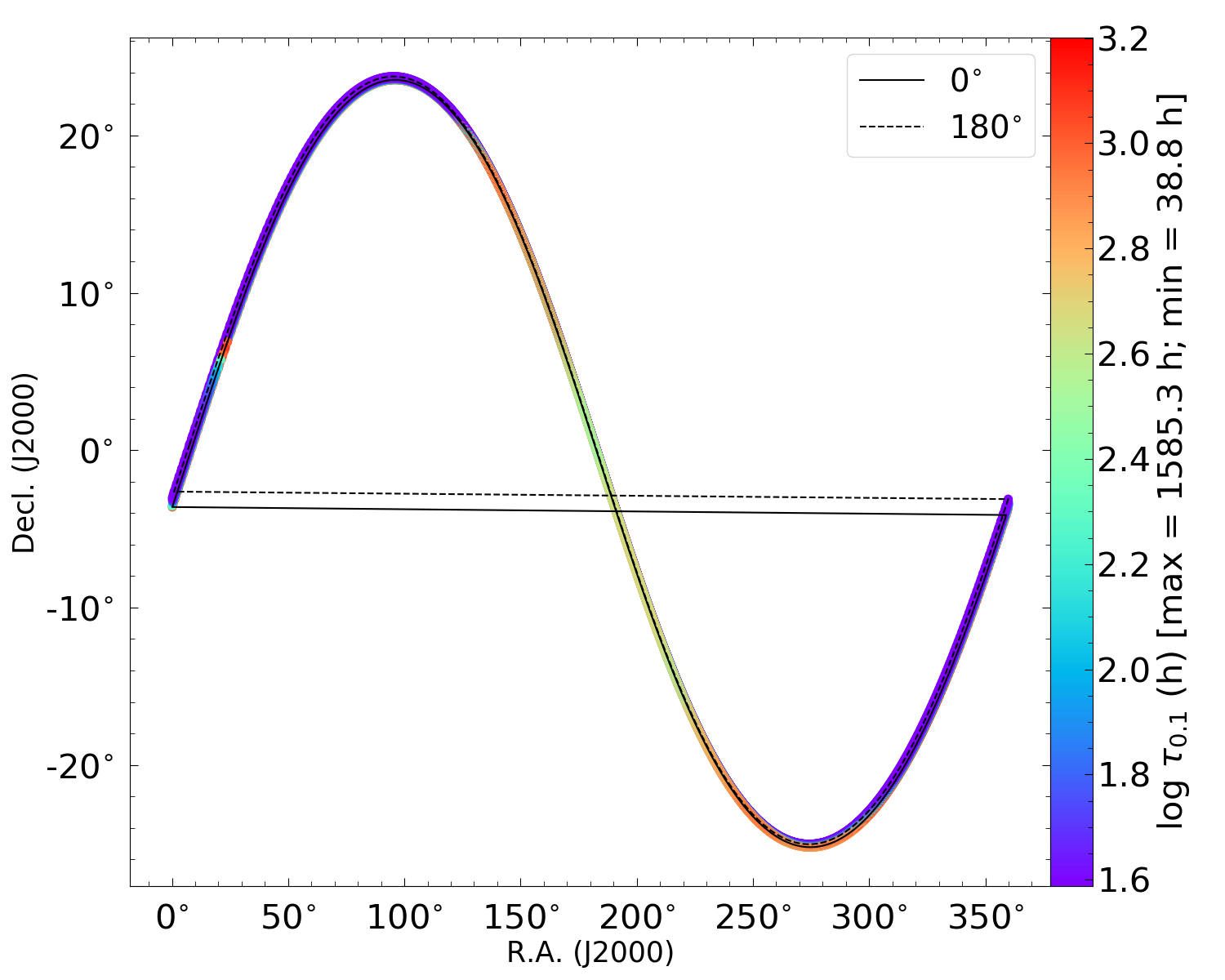}}
	\caption{$\tau_{1.0}$ and $\tau_{0.1}$ for Mars. The description of each map refers to Figure \ref{fig-times-Mercury}.}
	\label{fig-times-Mars}
\end{figure*}

\begin{figure*}[!htb]
	\centering
%	\subfigbottomskip=-0.3cm
%	\subfigcapskip=-0.3cm
	\subfigure[$\tau_{1.0}$ as a function of $\theta_{\mathrm{CES}}$ and $\phi_{\mathrm{CES}}$]{\includegraphics[height=0.354\textwidth]{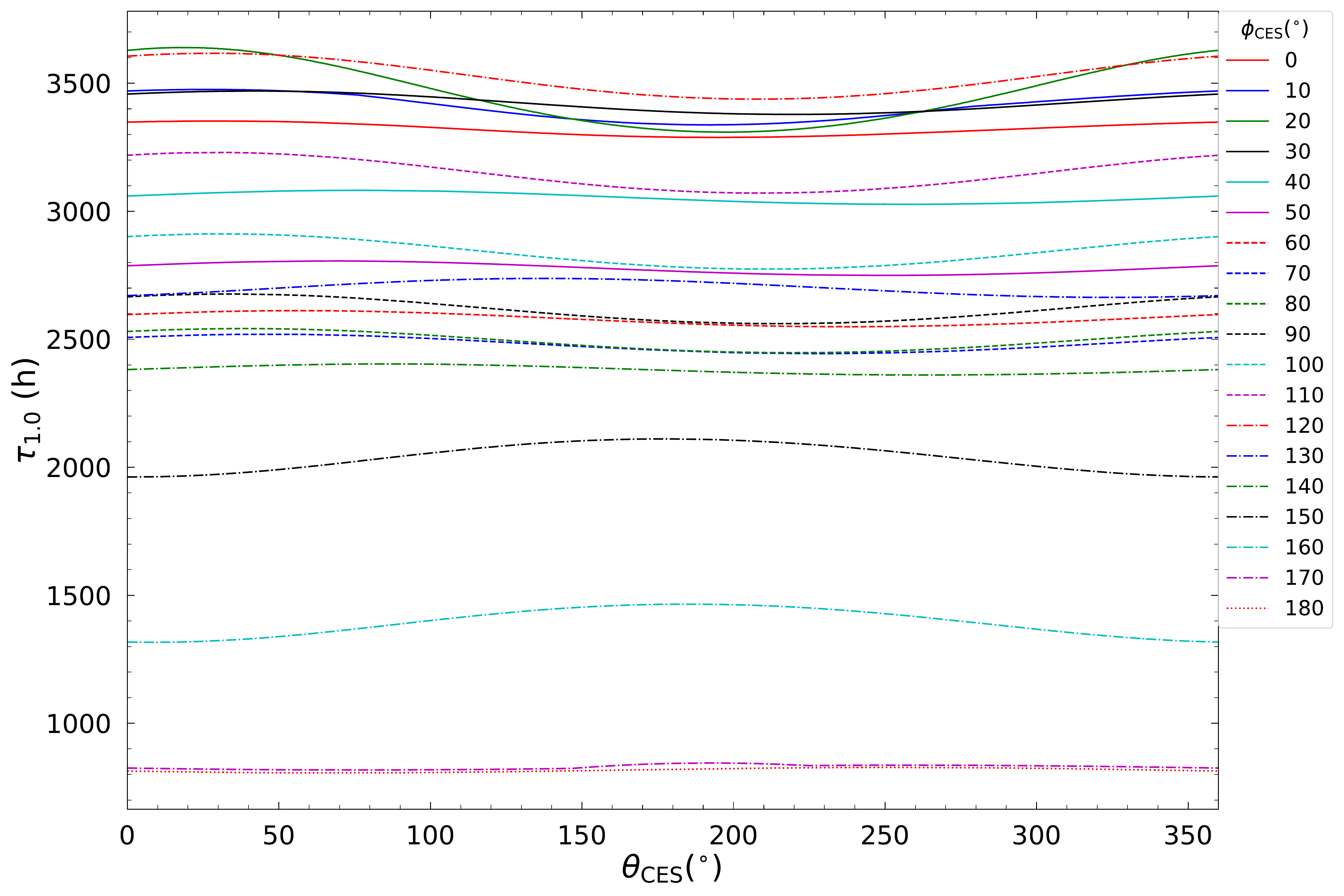}}
	\subfigure[Distribution of $\tau_{1.0}$]{\includegraphics[height=0.368\textwidth]{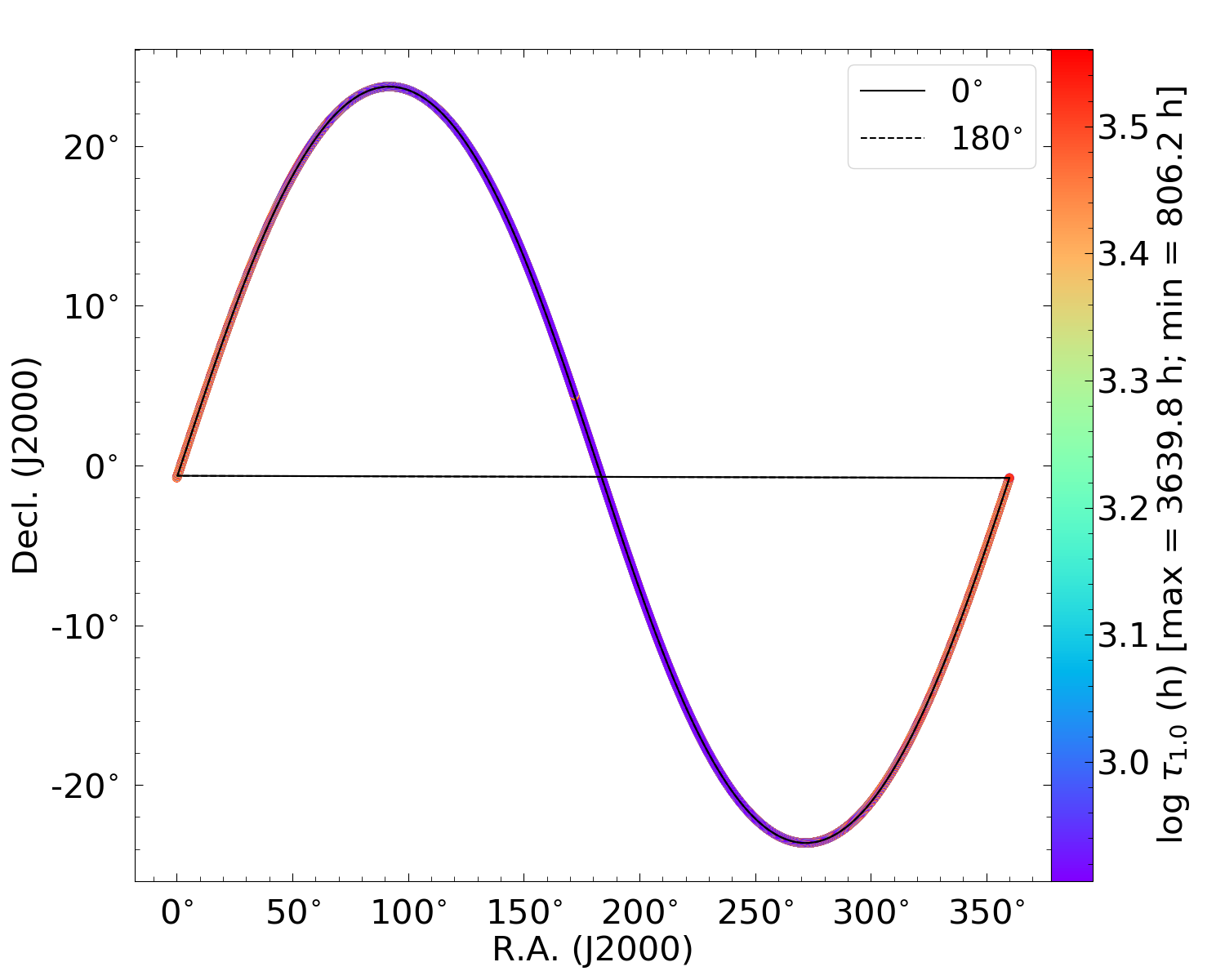}}	
	\subfigure[$\tau_{0.1}$ as a function of $\theta_{\mathrm{CES}}$ and $\phi_{\mathrm{CES}}$]{\includegraphics[height=0.346\textwidth]{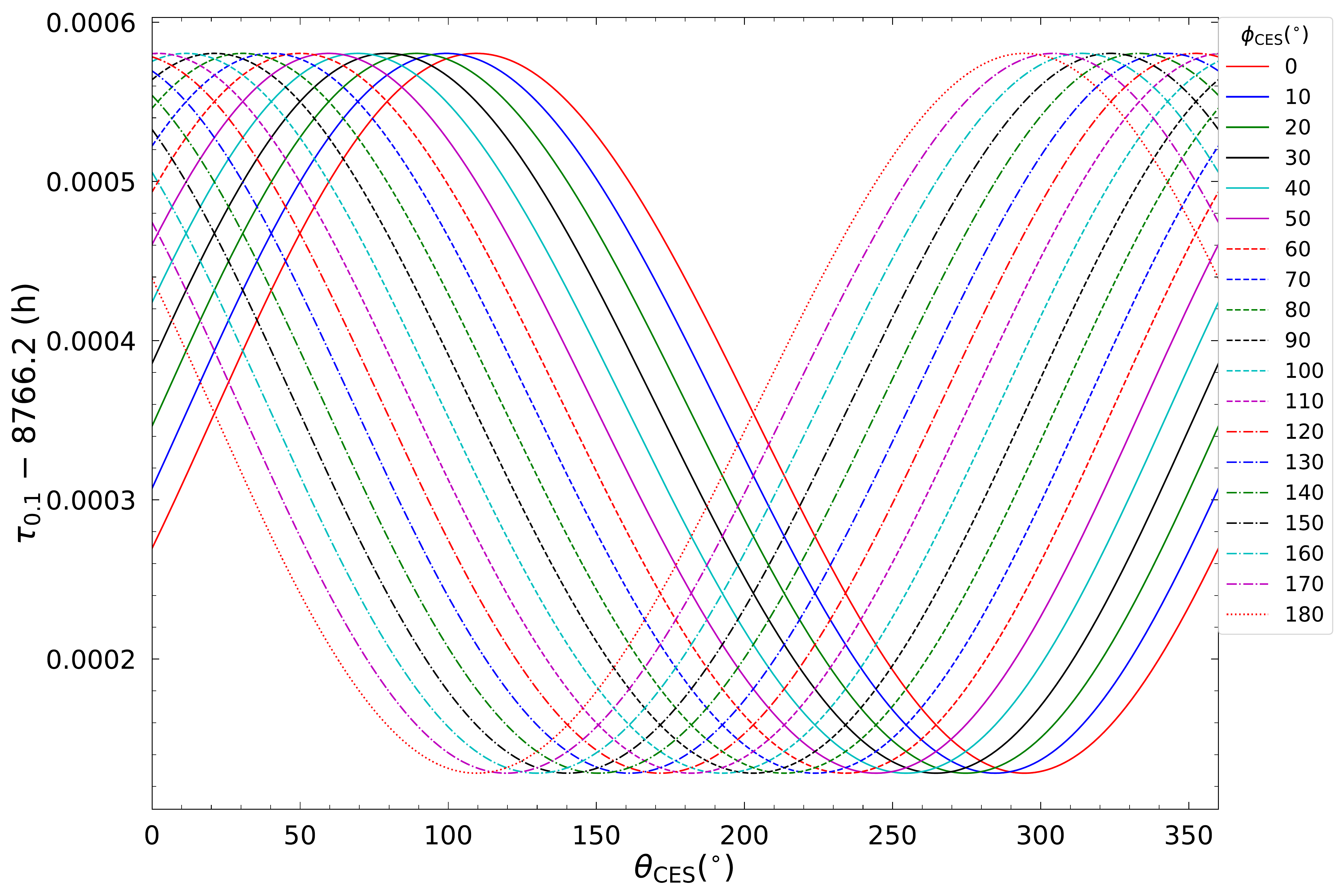}}
	\subfigure[Distribution of $\tau_{0.1}$]{\includegraphics[height=0.365\textwidth]{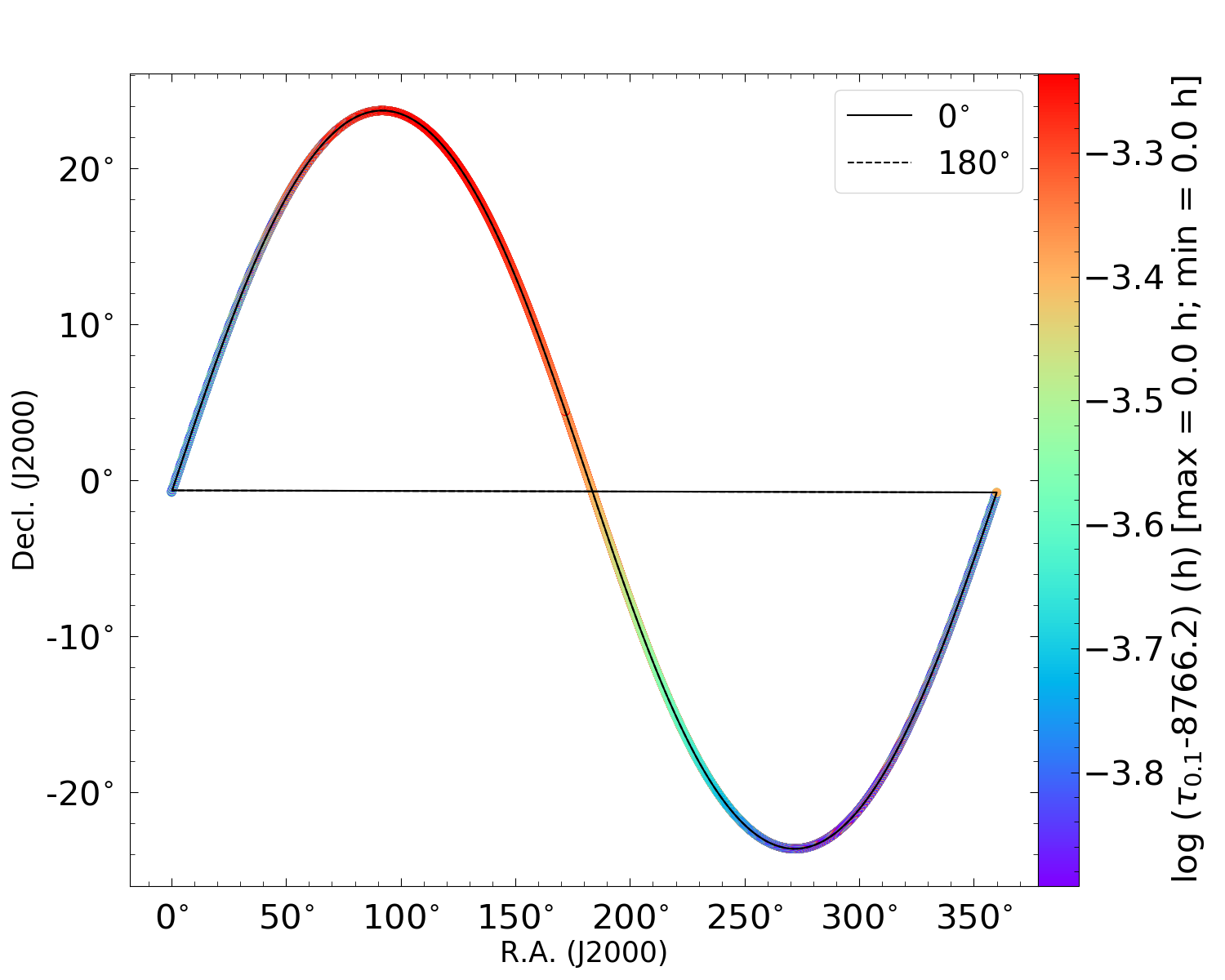}}
	\caption{$\tau_{1.0}$ and $\tau_{0.1}$ for Uranus. The description of each map refers to Figure \ref{fig-times-Mercury}.}
	\label{fig-times-Uranus}
\end{figure*}

\begin{figure*}[!htb]
	\centering
%	\subfigbottomskip=-0.3cm
%	\subfigcapskip=-0.3cm
	\subfigure[$\tau_{1.0}$ as a function of $\theta_{\mathrm{CES}}$ and $\phi_{\mathrm{CES}}$]{\includegraphics[height=0.354\textwidth]{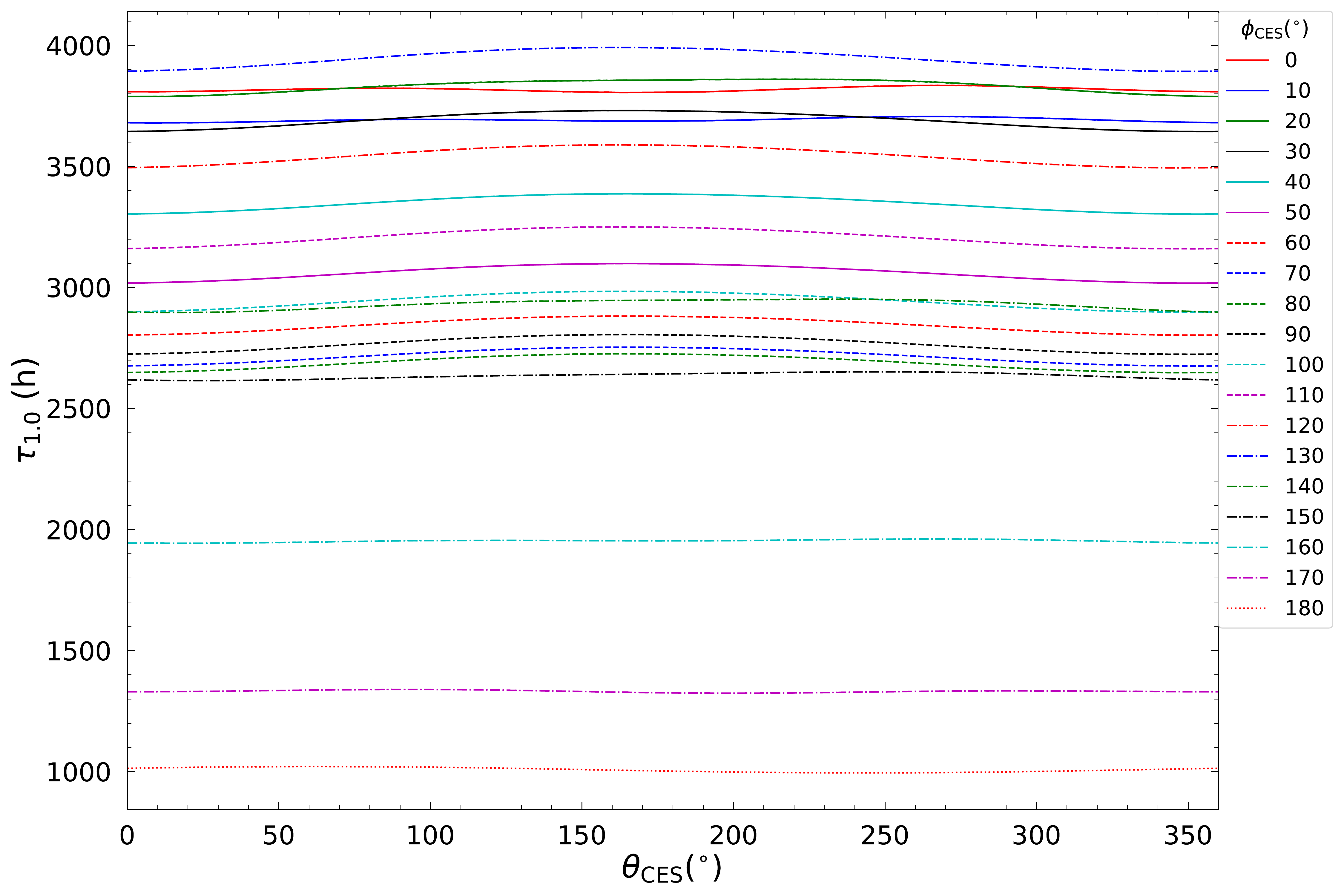}}
	\subfigure[Distribution of  $\tau_{1.0}$]{\includegraphics[height=0.368\textwidth]{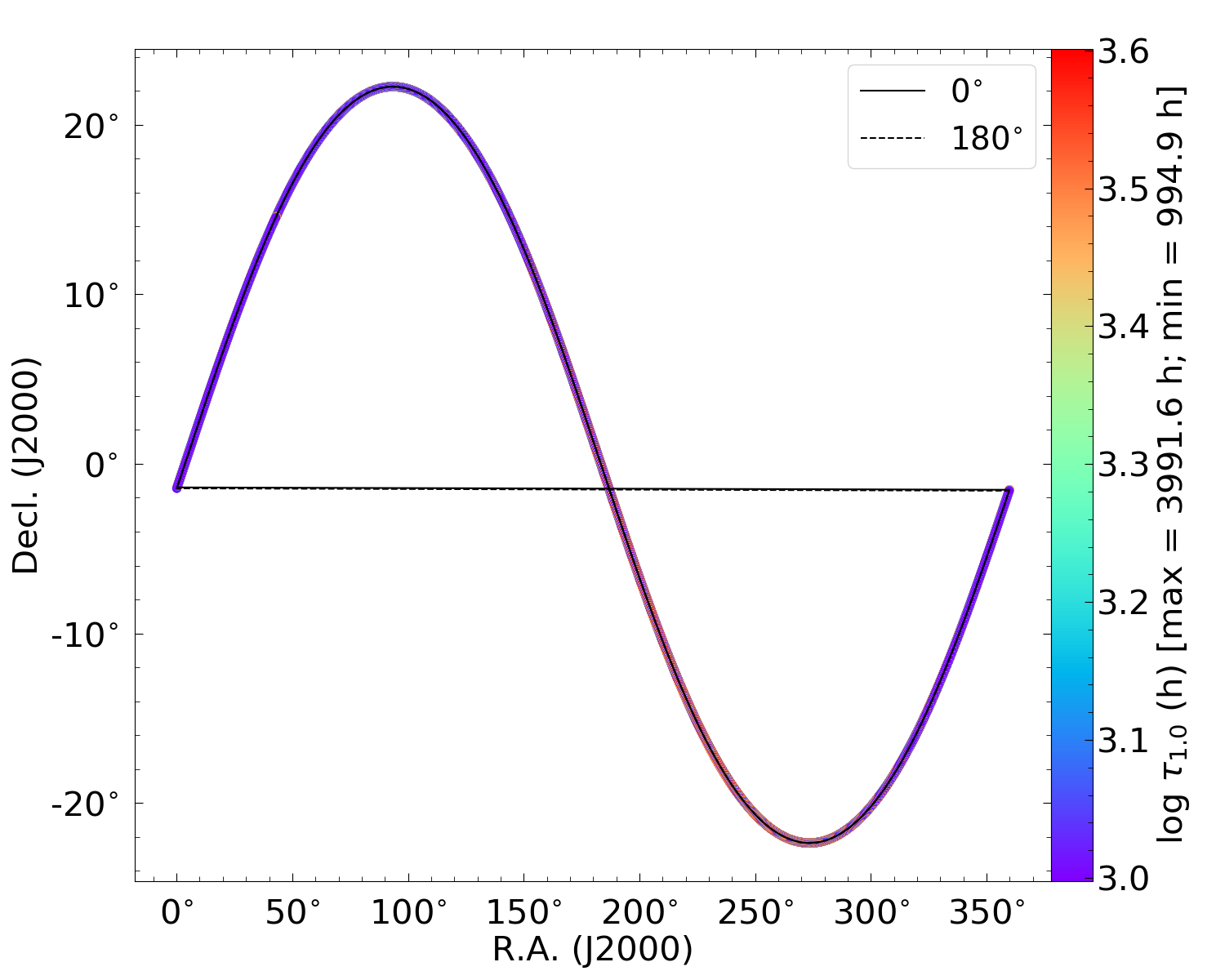}}	
	\subfigure[$\tau_{0.1}$ as a function of $\theta_{\mathrm{CES}}$ and $\phi_{\mathrm{CES}}$]{\includegraphics[height=0.35\textwidth]{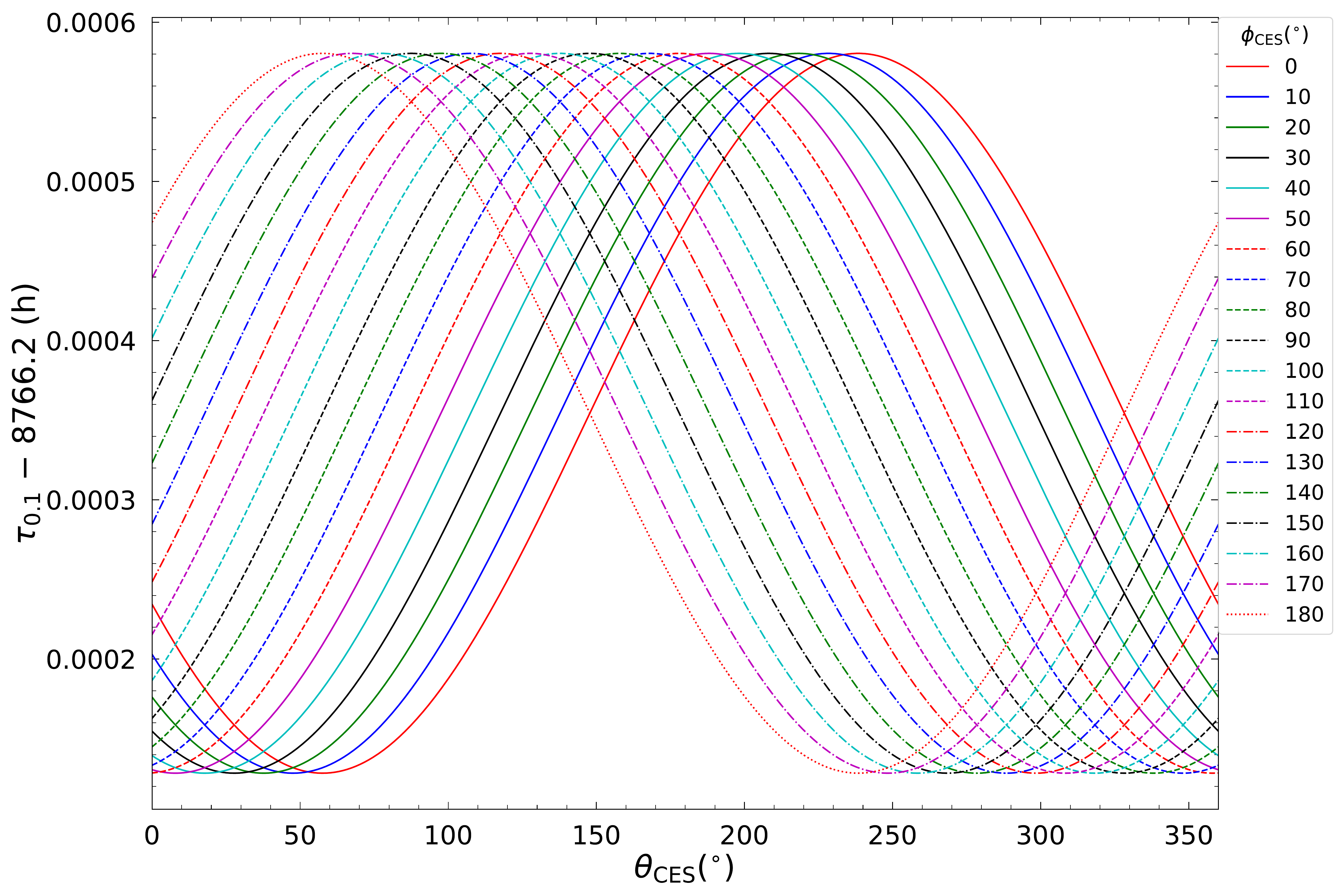}}
	\subfigure[Distribution of $\tau_{0.1}$]{\includegraphics[height=0.371\textwidth]{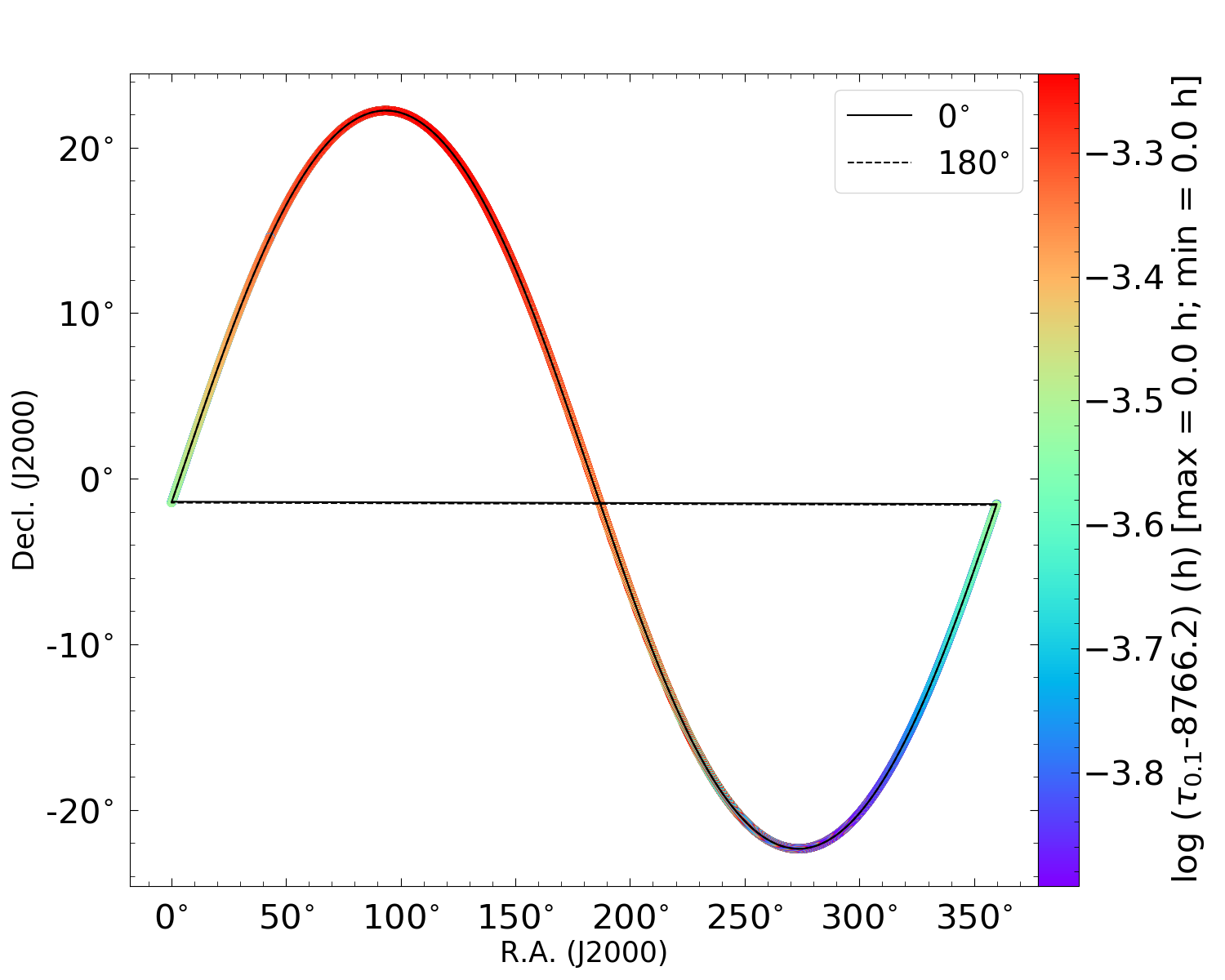}}
	\caption{$\tau_{1.0}$ and $\tau_{0.1}$ for Neptune. The description of each map refers to Figure \ref{fig-times-Mercury}.}
	\label{fig-times-Neptune}
\end{figure*}

This category includes Mars, Uranus, and Neptune. Unlike the category inside the Earth's orbit, the computed points of this category are a group of ribbons with R.A.(J2000) ranging from 0$^{\circ}$ to 360$^{\circ}$ (see Figures \ref{fig-times-Mars}--\ref{fig-times-Neptune}).

The effect of Mars is a little greater than that of Mercury (see Figure \ref{fig-times-Mars}). The minimum values of $\tau_{1.0}$ and $\tau_{0.1}$ are 3.9 and 38.8 hr, respectively. $\tau_{1.0}$ is less than 10 hr and $\tau_{0.1}$ less than 100 hr in most cases. The maximum values of $\tau_{1.0}$ and $\tau_{0.1}$ are 379.4 and 1585.3 hr, respectively. Large values of $\tau_{1.0}$ and $\tau_{0.1}$ appear at about 10--20$^{\circ}$ and about 0--40$^{\circ}$, respectively. Mars's gravitational field can affect astrometry over single-epoch observations at precision levels of both 1.0 and 0.1 $\mu$as, and will likely affect multi-epoch observations at a precision level of 0.1 $\mu$as, but will hardly affect multi-epoch observations at a precision level of 1.0 $\mu$as (see Table \ref{tab:duration-effect}). 

From Figures \ref{fig-times-Uranus} and \ref{fig-times-Neptune}, it can be seen that the effects of Uranus and Neptune are larger than that of Venus. The minimum value of $\tau_{1.0}$ is larger than 994.9 hr (i.e., exceeding one month). The value of $\tau_{0.1}$ is $\sim$ 1 yr, which is the period of revolution of Earth around the Sun, indicating that the obtained value reaches the calculation limit of this work. Therefore, for Uranus and Neptune, their gravitational fields can largely affect astrometry for both single-epoch and multi-epoch observations (see Table \ref{tab:duration-effect}). 

\subsubsection{Category of Others}\label{sec:imapact-regions-duration:other}

\begin{figure*}[!htb]
	\centering
%	\subfigbottomskip=-0.3cm
%	\subfigcapskip=-0.3cm
	\subfigure[$\tau_{1.0}$ as a function of $\theta_{\mathrm{CES}}$ and $\phi_{\mathrm{CES}}$]{\includegraphics[height=0.36\textwidth]{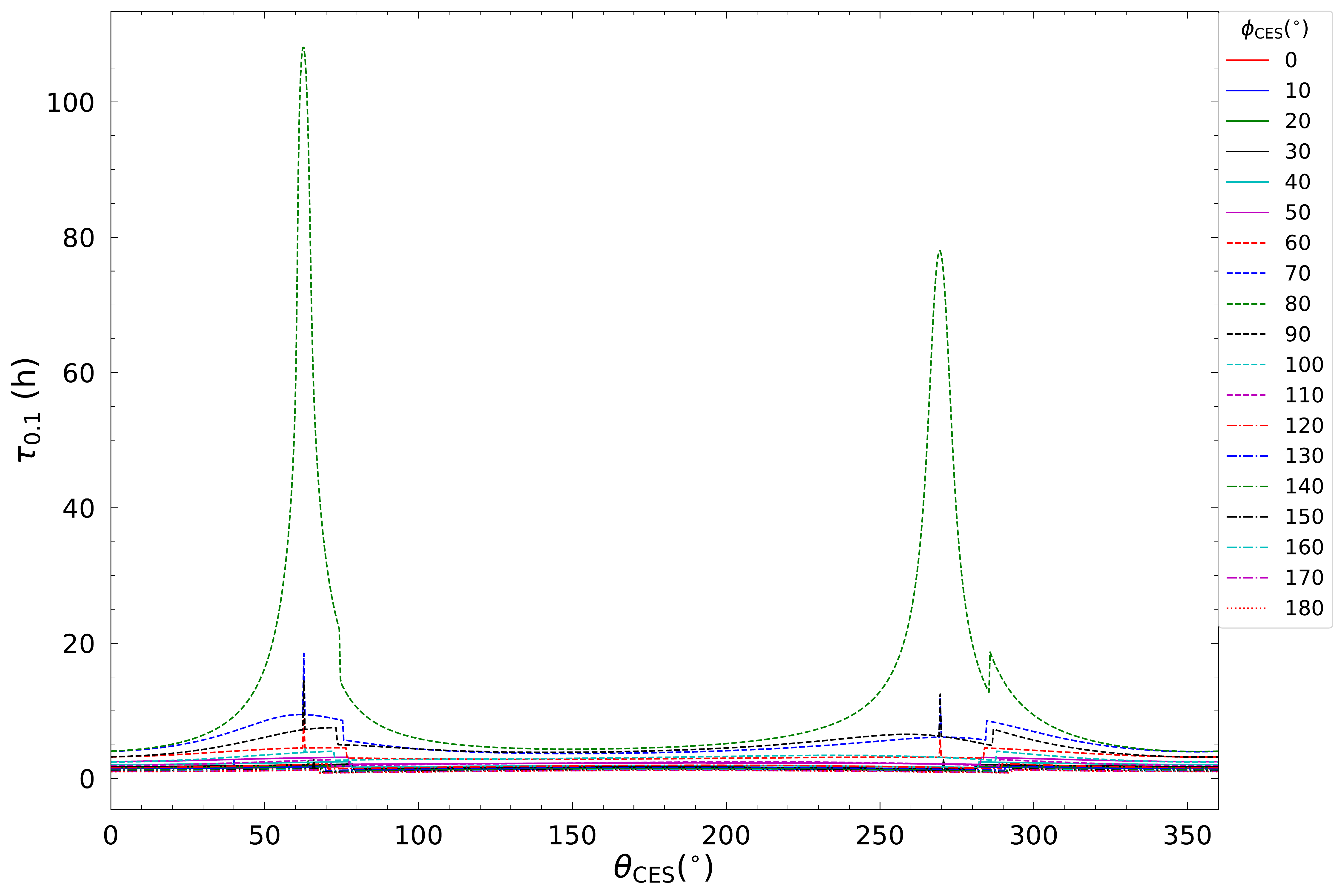}}
	\subfigure[Distribution of $\tau_{1.0}$]{\includegraphics[height=0.36\textwidth]{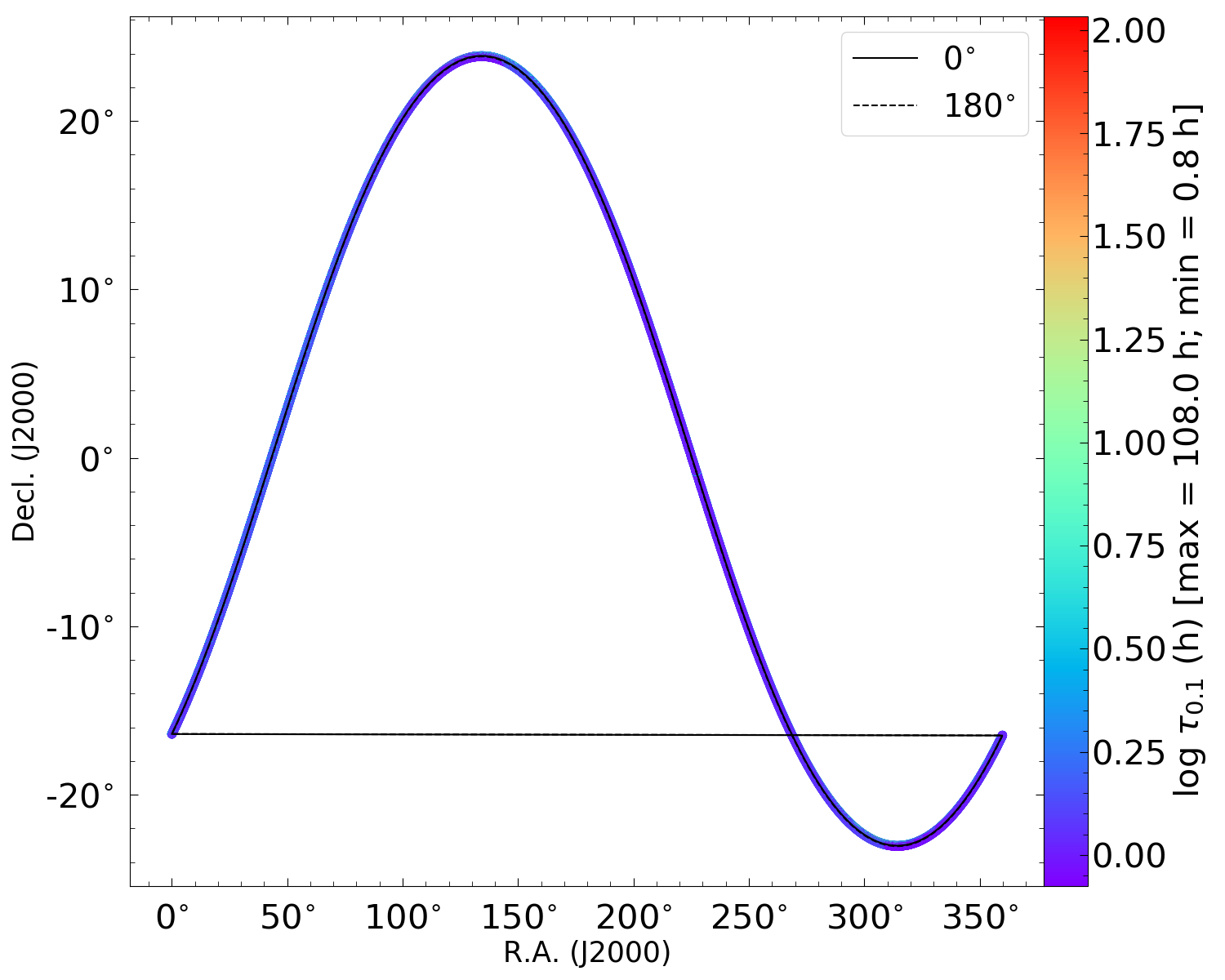}}
	\caption{$\tau_{0.1}$ for Pluto. The description of each map refers to Figure \ref{fig-times-Mercury}.}
	\label{fig-times-Pluto}
\end{figure*}

\begin{figure*}[!htb]
	\centering
%	\subfigbottomskip=-0.3cm
%	\subfigcapskip=-0.3cm
	\subfigure[$\tau_{1.0}$ as a function of $\theta_{\mathrm{CES}}$ and $\phi_{\mathrm{CES}}$]{\includegraphics[height=0.36\textwidth]{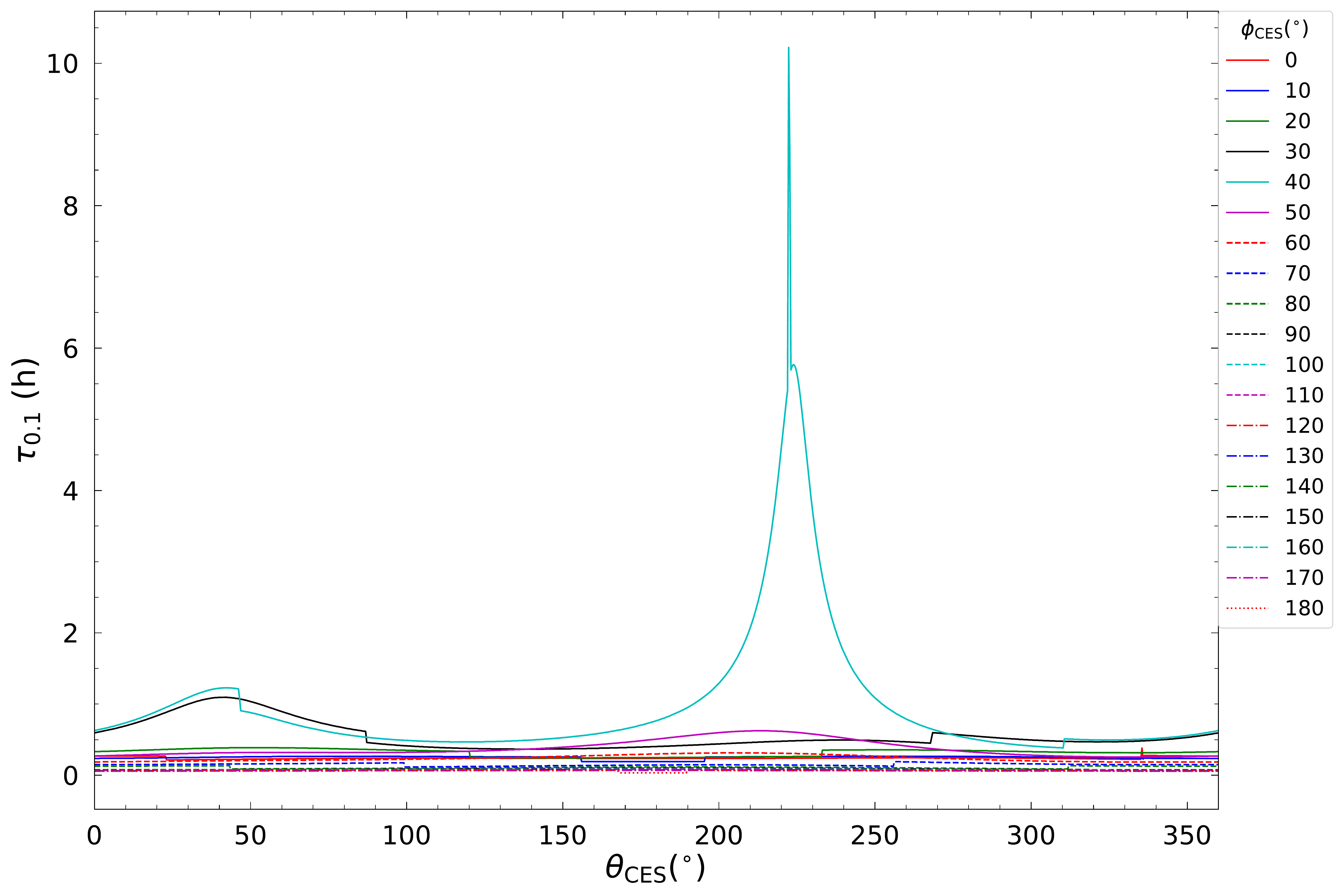}}
	\subfigure[Distribution of $\tau_{1.0}$]{\includegraphics[height=0.36\textwidth]{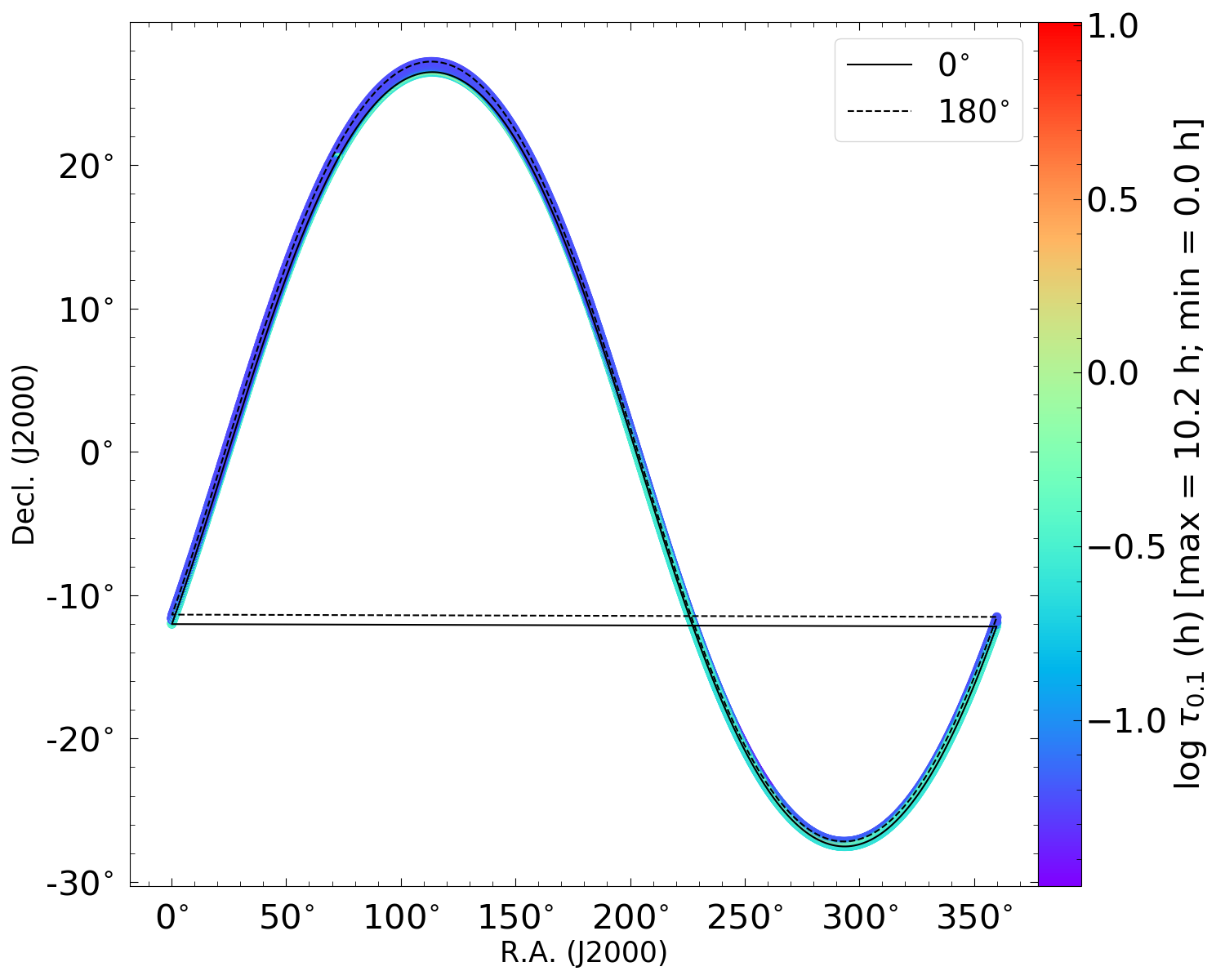}}
	\caption{$\tau_{0.1}$ for Ceres. The description of each map refers to Figure \ref{fig-times-Mercury}.}
	\label{fig-times-Ceres}
\end{figure*}

This category includes Pluto and Ceres. Objects in this category are also outside the Earth's orbit, and thus show similar trajectories as the category of planets outside the Earth's orbit, i.e., a group of ribbons with R.A.(J2000) ranging from 0$^{\circ}$ to 360$^{\circ}$ (see Figures \ref{fig-times-Pluto}--\ref{fig-times-Ceres}). For these two objects, the values of $\beta_{\mathrm{1.0, max}}$ are less than their angular radii seen from the Earth. Therefore, the gravitational fields of Pluto and Ceres do not affect SKA astrometry under a precision level of 1.0 $\mu$as, and we did not calculate $\tau_{1.0}$.

For Pluto, the maximum value of $\tau_{0.1}$ is 108.0 hr, but it only appears at $\phi_{\mathrm{CES}} \sim 80^{\circ}$ (see Figure \ref{fig-times-Pluto}). The minimum value of $\tau_{0.1}$ exceeds 8.0 min, and the value of $\tau_{0.1}$ exceeding 4.0 hr only appears at $\phi_{\mathrm{CES}}$ $\sim$ 60$^{\circ}$--90$^{\circ}$. 
Therefore, Pluto's gravitational field can largely affect astrometry over single-epoch observations at 9200 MHz, but will hardly affect single-epoch observations at 1400 MHz, and will hardly affect multi-epoch observations (see Table \ref{tab:duration-effect}). 
%Therefore, for Pluto, the effect of its gravitational field on SKA astrometry over single-epoch observations may be small at 1400 MHz but may be large at 9200 MHz. In addition, Pluto's gravitational field will hardly affect multi-epoch observations (see Table \ref{tab:duration-effect}). 

For Ceres, the maximum value of $\tau_{0.1}$ is 10.2 hr, but it only appears at $\phi_{\mathrm{CES}} \sim 40^{\circ}$ (see Figures \ref{fig-times-Ceres}).  
The groups of values of $\tau_{1.0}$ exceeding 8 min only appear at $\phi_{\mathrm{CES}}$ $\sim$ 0$^{\circ}$--60$^{\circ}$.
Therefore, its gravitational field will hardly affect SKA astrometry (see Table \ref{tab:duration-effect}). %The gravitational fields of asteroids other than Ceres do not affect SKA astrometry.
% 对于Pallas,其beta最大1角秒，只在其近心点成立，且此时地球在远心点，因此其最大时间不会超过Ceres的最长时间的1/4，因此，最多在极少情况下对9200 GHz可能有影响。粗略计算，在上述情况下，1角秒对应于823.9 km，Pallas总速度为约22.61 km/s，地球近圆轨道，按平均公转速度来算，地球总速度为约29.78 km/s，相对速度为约7.2 km，因此，对应时间估计约为115 s。这表明，即使对应于9200 GHz，Pallas也不会有影响。

\section{Summary and Conclusions}\label{sec-summary}

We have calculated the maximum deflection angle caused by 195 objects in the solar system, including the Sun, all planets, 177 satellites (contains the Moon), and eight asteroids with $GM > 0.1$ km$^{3}$ s$^{-2}$ (see Table \ref{tab:original-para}). An overview of the deflection angle caused by these objects is as follows (see Table \ref{tab:betas}).

\begin{enumerate}
	\item Twenty-one satellites and six asteroids can deflect light from CESs by more than 0.1 $\mu$as, and 14 satellites and one asteroid (i.e., Ceres) can bend light by more than 1.0 $\mu$as.
	\item Jupiter and Saturn can bend light by amount of 0.1 $\mu$as up to 100$^{\circ}$ and by amount of 1.0 $\mu$as up to dozens of degrees. The ranges of influence under 0.1 $\mu$as and 1.0 $\mu$as for the other planets (other than the Earth) and the Moon are $1.4^{\circ}$--63.6$^{\circ}$ and 0.1$^{\circ}$--7.1$^{\circ}$, respectively. But for the satellites and Ceres, the corresponding ranges of influence are all from a few arcseconds to a few hundred arcseconds.
\end{enumerate}

Further computations, regarding the zones and durations of perturbations caused by the celestial bodies, were made towards Mercury, Venus, Mars, Uranus, Neptune, Pluto, and Ceres, and the main results are as follows.

\begin{enumerate}
	\item The computed perturbation zones with deflection angles larger than 1.0 $\mu$as are ribbons with a width of several degrees or less, except for Venus whose corresponding ribbon width is $\sim$ 13$^{\circ}$--16$^{\circ}$. For deflection angles larger than 0.1 $\mu$as, the widths increase to more than a dozen degrees for Uranus and Neptune, increase to close to 80$^{\circ}$ for Venus, and remain almost unchanged for Mercury, Mars, Pluto, and Ceres (see Table \ref{tab:coverage} and Figure \ref{fig-coverage}).
	\item The computation of the durations of perturbations (see Figures 4--10) posed by the gravitational field of solar system objects indicates that the gravitational field of Ceres will hardly affect SKA astrometry, but that of Pluto may have a small effect on the astrometry of single-epoch observations under a precision level of 0.1 $\mu$as. The gravitational fields of Mercury and Mars may have a great influence on astrometry for single-epoch observations at precision levels of both 1.0 and 0.1 $\mu$as, and those of other objects can largely affect astrometry for both single and multi-epoch observations (see Table \ref{tab:duration-effect}).
\end{enumerate}

\acknowledgments

We would like to thank the anonymous referee for the helpful comments and suggestions that helped to improve the paper. This work was sponsored by the Natural Science Foundation of Jiangsu Province (Grants No. BK20210999), the Entrepreneurship and Innovation Program of Jiangsu Province, the NSFC Grants Nos. 12203104, 11933011 and 11873019, the Key Laboratory for Radio Astronomy, CAS. 

%\software{Astropy \citep{Astropy2013,Astropy2018}, Matplotlib \citep{Matplotlib2007}, Numpy \citep{Numpy2020}, Pandas \citep{Pandas2010, Pandas2021}.}

\clearpage
\appendix

Table \ref{tab:original-para} presents the maximum deflection angle caused by 195 objects in the solar system.

\setcounter{table}{0}
\renewcommand{\thetable}{A\arabic{table}}

\startlongtable
\begin{deluxetable}{lcccccccccc}
	\tablecolumns{11}
	%	\tablenum{1}
	\tabletypesize{\small}
	\setlength\tabcolsep{3pt}
	%	\tablewidth{0pt}
	\renewcommand{\arraystretch}{1.2}
	\tablecaption{Maximum Deflction Angle, $\alpha_\mathrm{max}$, under the Gravitational Field of Celestial Bodies in Solar System \label{tab:original-para}}
	%\begin{tabular}
	\tablehead{
		\colhead{Objects} & \colhead{Index} & \colhead{$GM$} & \colhead{$R_\mathrm{L}$}　& \colhead{$a$} & \colhead{$e$} & \colhead{$I$} &  \colhead{$\omega$} & \colhead{$\Omega$}  & \colhead{$R_\mathrm{sp}$} & \colhead{$\alpha_\mathrm{max}$} \\
		\colhead{} & \colhead{} & \colhead{(km$^3$ s$^{-2}$)} & \colhead{(km)} & \colhead{(au)} & \colhead{} & \colhead{($^{\circ}$)} & \colhead{($^{\circ}$)} & \colhead{($^{\circ}$)} & \colhead{} & \colhead{($\mu$as)}
	}
	\startdata
	Sun	&	1	&	1.33E+11	&	695700.0 	&	...	&	...	&	...	&	...	&	...	&	0.00 	&	1751190.33 	\\
	Mercury	&	2	&	2.20E+04	&	2439.4 	&	0.387 	&	0.206 	&	7.00 	&	29.13 	&	48.33 	&	0.00 	&	82.91 	\\
	Venus	&	3	&	3.25E+05	&	6051.8 	&	0.723 	&	0.007 	&	3.39 	&	54.92 	&	76.68 	&	0.00 	&	492.78 	\\
	Earth\tablenotemark{$\dagger$}	&	4	&	3.99E+05	&	6378.1 	&	1.000 	&	0.017 	&	0.00 	&	102.94 	&	0.00 	&	0.00 	&	543.05 	\\
	Moon	&	5	&	4.90E+03	&	1737.5 	&	0.003 	&	0.055 	&	...	&	...	&	...	&	0.00 	&	25.90 	\\
	Mars	&	6	&	4.28E+04	&	3389.5 	&	1.524 	&	0.093 	&	1.85 	&	-25.62 	&	49.56 	&	0.00 	&	116.00 	\\
	Phobos	&	7	&	7.11E-04	&	11.1 	&	...	&	...	&	...	&	...	&	...	&	0.00 	&	0.00 	\\
	Deimos	&	8	&	9.85E-05	&	6.2 	&	...	&	...	&	...	&	...	&	...	&	0.00 	&	0.00 	\\
	Jupiter	&	9	&	1.27E+08	&	69911.0 	&	5.203 	&	0.048 	&	1.30 	&	-85.75 	&	100.47 	&	0.00 	&	16635.22 	\\
	Ganymede	&	10	&	9.89E+03	&	2631.2 	&	...	&	...	&	...	&	...	&	...	&	0.00 	&	34.50 	\\
	Callisto	&	11	&	7.18E+03	&	2410.3 	&	...	&	...	&	...	&	...	&	...	&	0.00 	&	27.34 	\\
	Io	&	12	&	5.96E+03	&	1821.6 	&	...	&	...	&	...	&	...	&	...	&	0.00 	&	30.04 	\\
	Europa	&	13	&	3.20E+03	&	1560.8 	&	...	&	...	&	...	&	...	&	...	&	0.00 	&	18.84 	\\
	Himalia	&	14	&	4.50E-01	&	85.0 	&	...	&	...	&	...	&	...	&	...	&	0.02 	&	0.05 	\\
	Amalthea	&	15	&	1.38E-01	&	83.5 	&	...	&	...	&	...	&	...	&	...	&	0.00 	&	0.02 	\\
	Thebe	&	16	&	1.00E-01	&	49.3 	&	...	&	...	&	...	&	...	&	...	&	0.00 	&	0.02 	\\
	Elara	&	17	&	5.80E-02	&	43.0 	&	...	&	...	&	...	&	...	&	...	&	0.02 	&	0.01 	\\
	Pasiphae	&	18	&	2.00E-02	&	30.0 	&	...	&	...	&	...	&	...	&	...	&	0.04 	&	0.01 	\\
	Carme	&	19	&	8.80E-03	&	23.0 	&	...	&	...	&	...	&	...	&	...	&	0.04 	&	0.00 	\\
	Metis	&	20	&	8.00E-03	&	21.5 	&	...	&	...	&	...	&	...	&	...	&	0.00 	&	0.00 	\\
	Sinope	&	21	&	5.00E-03	&	19.0 	&	...	&	...	&	...	&	...	&	...	&	0.04 	&	0.00 	\\
	Lysithea	&	22	&	4.20E-03	&	18.0 	&	...	&	...	&	...	&	...	&	...	&	0.02 	&	0.00 	\\
	Ananke	&	23	&	2.00E-03	&	14.0 	&	...	&	...	&	...	&	...	&	...	&	0.04 	&	0.00 	\\
	Leda	&	24	&	7.30E-04	&	10.0 	&	...	&	...	&	...	&	...	&	...	&	0.02 	&	0.00 	\\
	Adrastea	&	25	&	5.00E-04	&	8.2 	&	...	&	...	&	...	&	...	&	...	&	0.00 	&	0.00 	\\
	Callirrhoe	&	26	&	5.80E-05	&	4.3 	&	...	&	...	&	...	&	...	&	...	&	0.04 	&	0.00 	\\
	Themisto	&	27	&	4.60E-05	&	4.0 	&	...	&	...	&	...	&	...	&	...	&	0.01 	&	0.00 	\\
	Praxidike	&	28	&	2.90E-05	&	3.4 	&	...	&	...	&	...	&	...	&	...	&	0.04 	&	0.00 	\\
	Megaclite	&	29	&	1.40E-05	&	2.7 	&	...	&	...	&	...	&	...	&	...	&	0.04 	&	0.00 	\\
	Iocaste	&	30	&	1.30E-05	&	2.6 	&	...	&	...	&	...	&	...	&	...	&	0.04 	&	0.00 	\\
	Kalyke	&	31	&	1.30E-05	&	2.6 	&	...	&	...	&	...	&	...	&	...	&	0.04 	&	0.00 	\\
	Taygete	&	32	&	1.10E-05	&	2.5 	&	...	&	...	&	...	&	...	&	...	&	0.04 	&	0.00 	\\
	Harpalyke	&	33	&	8.00E-06	&	2.2 	&	...	&	...	&	...	&	...	&	...	&	0.04 	&	0.00 	\\
	Aoede	&	34	&	6.00E-06	&	2.0 	&	...	&	...	&	...	&	...	&	...	&	0.04 	&	0.00 	\\
	Autonoe	&	35	&	6.00E-06	&	2.0 	&	...	&	...	&	...	&	...	&	...	&	0.04 	&	0.00 	\\
	Eukelade	&	36	&	6.00E-06	&	2.0 	&	...	&	...	&	...	&	...	&	...	&	0.04 	&	0.00 	\\
	Helike	&	37	&	6.00E-06	&	2.0 	&	...	&	...	&	...	&	...	&	...	&	0.04 	&	0.00 	\\
	Hermippe	&	38	&	6.00E-06	&	2.0 	&	...	&	...	&	...	&	...	&	...	&	0.04 	&	0.00 	\\
	S/2003 J5	&	39	&	6.00E-06	&	2.0 	&	...	&	...	&	...	&	...	&	...	&	0.04 	&	0.00 	\\
	Thyone	&	40	&	6.00E-06	&	2.0 	&	...	&	...	&	...	&	...	&	...	&	0.04 	&	0.00 	\\
	Chaldene	&	41	&	5.00E-06	&	1.9 	&	...	&	...	&	...	&	...	&	...	&	0.04 	&	0.00 	\\
	Isonoe	&	42	&	5.00E-06	&	1.9 	&	...	&	...	&	...	&	...	&	...	&	0.04 	&	0.00 	\\
	Aitne	&	43	&	3.00E-06	&	1.5 	&	...	&	...	&	...	&	...	&	...	&	0.04 	&	0.00 	\\
	Arche	&	44	&	3.00E-06	&	1.5 	&	...	&	...	&	...	&	...	&	...	&	0.04 	&	0.00 	\\
	Carpo	&	45	&	3.00E-06	&	1.5 	&	...	&	...	&	...	&	...	&	...	&	0.03 	&	0.00 	\\
	Erinome	&	46	&	3.00E-06	&	1.6 	&	...	&	...	&	...	&	...	&	...	&	0.04 	&	0.00 	\\
	Euanthe	&	47	&	3.00E-06	&	1.5 	&	...	&	...	&	...	&	...	&	...	&	0.04 	&	0.00 	\\
	Eurydome	&	48	&	3.00E-06	&	1.5 	&	...	&	...	&	...	&	...	&	...	&	0.04 	&	0.00 	\\
	Hegemone	&	49	&	3.00E-06	&	1.5 	&	...	&	...	&	...	&	...	&	...	&	0.04 	&	0.00 	\\
	Cyllene	&	50	&	1.00E-06	&	1.0 	&	...	&	...	&	...	&	...	&	...	&	0.04 	&	0.00 	\\
	Euporie	&	51	&	1.00E-06	&	1.0 	&	...	&	...	&	...	&	...	&	...	&	0.03 	&	0.00 	\\
	Herse	&	52	&	1.00E-06	&	1.0 	&	...	&	...	&	...	&	...	&	...	&	0.04 	&	0.00 	\\
	Kale	&	53	&	1.00E-06	&	1.0 	&	...	&	...	&	...	&	...	&	...	&	0.04 	&	0.00 	\\
	Kallichore	&	54	&	1.00E-06	&	1.0 	&	...	&	...	&	...	&	...	&	...	&	0.04 	&	0.00 	\\
	Kore	&	55	&	1.00E-06	&	1.0 	&	...	&	...	&	...	&	...	&	...	&	0.04 	&	0.00 	\\
	Mneme	&	56	&	1.00E-06	&	1.0 	&	...	&	...	&	...	&	...	&	...	&	0.04 	&	0.00 	\\
	Orthosie	&	57	&	1.00E-06	&	1.0 	&	...	&	...	&	...	&	...	&	...	&	0.04 	&	0.00 	\\
	Pasithee	&	58	&	1.00E-06	&	1.0 	&	...	&	...	&	...	&	...	&	...	&	0.04 	&	0.00 	\\
	S/2000 J11	&	59	&	1.00E-06	&	1.0 	&	...	&	...	&	...	&	...	&	...	&	0.02 	&	0.00 	\\
	S/2003 J10	&	60	&	1.00E-06	&	1.0 	&	...	&	...	&	...	&	...	&	...	&	0.04 	&	0.00 	\\
	S/2003 J15	&	61	&	1.00E-06	&	1.0 	&	...	&	...	&	...	&	...	&	...	&	0.04 	&	0.00 	\\
	S/2003 J16	&	62	&	1.00E-06	&	1.0 	&	...	&	...	&	...	&	...	&	...	&	0.04 	&	0.00 	\\
	S/2003 J18	&	63	&	1.00E-06	&	1.0 	&	...	&	...	&	...	&	...	&	...	&	0.03 	&	0.00 	\\
	S/2003 J19	&	64	&	1.00E-06	&	1.0 	&	...	&	...	&	...	&	...	&	...	&	0.04 	&	0.00 	\\
	S/2003 J2	&	65	&	1.00E-06	&	1.0 	&	...	&	...	&	...	&	...	&	...	&	0.05 	&	0.00 	\\
	S/2003 J23	&	66	&	1.00E-06	&	1.0 	&	...	&	...	&	...	&	...	&	...	&	0.04 	&	0.00 	\\
	S/2003 J3	&	67	&	1.00E-06	&	1.0 	&	...	&	...	&	...	&	...	&	...	&	0.03 	&	0.00 	\\
	S/2003 J4	&	68	&	1.00E-06	&	1.0 	&	...	&	...	&	...	&	...	&	...	&	0.04 	&	0.00 	\\
	S/2010 J1	&	69	&	1.00E-06	&	1.0 	&	...	&	...	&	...	&	...	&	...	&	0.04 	&	0.00 	\\
	S/2010 J2	&	70	&	1.00E-06	&	1.0 	&	...	&	...	&	...	&	...	&	...	&	0.04 	&	0.00 	\\
	S/2011 J1	&	71	&	1.00E-06	&	1.0 	&	...	&	...	&	...	&	...	&	...	&	0.04 	&	0.00 	\\
	S/2011 J2	&	72	&	1.00E-06	&	1.0 	&	...	&	...	&	...	&	...	&	...	&	0.04 	&	0.00 	\\
	Sponde	&	73	&	1.00E-06	&	1.0 	&	...	&	...	&	...	&	...	&	...	&	0.04 	&	0.00 	\\
	Thelxinoe	&	74	&	1.00E-06	&	1.0 	&	...	&	...	&	...	&	...	&	...	&	0.04 	&	0.00 	\\
	S/2003 J12	&	75	&	1.00E-07	&	0.5 	&	...	&	...	&	...	&	...	&	...	&	0.03 	&	0.00 	\\
	S/2003 J9	&	76	&	1.00E-07	&	0.5 	&	...	&	...	&	...	&	...	&	...	&	0.04 	&	0.00 	\\
	Saturn	&	77	&	3.79E+07	&	58232.0 	&	9.537 	&	0.054 	&	2.49 	&	-21.06 	&	113.66 	&	0.00 	&	5979.49 	\\
	Titan	&	78	&	8.98E+03	&	2574.7 	&	...	&	...	&	...	&	...	&	...	&	0.00 	&	32.01 	\\
	Rhea	&	79	&	1.54E+02	&	764.3 	&	...	&	...	&	...	&	...	&	...	&	0.00 	&	1.85 	\\
	Iapetus	&	80	&	1.21E+02	&	735.6 	&	...	&	...	&	...	&	...	&	...	&	0.00 	&	1.50 	\\
	Dione	&	81	&	7.31E+01	&	561.7 	&	...	&	...	&	...	&	...	&	...	&	0.00 	&	1.19 	\\
	Tethys	&	82	&	4.12E+01	&	533.0 	&	...	&	...	&	...	&	...	&	...	&	0.00 	&	0.71 	\\
	Enceladus	&	83	&	7.20E+00	&	252.1 	&	...	&	...	&	...	&	...	&	...	&	0.00 	&	0.26 	\\
	Mimas	&	84	&	2.50E+00	&	198.2 	&	...	&	...	&	...	&	...	&	...	&	0.00 	&	0.12 	\\
	Phoebe	&	85	&	5.53E-01	&	106.5 	&	...	&	...	&	...	&	...	&	...	&	0.01 	&	0.05 	\\
	Hyperion	&	86	&	3.73E-01	&	135.0 	&	...	&	...	&	...	&	...	&	...	&	0.00 	&	0.03 	\\
	Janus	&	87	&	1.26E-01	&	89.5 	&	...	&	...	&	...	&	...	&	...	&	0.00 	&	0.01 	\\
	Epimetheus	&	88	&	3.51E-02	&	58.1 	&	...	&	...	&	...	&	...	&	...	&	0.00 	&	0.01 	\\
	Prometheus	&	89	&	1.07E-02	&	43.1 	&	...	&	...	&	...	&	...	&	...	&	0.00 	&	0.00 	\\
	Pandora	&	90	&	9.24E-03	&	40.7 	&	...	&	...	&	...	&	...	&	...	&	0.00 	&	0.00 	\\
	Siarnaq	&	91	&	2.60E-03	&	20.0 	&	...	&	...	&	...	&	...	&	...	&	0.02 	&	0.00 	\\
	Albiorix	&	92	&	1.40E-03	&	16.0 	&	...	&	...	&	...	&	...	&	...	&	0.01 	&	0.00 	\\
	Helene	&	93	&	7.60E-04	&	17.6 	&	...	&	...	&	...	&	...	&	...	&	0.00 	&	0.00 	\\
	Paaliaq	&	94	&	5.50E-04	&	11.0 	&	...	&	...	&	...	&	...	&	...	&	0.01 	&	0.00 	\\
	Atlas	&	95	&	4.40E-04	&	15.1 	&	...	&	...	&	...	&	...	&	...	&	0.00 	&	0.00 	\\
	Pan	&	96	&	3.30E-04	&	14.1 	&	...	&	...	&	...	&	...	&	...	&	0.00 	&	0.00 	\\
	Ymir	&	97	&	3.30E-04	&	9.0 	&	...	&	...	&	...	&	...	&	...	&	0.02 	&	0.00 	\\
	Telesto	&	98	&	2.70E-04	&	12.4 	&	...	&	...	&	...	&	...	&	...	&	0.00 	&	0.00 	\\
	Kiviuq	&	99	&	2.20E-04	&	8.0 	&	...	&	...	&	...	&	...	&	...	&	0.01 	&	0.00 	\\
	Tarvos	&	100	&	1.80E-04	&	7.5 	&	...	&	...	&	...	&	...	&	...	&	0.02 	&	0.00 	\\
	Calypso	&	101	&	1.70E-04	&	10.7 	&	...	&	...	&	...	&	...	&	...	&	0.00 	&	0.00 	\\
	Ijiraq	&	102	&	8.00E-05	&	6.0 	&	...	&	...	&	...	&	...	&	...	&	0.01 	&	0.00 	\\
	Erriapus	&	103	&	5.10E-05	&	5.0 	&	...	&	...	&	...	&	...	&	...	&	0.01 	&	0.00 	\\
	Narvi	&	104	&	2.30E-05	&	3.5 	&	...	&	...	&	...	&	...	&	...	&	0.02 	&	0.00 	\\
	Skathi	&	105	&	2.10E-05	&	4.0 	&	...	&	...	&	...	&	...	&	...	&	0.01 	&	0.00 	\\
	Mundilfari	&	106	&	1.40E-05	&	3.5 	&	...	&	...	&	...	&	...	&	...	&	0.02 	&	0.00 	\\
	Suttungr	&	107	&	1.40E-05	&	3.5 	&	...	&	...	&	...	&	...	&	...	&	0.02 	&	0.00 	\\
	Thrymr	&	108	&	1.40E-05	&	3.5 	&	...	&	...	&	...	&	...	&	...	&	0.02 	&	0.00 	\\
	Daphnis	&	109	&	5.20E-06	&	3.8 	&	...	&	...	&	...	&	...	&	...	&	0.00 	&	0.00 	\\
	Pallene	&	110	&	2.20E-06	&	2.5 	&	...	&	...	&	...	&	...	&	...	&	0.00 	&	0.00 	\\
	Methone	&	111	&	6.00E-07	&	1.6 	&	...	&	...	&	...	&	...	&	...	&	0.00 	&	0.00 	\\
	Polydeuces	&	112	&	3.00E-07	&	1.3 	&	...	&	...	&	...	&	...	&	...	&	0.00 	&	0.00 	\\
	Anthe	&	113	&	1.00E-07	&	0.9 	&	...	&	...	&	...	&	...	&	...	&	0.00 	&	0.00 	\\
	Aegaeon	&	114	&	4.00E-09	&	0.3 	&	...	&	...	&	...	&	...	&	...	&	0.00 	&	0.00 	\\
	Aegir	&	115	&	0.00E+00	&	3.0 	&	...	&	...	&	...	&	...	&	...	&	0.02 	&	0.00 	\\
	Bebhionn	&	116	&	0.00E+00	&	3.0 	&	...	&	...	&	...	&	...	&	...	&	0.01 	&	0.00 	\\
	Bergelmir	&	117	&	0.00E+00	&	3.0 	&	...	&	...	&	...	&	...	&	...	&	0.02 	&	0.00 	\\
	Bestla	&	118	&	0.00E+00	&	3.5 	&	...	&	...	&	...	&	...	&	...	&	0.02 	&	0.00 	\\
	Farbauti	&	119	&	0.00E+00	&	2.5 	&	...	&	...	&	...	&	...	&	...	&	0.02 	&	0.00 	\\
	Fenrir	&	120	&	0.00E+00	&	2.0 	&	...	&	...	&	...	&	...	&	...	&	0.02 	&	0.00 	\\
	Fornjot	&	121	&	0.00E+00	&	3.0 	&	...	&	...	&	...	&	...	&	...	&	0.02 	&	0.00 	\\
	Greip	&	122	&	0.00E+00	&	3.0 	&	...	&	...	&	...	&	...	&	...	&	0.02 	&	0.00 	\\
	Hati	&	123	&	0.00E+00	&	3.0 	&	...	&	...	&	...	&	...	&	...	&	0.02 	&	0.00 	\\
	Hyrrokkin	&	124	&	0.00E+00	&	3.0 	&	...	&	...	&	...	&	...	&	...	&	0.02 	&	0.00 	\\
	Jarnsaxa	&	125	&	0.00E+00	&	3.0 	&	...	&	...	&	...	&	...	&	...	&	0.02 	&	0.00 	\\
	Kari	&	126	&	0.00E+00	&	3.0 	&	...	&	...	&	...	&	...	&	...	&	0.02 	&	0.00 	\\
	Loge	&	127	&	0.00E+00	&	3.0 	&	...	&	...	&	...	&	...	&	...	&	0.02 	&	0.00 	\\
	S/2004 S12	&	128	&	0.00E+00	&	2.5 	&	...	&	...	&	...	&	...	&	...	&	0.02 	&	0.00 	\\
	S/2004 S13	&	129	&	0.00E+00	&	3.0 	&	...	&	...	&	...	&	...	&	...	&	0.02 	&	0.00 	\\
	S/2004 S17	&	130	&	0.00E+00	&	2.0 	&	...	&	...	&	...	&	...	&	...	&	0.02 	&	0.00 	\\
	S/2004 S7	&	131	&	0.00E+00	&	3.0 	&	...	&	...	&	...	&	...	&	...	&	0.02 	&	0.00 	\\
	S/2006 S1	&	132	&	0.00E+00	&	3.0 	&	...	&	...	&	...	&	...	&	...	&	0.02 	&	0.00 	\\
	S/2006 S3	&	133	&	0.00E+00	&	2.5 	&	...	&	...	&	...	&	...	&	...	&	0.02 	&	0.00 	\\
	S/2007 S2	&	134	&	0.00E+00	&	3.0 	&	...	&	...	&	...	&	...	&	...	&	0.01 	&	0.00 	\\
	S/2007 S3	&	135	&	0.00E+00	&	2.0 	&	...	&	...	&	...	&	...	&	...	&	0.02 	&	0.00 	\\
	Skoll	&	136	&	0.00E+00	&	3.0 	&	...	&	...	&	...	&	...	&	...	&	0.01 	&	0.00 	\\
	Surtur	&	137	&	0.00E+00	&	3.0 	&	...	&	...	&	...	&	...	&	...	&	0.02 	&	0.00 	\\
	Tarqeq	&	138	&	0.00E+00	&	3.0 	&	...	&	...	&	...	&	...	&	...	&	0.01 	&	0.00 	\\
	Uranus	&	139	&	5.79E+06	&	25362.0 	&	19.189 	&	0.047 	&	0.77 	&	96.94 	&	74.02 	&	0.00 	&	2097.18 	\\
	Titania	&	140	&	2.28E+02	&	788.9 	&	...	&	...	&	...	&	...	&	...	&	0.00 	&	2.66 	\\
	Oberon	&	141	&	1.92E+02	&	761.4 	&	...	&	...	&	...	&	...	&	...	&	0.00 	&	2.32 	\\
	Ariel	&	142	&	8.64E+01	&	578.9 	&	...	&	...	&	...	&	...	&	...	&	0.00 	&	1.37 	\\
	Umbriel	&	143	&	8.15E+01	&	584.7 	&	...	&	...	&	...	&	...	&	...	&	0.00 	&	1.28 	\\
	Miranda	&	144	&	4.40E+00	&	235.8 	&	...	&	...	&	...	&	...	&	...	&	0.00 	&	0.17 	\\
	Puck	&	145	&	1.93E-01	&	81.0 	&	...	&	...	&	...	&	...	&	...	&	0.00 	&	0.02 	\\
	Sycorax	&	146	&	1.80E-01	&	75.0 	&	...	&	...	&	...	&	...	&	...	&	0.00 	&	0.02 	\\
	Portia	&	147	&	1.12E-01	&	70.0 	&	...	&	...	&	...	&	...	&	...	&	0.00 	&	0.01 	\\
	Juliet	&	148	&	3.72E-02	&	53.0 	&	...	&	...	&	...	&	...	&	...	&	0.00 	&	0.01 	\\
	Belinda	&	149	&	2.38E-02	&	45.0 	&	...	&	...	&	...	&	...	&	...	&	0.00 	&	0.00 	\\
	Cressida	&	150	&	2.29E-02	&	41.0 	&	...	&	...	&	...	&	...	&	...	&	0.00 	&	0.01 	\\
	Caliban	&	151	&	2.00E-02	&	36.0 	&	...	&	...	&	...	&	...	&	...	&	0.00 	&	0.01 	\\
	Rosalind	&	152	&	1.70E-02	&	36.0 	&	...	&	...	&	...	&	...	&	...	&	0.00 	&	0.00 	\\
	Desdemona	&	153	&	1.19E-02	&	35.0 	&	...	&	...	&	...	&	...	&	...	&	0.00 	&	0.00 	\\
	Prospero	&	154	&	6.60E-03	&	25.0 	&	...	&	...	&	...	&	...	&	...	&	0.01 	&	0.00 	\\
	Bianca	&	155	&	6.20E-03	&	27.0 	&	...	&	...	&	...	&	...	&	...	&	0.00 	&	0.00 	\\
	Setebos	&	156	&	5.80E-03	&	24.0 	&	...	&	...	&	...	&	...	&	...	&	0.01 	&	0.00 	\\
	Ophelia	&	157	&	3.60E-03	&	21.4 	&	...	&	...	&	...	&	...	&	...	&	0.00 	&	0.00 	\\
	Cordelia	&	158	&	3.00E-03	&	20.1 	&	...	&	...	&	...	&	...	&	...	&	0.00 	&	0.00 	\\
	Stephano	&	159	&	1.70E-03	&	16.0 	&	...	&	...	&	...	&	...	&	...	&	0.00 	&	0.00 	\\
	Perdita	&	160	&	1.20E-03	&	13.0 	&	...	&	...	&	...	&	...	&	...	&	0.00 	&	0.00 	\\
	Mab	&	161	&	6.00E-04	&	12.0 	&	...	&	...	&	...	&	...	&	...	&	0.00 	&	0.00 	\\
	Francisco	&	162	&	5.60E-04	&	11.0 	&	...	&	...	&	...	&	...	&	...	&	0.00 	&	0.00 	\\
	Ferdinand	&	163	&	4.20E-04	&	10.0 	&	...	&	...	&	...	&	...	&	...	&	0.01 	&	0.00 	\\
	Margaret	&	164	&	4.20E-04	&	10.0 	&	...	&	...	&	...	&	...	&	...	&	0.01 	&	0.00 	\\
	Trinculo	&	165	&	3.10E-04	&	9.0 	&	...	&	...	&	...	&	...	&	...	&	0.00 	&	0.00 	\\
	Cupid	&	166	&	2.00E-04	&	9.0 	&	...	&	...	&	...	&	...	&	...	&	0.00 	&	0.00 	\\
	Neptune	&	167	&	6.84E+06	&	24622.0 	&	30.070 	&	0.009 	&	1.77 	&	-86.82 	&	131.78 	&	0.00 	&	2548.39 	\\
	Triton	&	168	&	1.43E+03	&	1353.4 	&	...	&	...	&	...	&	...	&	...	&	0.00 	&	9.68 	\\
	Proteus	&	169	&	3.36E+00	&	210.0 	&	...	&	...	&	...	&	...	&	...	&	0.00 	&	0.15 	\\
	Nereid	&	170	&	2.06E+00	&	170.0 	&	...	&	...	&	...	&	...	&	...	&	0.00 	&	0.11 	\\
	Larissa	&	171	&	3.30E-01	&	97.0 	&	...	&	...	&	...	&	...	&	...	&	0.00 	&	0.03 	\\
	Galatea	&	172	&	2.50E-01	&	88.0 	&	...	&	...	&	...	&	...	&	...	&	0.00 	&	0.03 	\\
	Despina	&	173	&	1.40E-01	&	75.0 	&	...	&	...	&	...	&	...	&	...	&	0.00 	&	0.02 	\\
	Thalassa	&	174	&	2.50E-02	&	41.0 	&	...	&	...	&	...	&	...	&	...	&	0.00 	&	0.01 	\\
	Naiad	&	175	&	1.30E-02	&	33.0 	&	...	&	...	&	...	&	...	&	...	&	0.00 	&	0.00 	\\
	Halimede	&	176	&	1.20E-02	&	31.0 	&	...	&	...	&	...	&	...	&	...	&	0.00 	&	0.00 	\\
	Neso	&	177	&	1.10E-02	&	30.0 	&	...	&	...	&	...	&	...	&	...	&	0.01 	&	0.00 	\\
	Sao	&	178	&	4.50E-03	&	22.0 	&	...	&	...	&	...	&	...	&	...	&	0.01 	&	0.00 	\\
	Laomedeia	&	179	&	3.90E-03	&	21.0 	&	...	&	...	&	...	&	...	&	...	&	0.01 	&	0.00 	\\
	Psamathe	&	180	&	3.30E-03	&	20.0 	&	...	&	...	&	...	&	...	&	...	&	0.01 	&	0.00 	\\
	S/2004 N1	&	181	&	3.00E-04	&	9.0 	&	...	&	...	&	...	&	...	&	...	&	0.00 	&	0.00 	\\
	Pluto	&	182	&	8.70E+02	&	1188.3 	&	39.482 	&	0.249 	&	17.14 	&	113.76 	&	110.30 	&	0.00 	&	6.72 	\\
	Charon	&	183	&	1.02E+02	&	603.6 	&	...	&	...	&	...	&	...	&	...	&	0.00 	&	1.56 	\\
	Hydra	&	184	&	6.50E-03	&	30.5 	&	...	&	...	&	...	&	...	&	...	&	0.00 	&	0.00 	\\
	Nix	&	185	&	1.30E-03	&	23.0 	&	...	&	...	&	...	&	...	&	...	&	0.00 	&	0.00 	\\
	Kerberos	&	186	&	1.10E-03	&	14.0 	&	...	&	...	&	...	&	...	&	...	&	0.00 	&	0.00 	\\
	Styx	&	187	&	0.00E+00	&	10.0 	&	...	&	...	&	...	&	...	&	...	&	0.00 	&	0.00 	\\
	Ceres	&	188	&	6.26E+01	&	469.7 	&	2.766 	&	0.078 	&	10.59 	&	-6.53 	&	80.27 	&	0.00 	&	1.22 	\\
	Pallas	&	189	&	1.43E+01	&	272.5 	&	2.774 	&	0.230 	&	34.90 	&	137.52 	&	172.92 	&	0.00 	&	0.48 	\\
	Vesta	&	190	&	1.73E+01	&	262.7 	&	2.362 	&	0.088 	&	7.14 	&	47.21 	&	103.81 	&	0.00 	&	0.60 	\\
	Hygiea	&	191	&	7.00E+00	&	203.6 	&	3.142 	&	0.112 	&	3.83 	&	29.26 	&	283.19 	&	0.00 	&	0.32 	\\
    Interamnia	&	192	&	5.00E+00	&	153.2 	&	3.056 	&	0.155 	&	17.31 	&	-185.44 	&	280.28 	&	0.00 	&	0.30 	\\
    Psyche	&	193	&	1.53E+00	&	113.0 	&	2.924 	&	0.134 	&	3.10 	&	79.12 	&	150.03 	&	0.00 	&	0.12 	\\
    Kalliope	&	194	&	4.91E-01	&	83.8 	&	2.909 	&	0.098 	&	13.70 	&	291.55 	&	66.00 	&	0.00 	&	0.05 	\\
    Camilla	&	195	&	7.48E-01	&	105.2 	&	3.489 	&	0.065 	&	10.00 	&	133.37 	&	172.58 	&	0.00 	&	0.07 	\\
	\enddata
	\tablecomments{(1) $GM$, no satellites provides error except Styx.
		(2) $R_\mathrm{L}$ is the radius of the Objects.
		(3) $a$ and $e$ are the semi-major axis and eccentricity, respectively.
		(4) $I$, $\omega$ and $\Omega$ are inclination, preihelion, and longitude of ascending node, respectively.
		(5) $R_\mathrm{sp}$: 0 denotes $< 0.005$ except the sun, planets, pluto and the Moon.
		(6) $\alpha_{\mathrm{max}}$: 0.00 denotes $< 0.005$.}
	\tablenotetext{\dagger}{The orbital elements are from the Earth-Moon barycenter.}
\end{deluxetable}
	
\end{CJK*}
\end{document}